\documentclass[aps,groupedaddress,prd,nofootinbib,twocolumn]{revtex4}

\usepackage{amsmath,amssymb,bm}
\usepackage{graphicx}
\usepackage{epstopdf}
\usepackage{amsfonts}
\usepackage{amssymb}
\usepackage{amsbsy}
\usepackage{amsmath}
\usepackage{latexsym}
\usepackage{bbm}
\usepackage{mathtools}
\usepackage{sansmath}
\usepackage{mathrsfs} 
\usepackage{subfigure}
\usepackage{lipsum}
\usepackage{float}
\usepackage{color}
\usepackage{enumitem}
\usepackage{comment}
\usepackage{csquotes}								
\usepackage{physics}
\usepackage{accents}
\usepackage{pgfplots}
\usepackage{hyperref}
\usepackage{accents}
\usepackage{xcolor}
\hypersetup{
    colorlinks,
    linkcolor={blue!100},
    citecolor={red!100},
    urlcolor={blue!80!black}
}

\usepackage[plain]{algorithm}
\usepackage{algpseudocode}

\newcommand{\1}{\hat{I}}
\newcommand{\CQ}{\text{\tiny{CQ}}}
\newcommand{\QQ}{\text{\tiny{QQ}}}
\newcommand{\CC}{\text{\tiny{CC}}}
\newcommand{\PS}{\text{\tiny{PS}}}
\newcommand{\eff}{\text{\tiny{eff}}}

\newcommand{\zero}{(0)}

\newcommand{\one}{(1)}
\newcommand{\VN}{\text{\tiny{VN}}}
\newcommand{\lin}{\text{\tiny{LIN}}}
\newcommand{\T}{\text{\tiny{T}}}

\newcommand{\INT}[1]{\underaccent{\tilde}{#1}}

\newcommand{\n}{n}
\newcommand{\m}{\mu}

\setlist[itemize]{noitemsep}
\setlist[enumerate]{noitemsep}

\def\be{\begin{equation}}
\def\ee{\end{equation}}
\def\bea {\begin{eqnarray}}
\def\eea {\end{eqnarray}}
\def\nn {\nonumber}

\newcommand{\frob}[1]{\Vert{#1}\Vert_{\text{F}}} 
\newcommand{\frobinner}[2]{\langle{#1},{#2}\rangle_{\text{F}}}

\newcommand{\di}{\partial}

\begin{document}
 
\title{Motivating semiclassical gravity: a classical-quantum approximation \\
for bipartite quantum systems}

\author{Viqar Husain} \email{vhusain@unb.ca} 
\author{Irfan Javed} \email{i.javed@unb.ca}
\author{Sanjeev S.\ Seahra} \email{ssseahra@unb.ca}
\author{Nomaan X} \email{nomaan.math@unb.ca}

\affiliation{Department of Mathematics and Statistics, University of New Brunswick, Fredericton, NB, Canada E3B 5A3}
 
\begin{abstract}

We  derive a ``classical-quantum'' approximation scheme for a broad class of bipartite quantum systems from fully quantum dynamics. In this approximation, one subsystem evolves via classical equations of motion with quantum corrections, and the other subsystem evolves quantum mechanically with equations of motion informed by the evolving classical degrees of freedom.  Using perturbation theory, we derive an estimate for the growth rate of entanglement of the subsystems and deduce a ``scrambling time''---the time required for the subsystems to become significantly entangled from an initial product state.  We argue that a necessary condition for the validity of the classical-quantum approximation is consistency of initial data with the generalized Bohr correspondence principle.  We illustrate the general formalism by numerically studying the fully quantum, fully classical, and classical-quantum dynamics of a system of two oscillators with nonlinear coupling. This system exhibits parametric resonance, and we show that quantum effects quench parametric resonance at late times.  Lastly, we present a curious late-time scaling relation between the average value of the von Neumann entanglement of the interacting oscillator system and its total energy: $S\sim 2/3 \ln E$.

\end{abstract}

\maketitle

\tableofcontents

\section{Introduction}\label{sec:introduction}

At present, there is no consensus on a quantum theory of gravity, not even an approach to one, with several ideas in circulation; see, e.g.,  \cite{Carlip:2015asa,Loll:2019rdj,Aastrup:2012jj,Bodendorfer:2016uat} for recent reviews. If  quantum gravity is assumed to have the structure of usual quantum mechanics, then its Hilbert space would be a tensor product of the Hilbert spaces of gravity and matter. As such, it may be thought of as a ``bipartite" system. The question that arises is whether there are viable approximations of the unknown theory of quantum gravity where matter is quantum and spacetime is classical. Two such approximations are known, quantum field theory (QFT) in curved spacetime  and the semiclassical Einstein equation. It is, however, not clear how these approximations might arise from a theory of quantum gravity or what their domains of validity are. 

The study of QFT on fixed curved spacetimes has been a subject of interest for decades. The area is a natural outgrowth of QFT on flat spacetime with pioneering applications to black holes and cosmology \cite{Parker:1969aa,DEWITT1975295,birrell_davies_1982,fulling_1989,Ford:1997hb}. In this paradigm,  equations of motion are of the form 
\begin{subequations}\label{eq:semiclassical system 1}
    \begin{align}
        G_{\alpha\beta}(g) & = 0, \label{eq:Einstein equation 1} \\
        i \,\di_t |\psi\rangle & = \hat{H}_\psi(g) |\psi\rangle, \label{eq:quantum EOM 1}
     \end{align}
\end{subequations}
where (\ref{eq:Einstein equation 1}) is the vacuum Einstein equation and (\ref{eq:quantum EOM 1}) is the Schr\"odinger picture functional evolution equation for the state vector $|\psi\rangle$ of the quantum field on the curved background $g$.\footnote{In the more utilized Heisenberg picture, (\ref{eq:quantum EOM 1}) would be replaced by the Heisenberg equation of motion $d\hat{O}/dt = i[\hat{H}_\psi(g),\hat{O}] + \di_t \hat{O}$ for an arbitrary operator $\hat{O}$.} The matter Hamiltonian operator $\hat{H}_\psi(g)$ is parametrized by the classical metric tensor $g$; therefore, the first equation is solved for $g$ first, followed by the second for $|\psi\rangle$. (The usual textbook Heisenberg picture procedure involves solving the free field equation on a given background and quantizing the mode expansion to construct the Hamiltonian operator.) 

In this approach, it is apparent that the quantum field has no dynamical effect on spacetime, i.e., no ``backreaction" \cite{Brout:1995aa}. The original Hawking radiation derivation \cite{Hawking:1975vcx} relied on these equations, and the calculation of the growth of primordial perturbations during inflation (see, e.g., \cite{Baumann:2009ds}) uses a modified version of (\ref{eq:semiclassical system 1}), which adds a classical homogeneous scalar stress-energy tensor to the right-hand side of (\ref{eq:Einstein equation 1}).

From the perspective of quantum gravity, both matter and spacetime geometry are expected to be quantum fields, and Eqs. (\ref{eq:semiclassical system 1}) are widely regarded as an emergent approximation to the fully quantum equations of motion of a bipartite system with Hilbert space ${\cal H}_\text{gravity}\otimes {\cal H}_\text{matter}$.

It is instructive to recall how similar approximations are used in nongravitational systems.  For example, in quantum chemistry, energy eigenstates of molecules are determined by considering a subsystem of atomic nuclei coupled to a subsystem of electrons. Since nuclei are much more massive than electrons, a common approximation is to consider the electrons as quantum particles moving in the electric field of atomic nuclei, which are modeled as very slowly moving classical objects. ``Heavy'' nuclei and the ``light'' electrons may be viewed, respectively, as analogous to spacetime geometry and matter fields in (\ref{eq:semiclassical system 1}). Another example is the dynamics of ultracold neutrons moving in Earth's gravitational field \cite{Abele:2009dw}, where quantum mechanics describes neutrons moving in the effectively nondynamical Newtonian gravitational field of the Earth. 

Both the molecular and ultracold neutron calculations rely on the Born-Oppenheimer approximation \cite{https://doi.org/10.1002/andp.19273892002}, where the fully quantum dynamical equations are expanded in terms of large parameters (the nuclear and Earth masses). Similarly, several authors have used the Born-Oppenheimer approximation to derive equations (\ref{eq:semiclassical system 1}) from the Wheeler-DeWitt equation in canonical quantum gravity \cite{lapchinski1979canonical,PhysRevD.44.1067,TPadmanabhan_1989,singh1989notes} with the relevant large expansion parameter being the Planck mass $M_\text{Pl}$.  

Despite this theoretical grounding, there are serious deficiencies in the structure of (\ref{eq:semiclassical system 1}). Among the more important ones is the lack of conservation of energy. Since the metric appears as an external source in (\ref{eq:quantum EOM 1}), it can drive particle creation in the quantum field $|\psi\rangle$ without a commensurate reduction in the energy stored in the spacetime geometry. The same is true for the Born-Oppenheimer approximation as applied to molecules or ultracold neutrons falling near Earth's surface, but in such cases, it is physically sensible to ignore the (negligible) energy transfer between light and heavy objects. However, for certain gravitational calculations, it is important to follow energy exchange between subsystems. The prime example is the Hawking effect, where the energy carried away by the quantum fields to future null infinity reduces the mass of a black hole in the famous ``black hole evaporation" process; this becomes especially important in the late stages of Hawking evaporation when the rate of mass loss of a black hole is conjectured to rapidly increase, a theoretical scenario where Eqs. (\ref{eq:semiclassical system 1}) clearly fail; there is no analogous prediction in quantum chemistry or for neutrons in Earth's gravitational field. 

These considerations suggest a modification of Eqs.\ (\ref{eq:semiclassical system 1}) to allow the quantum state of matter to have a dynamical effect on spacetime geometry:
\begin{subequations}\label{eq:semiclassical system 2}
    \begin{align}
        G_{\alpha\beta}(g)   & = M_\text{Pl}^{-2} \langle \psi| \hat{T}_{\alpha\beta}(g)|\psi \rangle, \label{eq:Einstein equation 2} \\
        i \,\di_t |\psi\rangle & = \hat{H}_\psi(g) |\psi\rangle. \label{eq:quantum EOM 2}
     \end{align}
\end{subequations}
The new feature is the expectation value of the stress-energy tensor in (\ref{eq:Einstein equation 2}); this makes the coupled set of equations nonlinear in the matter state. The first of these equations was proposed by Moller and Rosenfeld \cite{Lichnerowicz:1962gnh,Rosenfeld}; the second was added in a discussion of whether gravity should be quantized based on the validity of the pair of equations \cite{Page:1981aj}. 

While equations (\ref{eq:semiclassical system 2}) have obvious intuitive appeal, they also present many conceptual and practical difficulties. One issue becomes apparent when one tries to solve the equations perturbatively by expanding the metric and state vector in powers of $M_\text{Pl}^{-2}$.  While such an expansion reproduces (\ref{eq:semiclassical system 1}) at zeroth order, the next-to-leading order corrections to the Einstein tensor $G_{\alpha\beta}^{(1)}$ obey
\begin{equation}\label{eq:corrected Einstein equation}
    G_{\alpha\beta}^{(1)} = M_\text{Pl}^{-2} \langle \psi^{(0)} | \hat{T}_{\alpha\beta}(g^{(0)})|\psi^{(0)} \rangle.
\end{equation}
Here, $g^{(0)}$ and $|\psi^{(0)}\rangle$ are solutions to the zeroth order equations. For the simple situation where the zeroth order solution represents a free quantum field on Minkowski spacetime, it is easy to see that the right-hand side of equation (\ref{eq:corrected Einstein equation}) is formally infinite due to zero point fluctuations. Attempting to remedy this divergence by imposing a Planck scale cutoff generates a huge effective cosmological constant that is inconsistent with observations.  

In order to rescue the theory, one can add large and/or infinite counterterms to the Einstein equations or attempt to regulate $\langle T_{\alpha\beta} \rangle$ via point-splitting or other techniques. However, these methods often do not generalize to situations where the zeroth order spacetime solution is not flat.  For example, when the point-splitting regularization procedure is applied to $\langle \psi| \hat{T}_{\alpha\beta}(g)|\psi \rangle$ in a curved spacetime, the result generally involves higher derivatives of the metric and therefore leads to higher order theories of gravity that are prone to various pathologies \cite{Ford:1997hb}.  

Additional computational difficulties are encountered in the perturbative approach to (\ref{eq:semiclassical system 2}) if higher order terms are included in the expansion due to corrections to the state vector, mode function inner products, and the metric. There have been other (nonperturbative) critiques of (\ref{eq:semiclassical system 2}): for example, if the state $|\psi\rangle$ is a superposition of quantum states, representing, for instance, masses localized in different spatial locations, then according to (\ref{eq:Einstein equation 2}), the gravitational field corresponds to a weighted sum of the locations of the particle in the superposition, a feature that predicts a rapid change in spacetime geometry if the quantum superposition is measured  \cite{Page:1981aj}. Also, the nonlinearity of  (\ref{eq:semiclassical system 2}) in the quantum state $|\psi\rangle$ means that the principle of quantum superposition is lost. 

Given these objections, we see if Eqs.\ (\ref{eq:semiclassical system 2}) are derivable as controlled approximations from first principles, similar to the Born-Oppenheimer derivation of (\ref{eq:semiclassical system 1}) from the Wheeler-DeWitt equation. There is some reason for optimism as equations (\ref{eq:semiclassical system 1}) are the $M_\text{Pl} \rightarrow \infty$ limit of (\ref{eq:semiclassical system 2}) as noted above. This raises the possibility that (\ref{eq:semiclassical system 2}) could emerge from the Born-Oppenheimer approximation if higher order terms in $M_\text{Pl}^{-2}$ were retained. If this were to work for Wheeler-DeWitt quantum gravity, it should also work for simpler quantum systems, something which leads to the following question: does application of the higher order Born-Oppenheimer approximation applied to any bipartite quantum system with a ``heavy" and a ``light" component lead to something similar to equations (\ref{eq:semiclassical system 2})?

\citet{singh1989notes} investigated this question for a simple bipartite quantum mechanical system, finding out that the answer was a qualified ``no." Specifically, they showed that the only way to recover the analogue of (\ref{eq:semiclassical system 2}) using the Born-Oppenheimer approximation was to replace certain quantities in the classical equations of motion of the heavy degree of freedom with their expectation values in an \emph{ad hoc} manner; an extension of this analysis to the Wheeler-DeWitt equation gives a similar result for gravity. \citet{PhysRevD.44.1067} also studied the higher order Born-Oppenheimer approximation for the Wheeler-DeWitt equation. As in \cite{singh1989notes}, corrections to equations (\ref{eq:semiclassical system 1}) are not of the form of (\ref{eq:semiclassical system 2}). \citet{PhysRevD.45.2044} applied the same higher order Born-Oppenheimer formalism to scalar quantum electrodynamics to study the pair production of particles in a semiclassical electric field, including backreaction effects.

In this paper, we derive the analogue of equation (\ref{eq:semiclassical system 2}) for a broad class of bipartite quantum mechanical systems using assumptions that are related to, but logically distinct from, Born-Oppenheimer. More specifically, our ``classical-quantum'' approximation relies on the following hypotheses.
\begin{enumerate}
    \item The bipartite quantum system starts in a product state.
    \item One of the subsystems is in a semiclassical state.
    \item The coupling between the subsystems is weak.
\end{enumerate}
Under these conditions, we find that approximate equations of motion qualitatively similar to (\ref{eq:semiclassical system 1}) hold for a \emph{finite period of time}. That is, for some interval after $t=0$, the equations of motion of subsystem 1 are classical and sourced by expectation values of the quantum state of subsystem 2, and the classical configuration variables of subsystem 1 appear in the quantum equations of motion of subsystem 2 as time-dependent functions.

One of the key features of this approximation is that it is possible to derive an effective Hamiltonian for each subsystem from the equations of motion of the reduced density matrices of the fully quantum formalism (referred to here as ``quantum-quantum"). In our derivation, expectation values appear naturally in the approximate equation of motion of subsystem 1 and are therefore nonlinear in the quantum state of subsystem 2.  

The main assumption in our derivation of the classical-quantum approximation is the choice of initial state of the bipartite quantum system: this is a product state, $|\varphi_1 \rangle \otimes |\varphi_2 \rangle$, where $|\varphi_1 \rangle$ is a semiclassical state of subsystem 1 in isolation and $|\varphi_2 \rangle$  is any state of subsystem 2 in isolation. Under evolution with a Hamiltonian of the form $\hat{H}_{1} + \hat{H}_{2} + \hat{H}_\text{int}$, such a state becomes entangled and $|\varphi_1 \rangle$ begins to lose its semiclassicality. This suggests two indepedendent time scales:  the characteristic timescale for the growth of entanglement (which we call the ``scrambling time") and the characteristic timescale for the decline of semiclassicality for subsystem 1 (referred to as the ``Ehrenfest time" in some of the literature). 

We derive bounds on the scrambling time using perturbation theory. This provides a time interval estimate  for the validity of the classical-quantum approximation. We also show that the differences between trajectories of observables in the quantum-quantum and classical-quantum systems is correlated to the deviation of subsystem 1 expectation values from their classical counterparts. That is, subsystem 1 must adhere closely to the generalized Bohr correspondence principle (in which classical and quantum predictions agree) for the classical-quantum approximation to be valid. 

To illustrate the behaviour of the classical-quantum approximation, we numerically study a system of two harmonic oscillators with nonlinear coupling.  The results of the classical-quantum approximation are compared to the quantum-quantum and fully classical system trajectories.  This analysis builds on recent work by some of us on a hybrid classical-quantum treatment of a tripartite system of an oscillator coupled to spins \cite{Husain:2022kaz}. We find that there is general good agreement between the classical-quantum and quantum-quantum calculations up to the scrambling time. An interesting classical feature of the coupled oscillator system is that it exhibits parametric resonance for certain interaction potentials and parameter choices. This effect is associated with the efficient energy exchange between the oscillators; we find that it can persist in both the classical-quantum and quantum-quantum calculations on timescales shorter than the scrambling time for certain choices of parameters. (We are unaware of any other numerical simulations of the fully quantum dynamics of an oscillatory system exhibiting parametric resonance in the literature.) We also note that particle creation driven by parametric resonance phenomenon may play a role in the preheating phases of cosmological evolution after inflation \cite{Kofman:1997yn}; other authors have recently considered the semiclassical Einstein equation in this context \cite{rathorea2022validity}. Finally, we investigate the long time evolution of the von Neumann entropy using numeric simulations in the quantum-quantum case. We find an approximate (and slightly mysterious) scaling relation between the long term average of the the entanglement entropy and the total energy of the system: $S_{\VN}\sim \frac{2}{3} \ln E$.

The layout of the paper is as follows: in \S\ref{sec:general framework}, we introduce the bipartite system and relevant measures of entanglement;  in \S\ref{sec:approximations of quantum dynamics}, we describe several approximation schemes for the quantum-quantum equations of motion, derive a ``classical-quantum" approximation, and show that it conserves probability and energy;  in \S\ref{sec:perturbative}, we use perturbation theory to investigate the regime of validity of classical-quantum approximation; in \S\ref{sec:nonlinearly coupled oscillators}, we apply the approximations of previous section to the specific case of two harmonic oscillators with a nonlinear coupling; in \S\ref{sec:conclusions}, we present our conclusions and discuss some consequences for gravitational systems. The appendices contain several technical results and reviews of useful formulae.  

\section{General Framework}\label{sec:general framework}

\subsection{Hamiltonian}\label{sec:general Hamiltonian}

Consider a system with two coupled degrees of freedom (each of which we refer to as a ``subsystem''). The classical Hamiltonian is
\begin{multline}\label{eq:full classical Hamiltonian}
	\mathcal{H}(q_{1},p_{1},q_{2},p_{2}) = \mathcal{H}_{1}(q_{1},p_{1})+  \mathcal{H}_{2}(q_{2},p_{2}) \\ + \lambda \mathcal{V}_{1}(q_{1},p_{1})\mathcal{V}_{2}(q_{2},p_{2}).
\end{multline}
Here, $(q_{1},p_{1})$ and $(q_{2},p_{2})$ are canonical phase space coordinates for each subsystem, $\mathcal{V}_{1}$ and $\mathcal{V}_{2}$ govern the interaction between the two subsystems, and $\lambda \ge 0$ is a dimensionless coupling constant.  Note that the classical Hamiltonian is assumed to carry no explicit dependance on time.

The quantum version of this system is bipartite with Hilbert space $\mathscr{H} = \mathscr{H}_{1} \otimes \mathscr{H}_{2}$. Typically, we would expect the dimension of each of the subspaces to be infinite. For the purposes of numerical calculation, however, we will often consider finite truncations of each subspace defined by restricted basis sets such that\footnote{This kind of truncation is common in quantum chemistry, for example.}
\begin{equation}
	\text{dim}(\mathscr{H}_{1}) = d_{1}, \quad \text{dim}(\mathscr{H}_{2}) = d_{2}.
\end{equation}
The system Hamiltonian is of the form
\begin{equation}
	\hat{H} = \hat{H}_{1} \otimes \1_{2} + \1_{1} \otimes \hat{H}_{2} + \lambda \, \hat{V}_{1} \otimes \hat{V}_{2},
\end{equation}
where $\1_{1}$ and $\1_{2}$ are the identity operators on each subspace, $\hat{H}_{1}$ is the operator version of $\mathcal{H}_{1}$, $\hat{V}_{1}$ is the operator version of $\mathcal{V}_{1}$, etc.  All the operators are assumed to be Hermitian.

\subsection{Fully quantum-quantum (QQ) dynamics}

We assume that the energy eigenvalue problem for each of the subsystem Hamiltonians is easily solvable:
\begin{equation}
	\hat{H}_{1} |\n \rangle = E_{\n} | \n \rangle, \quad \hat{H}_{2} |\m \rangle = E_{\m} |\m\rangle.
\end{equation}
Note that our notation is such that the indices $\n,\n'=0,1,2\ldots d_{1}-1$ will always be associated with subsystem 1 and the indices $\m,\m'=0,1,2\ldots d_{2}-1$ will always be associated with subsystem 2. Then, a complete basis for $\mathcal{H}$ is given by the $d_{1}\times d_{2}$ basis states
\begin{equation}
	|\n,\m\rangle = |\n\rangle \otimes |\m\rangle, \quad \langle \n',\m' | \n,\m \rangle = \delta_{\n\n'} \delta_{\m\m'}.
\end{equation}
We write the full state vector of the system in the Schr\"odinger representation as
\begin{equation}\label{eq:QQ state expansion}
	|\psi(t)\rangle = \sum_{\n\m} z_{\n,\m}(t)|\n,\m\rangle.
\end{equation}
Inserting this into the Schr\"odinger equation
\begin{equation}
	i \di_{t} |\psi(t) \rangle = \hat{H} |\psi(t)\rangle
\end{equation}
yields
\begin{equation}
	i \di_{t} z_{\n',\m'}(t) = \sum_{\n\m} \langle \n',\m' | \hat{H} | \n,\m \rangle z_{\n,\m}(t).
\end{equation}
We organize the expansion coefficients into a matrix:
\begin{gather}\nonumber
	Z(t) = \begin{pmatrix*}[c]
	 z_{0,0}(t) & z_{0,1}(t) & \cdots & z_{0,d_{2}-1}(t) \\
	 z_{1,0}(t) & z_{1,1}(t) & \cdots & z_{1,d_{2}-1}(t) \\
	 \vdots & \vdots & \ddots & \vdots \\
	 z_{d_{1}-1,0}(t) & z_{d_{1}-1,1}(t) & \cdots & z_{d_{1}-1,d_{2}-1}(t) \\
	\end{pmatrix*}.
\end{gather}
Inserting the expansion (\ref{eq:QQ state expansion}) into the Schrodinger equation yields:
\begin{align}\label{eq:QQ EOM general}
	i \di_{t} Z = H_{1} Z + ZH_{2} + \lambda V_{1} Z V_{2}^{\T} ,
\end{align}
where the $H_{1}$, $H_{2}$, $V_{1}$, and $V_{2}$ square matrices have entries
\begin{align}\nonumber
	(H_{1})_{\n'\n} & = \langle \n' | \hat{H}_{1} | \n \rangle, & (V_{1})_{\n'\n} & = \langle \n' | \hat{V}_{1} | \n \rangle, \\
	(H_{2})_{\m'\m} & = \langle \m' | \hat{H}_{2} | \m \rangle, & (V_{2})_{\m'\m} & = \langle \m' | \hat{V}_{2} | \m \rangle.
\end{align}
Note that since $|\n\rangle$ and $|\m\rangle$ are eigenvectors of $\hat{H}_{1}$ and $\hat{H}_{2}$, respectively, $H_{1}$ and $H_{2}$ are real diagonal matrices. Furthermore, $V_{1}$ and $V_{2}$ are Hermitian matrices due to the self-adjointness of the associated operators. Equation (\ref{eq:QQ EOM general}) represents $d_{1} \times d_{2}$ complex linear differential equations whose solutions completely specify the quantum dynamics of the system. We also note that the normalization condition $\langle \psi(t) | \psi(t) \rangle = 1$ implies that $Z$ satisfies
\begin{equation}
	1 = \Tr[Z^{\dagger}(t)Z(t)] = \frob{Z(t)}^{2},
\end{equation}
where $\frob{\cdots}$ is the Frobenius norm for complex rectangular matrices defined in Appendix \ref{sec:Frobenius}. It is easy to confirm that the conservation of $\frob{Z(t)}$ is guaranteed by the equation of motion (\ref{eq:QQ EOM general}).

Using the expansion (\ref{eq:QQ state expansion}), we can find the expectation values of operators that only act on either of the Hilbert subspaces of system 1 or system 2:
\begin{subequations}
\begin{align}\nonumber
	\langle \hat{A}_{1} \otimes \1_{2} \rangle & = \sum_{\n\n'\m} z^{*}_{\n'\m} \langle \n' | \hat{A}_{1}|\n \rangle z_{\n\m} \\ & = \Tr (\rho_{1}A_{1}), \\ \nonumber
	\langle  \1_{1} \otimes \hat{A}_{2}\rangle & = \sum_{\n\m'\m} z^{*}_{\n\m'} \langle \m' | \hat{A}_{2}|\m \rangle z_{\n\m} \\ & = \Tr (\rho_{2}A_{2}),
\end{align}
\end{subequations}
where $A_{1}$ and $A_{2}$ are the matrix representations of each operator in the eigenbasis of $\hat{H}_{1}$ and $\hat{H}_{2}$, respectively; $\1_{1}$ and $\1_{2}$ are identity operators; and $\rho_{1}$ and $\rho_{2}$ are the reduced density matrices
\begin{equation}\label{eq:reduced density}
	\rho_{1}(t) = Z(t)Z^{\dagger}(t), \quad \rho_{2}(t) = Z^{\T}(t)Z^{*}(t).
\end{equation}
With (\ref{eq:QQ EOM general}), it is possible to write down evolution equations for both reduced density matrices. We find
\begin{subequations}\label{eq:QQ density matrix evolution general}
\begin{align}
i\di_{t}{\rho}_{1} & = [H_{1},\rho_{1}]  + \lambda  [V_{1},Z V_{2}^{\T}Z^{\dagger}], \\
i\di_{t}{\rho}_{2} & = [H_{2},\rho_{2}]  + \lambda  [V_{2},Z^{\T}V_{1}^{\T}Z^{*}].
\end{align}
\end{subequations}
Note that equations (\ref{eq:QQ density matrix evolution general}) do not represent a closed system of differential equations for $\rho_{1}$ and $\rho_{2}$; i.e., to solve these for $\rho_{1}$ and $\rho_{2}$, one must also solve (\ref{eq:QQ EOM general}) for $Z$.

Before moving on, we note that eigenstates of the full quantum Hamiltonian are defined by
\begin{equation}
	\hat{H} |\psi_{j}\rangle = E_{j} | \psi_{j} \rangle.
\end{equation}
We write as above
\begin{equation}
	|\psi_{j} \rangle = \sum_{\n\m} z_{\n\m,j} |
 \n,\m\rangle, \quad |\n,\m\rangle = |\n\rangle \otimes |\m\rangle,
\end{equation}
where $|\n\rangle$ and $|\m\rangle$ are eigenstates of $\hat{H}_{1}$ and $\hat{H}_{2}$, respectively. Then, we find that
\begin{align}\label{eq:QQ eigenvalue problem}
	E_{j} Z_{j} = H_{1} Z_{j} + Z_{j}H_{2} + \lambda V_{1} Z_{j} V_{2}^{\T},
\end{align}
where $Z_{j}$ is the matrix formed out of the $z_{n\mu,j}$ coefficients. We can choose the $Z_{j}$ to enforce $\langle \psi_{j} | \psi_{j'} \rangle = \delta_{jj'}$, which implies the orthogonality relation
\begin{equation}
	\frobinner{Z_{j}}{Z_{j'}} = \delta_{jj'},
\end{equation}
where the Frobenius inner product $\frobinner{\cdots}{\cdots}$ is defined in Appendix \ref{sec:Frobenius}.  If we retain the upper left-hand corner of all matrices, (\ref{eq:QQ eigenvalue problem}) becomes a finite dimensional eigenvalue problem that can be solved numerically.  

Any solution of the Schr\"odinger equation can be decomposed in terms of the energy eigenbasis as
\begin{equation}
	|\psi(t) \rangle = \sum_{j} c_{j} e^{-i E_{j}t} |\psi_{j}\rangle, \quad c_{j} = \langle \psi_{j} | \psi(0) \rangle
\end{equation}
and expanded in the basis $|n,\mu\rangle$ by writing
\begin{eqnarray}
|\psi(t) \rangle &=& \sum_{jn\mu} c_{j} e^{-i E_{j}t} |n,\mu\rangle \langle n,\mu|\psi_{j}\rangle \nn\\
&=&  \sum_{jn\mu} c_{j} e^{-i E_{j}t} z_{n\mu,j} |n,\mu\rangle\nn\\
&=& \sum_{n\mu} z_{n\mu}(t) |n,\mu\rangle,
\end{eqnarray}
where
\begin{equation}
    z_{n\mu}(t)= \sum_{j} c_{j} e^{-i E_{j}t} z_{n\mu,j}.
\end{equation}
Therefore, the solution of the dynamical system (\ref{eq:QQ EOM general})  in terms of the energy eigenstates is
\begin{equation}\label{eq:QQ eigenfunction solution}
	Z(t)  = \sum_{j} c_{j} e^{-i E_{j}t} Z_{j}, \quad
	c_{j} = \frobinner{Z_{j}}{Z(0)}.
\end{equation}
In practice, it is usually more computationally efficient to calculate $Z(t)$ using this formula rather than via direct numerical solution of a finite truncation of the dynamical system (\ref{eq:QQ EOM general}).

\subsection{Measures of entanglement}

We are primarily concerned with the situation where the system is prepared in an unentangled product state at $t = 0$.  As time progresses, we intuitively expect that the interaction will cause the entanglement of the system to grow. To quantify this process, we find it useful to deploy various measures of entanglement entropy. The simplest measure of entanglement are the purities $\gamma_{1} = \Tr(\rho_{1}^{2})$ and $\gamma_{2} = \Tr(\rho_{2}^{2})$ of the quantum state of subsystems 1 and 2, respectively. With the definitions of the reduced density matrices (\ref{eq:reduced density}), it is easy to show that each subsystem has the same purity $\gamma = \gamma_{1} = \gamma_{2}$; i.e.,
\begin{equation}
	\gamma = \frob{\rho_{1}}^{2} =  \frob{\rho_{2}}^{2} = \frob{ZZ^{\dagger}}^{2}.
\end{equation}
Via the properties listed in Appendix \ref{sec:Frobenius}, it is easy to confirm $\gamma \le 1$.  Furthermore, if either $\rho_{1}$ or $\rho_{2}$ represents a product state (e.g., if $\rho_{1} = \mathbf{w}\mathbf{w}^{\dagger}$, where $\mathbf{w}$ is column vector\footnote{In this paper, all matrices labelled by lowercase lettters in boldface type (e.g.,\ $ \mathbf{u}$) are assumed to be column vectors normalized under the usual inner product (equivalent to the Frobenius norm); i.e., $\mathbf{u}^{\dagger} \mathbf{u} = \frob{\mathbf{u}}^{2}=1$.}), then this inequality is saturated and $\gamma = 1$. Closely related to the concept of purity is the linear entanglement entropy
\begin{equation}
	S_\lin = 1 - \gamma = 1 - \frob{\rho_{1,2}}^{2},
\end{equation}
which satisfies $S_\lin = 0$ whenever the subsystems are described by pure states. Another popular measure of entanglement is the von Neumann entropy
\begin{equation}
	S_{\VN} = -\Tr(\rho_{1} \ln \rho_{1}) = -\Tr(\rho_{2} \ln \rho_{2}).
\end{equation}
Like the linear entropy, $S_{\VN} = 0$ whenever each subsystem is described by a pure state.

The nonzero eigenvalues of $\rho_{1} = ZZ^{\dagger}$ are the same as the nonzero eigenvalues of $\rho_{2} = Z^{\T}Z^{*}$. Furthermore, each of these nonzero eigenvalues are the square of one of the singular values of $Z$ or $Z^{\T}$, which we denote by $\{\chi_{i}\}$. Both the linear entropy and the von Neumann entropy can be expressed in terms of the singular values of $Z$: 
\begin{align}
	S_\lin = 1 - \sum_{i} \chi_{i}^{4}, \quad
	\label{eq:entanglement entropy} S_{\VN} = -\sum_{i} \chi_{i}^{2} \ln \chi_{i}^{2}.
\end{align}
Since $\Tr(\rho_{1})=1$ and each of the $\chi_{i}$ must be nonnegative real numbers, we have
\begin{equation}
	\sum_{i} \chi_{i}^{2} = 1 \quad \Rightarrow \quad \chi_{i} \in [0,1],
\end{equation}
for which it is easy to show that
\begin{equation}
	S_\lin \le S_{\VN};
\end{equation}
that is, the linear entropy provides a lower bound for the von Neumann entropy.  Further inequalities can be obtained by noting that both $S_\lin$ and $S_{\VN}$ are largest for maximally mixed states, which have
\begin{equation}
	\chi_{i}^{2} = \frac{1}{d}, \quad d = \text{min}(d_{1},d_{2}).
\end{equation}
This leads to
\begin{equation}
	S_\lin \le 1 - \frac{1}{d^{2}}, \quad S_{\VN} \le \ln d.
\end{equation}
From this, we see that $S_\lin \in [0,1]$ for all $d$. Conversely, $S_{\VN}$ is not bounded from above in the infinite dimensional ($d \to \infty$) case.

\section{Approximations of Quantum Dynamics}\label{sec:approximations of quantum dynamics}

\subsection{Product State Approximation}\label{sec:product state approximation}

Suppose that we prepare the system in a product state at $t=0$ and the coupling parameter $\lambda$ is small. Then, we expect that for some period of time the system remains in a nearly product state; i.e.\ the entanglement will be ``small''. However, as discussed in the previous subsection, there are multiple ways to quantify the entanglement of a quantum system, so it is not immediately obvious how to define ``small'' entanglement. One possibility is to say that a quantum state is nearly pure if its entanglement entropy is much less than that of a maximally entangled state.  When this definition is applied to the linear and von Neumann entropies, we find two possible conditions for defining a nearly product state:
\begin{equation}\label{eq:nearly product state conditions}
	1- \gamma = S_\lin \ll 1 - \frac{1}{d^{2}}, \quad S_{\VN} \ll \ln d.
\end{equation}
Both of these appear to be viable options for finite $d$, but we see that the second inequality becomes uninformative when $d \to \infty$. Also, the $1/d^{2}$ term in the first inequality is not of much practical importance, so it appears the most useful operational definition of a nearly product state is a state for which $S_\lin \ll 1$; however, we will analyze both the conditions (\ref{eq:nearly product state conditions}) below.

Over the time interval that the system is in a nearly product state (i.e.\ $S_\lin \ll 1$), we can write
\begin{equation}\label{eq:lambda expansion of Z}
	Z = \mathbf{u}_{1}\mathbf{u}_{2}^{\T} + \lambda \, \delta Z,  \quad \delta Z = \mathcal{O}(\lambda^{0}),
\end{equation}
where $\mathbf{u}_{1}=\mathbf{u}_{1}(t)$ and $\mathbf{u}_{2}=\mathbf{u}_{2}(t)$ are column vectors of dimension $d_{1}$ and $d_{2}$, respectively. Under this assumption, the reduced density matrices take the form
\begin{subequations}\label{eq: lambda expansion of rho}
\begin{align}
	\rho_{1} & =  \mathbf{u}_{1}\mathbf{u}_{1}^{\dagger}  + \lambda \, \delta\rho_{1},  \\
	\rho_{2} & =  \mathbf{u}_{2}\mathbf{u}_{2}^{\dagger}  + \lambda \, \delta\rho_{2}.
\end{align}
\end{subequations}
The expansions (\ref{eq:lambda expansion of Z}) and (\ref{eq: lambda expansion of rho}) allow us to expand (\ref{eq:QQ density matrix evolution general}) as
\begin{subequations}\label{eq:product state density matrix evolution}
\begin{align}
i\di_{t}{\rho}_{1} & = [H^{\eff}_{1},\rho_{1}] + \mathcal{O}(\lambda^{2}),  \\
i\di_{t}{\rho}_{2} & = [H^{\eff}_{2},\rho_{2}] + \mathcal{O}(\lambda^{2}),
\end{align}
\end{subequations}
with
\begin{subequations}
\begin{align}
H^{\eff}_{1} & = H_{1} + \lambda  \Tr(\rho_{2} V_{2}) V_{1},  \\
H^{\eff}_{2} & = H_{2} + \lambda  \Tr(\rho_{1} V_{1}) V_{2}.
\end{align}
\end{subequations}

If we drop the $\mathcal{O}(\lambda^{2})$ terms, (\ref{eq:product state density matrix evolution}) forms a closed set of differential equations for the density matrices and are identical to the von Neumann equations of two quantum systems with effective Hamiltonian matrices and an unusual coupling. Since each of the effective Hamiltonian matrices is Hermitian, we are guaranteed that if each subsystem is in a pure state at $t=0$, then they will remain in a pure state at later times. Hence, we may write
\begin{equation}\label{eq:pure state representation of rho}
	\hat{\rho}_{1}(t) = |\varphi_{1}(t)\rangle \langle \varphi_{1}(t) |, \quad \hat{\rho}_{2}(t) = |\varphi_{2}(t)\rangle \langle \varphi_{2}(t) |,
\end{equation}
with $|\varphi_{1}(t)\rangle \in \mathscr{H}_{1}$ and $|\varphi_{2}(t)\rangle \in \mathscr{H}_{2}$. This identification means that
\begin{equation}
	\Tr(\rho_{1} V_{1})  =  \langle \varphi_{1}|\hat{V}_{1}|\varphi_{1}\rangle, \quad \Tr(\rho_{2} V_{2})  = \langle  \varphi_{2}|\hat{V}_{2}|\varphi_{2}\rangle.
\end{equation}
Hence, effective Hamiltonian operators are given by
\begin{subequations}\label{eq:product state Hamiltonian operators}
\begin{align}
\label{eq:product state Hamiltonian operators 1} \hat{H}^{\eff}_{1} & = \hat{H}_{1} + \lambda \langle \varphi_{2}|\hat{V}_{2}|\varphi_{2}\rangle \hat{V}_{1},  \\
\label{eq:product state Hamiltonian operators 2} \hat{H}^{\eff}_{2} & = \hat{H}_{2} + \lambda \langle \varphi_{1}|\hat{V}_{1}|\varphi_{1}\rangle \hat{V}_{2}.
\end{align}
\end{subequations}
Then, to leading order in $\lambda$ equations (\ref{eq:product state density matrix evolution}) are consistent with the Schr\"odinger equations
\begin{equation}\label{eq:product state Schrodinger equations}
	i \di_{t} |\varphi_{1}(t)\rangle = \hat{H}^{\eff}_{1} |\varphi_{1}(t)\rangle, \quad i \di_{t} |\varphi_{2}(t)\rangle = \hat{H}^{\eff}_{2} |\varphi_{2}(t)\rangle.
\end{equation}
We decompose each state as follows:
\begin{equation}\label{eq:state expansion}
|\varphi_{1}(t) \rangle = \sum_{n}w_{n}(t) |n\rangle, \quad |\varphi_{2}(t)\rangle = \sum_{m}z_{m}(t)|m\rangle.
\end{equation}
Note that (\ref{eq: lambda expansion of rho}), (\ref{eq:pure state representation of rho}) and (\ref{eq:state expansion}) imply that
\begin{subequations}
\begin{align}
	\mathbf{w}\mathbf{w}^{\dagger} & = \mathbf{u}_{1}\mathbf{u}_{1}^{\dagger}  + \mathcal{O}(\lambda),\\
	\mathbf{z}\mathbf{z}^{\dagger} & = \mathbf{u}_{2}\mathbf{u}_{2}^{\dagger}  + \mathcal{O}(\lambda).
\end{align}
\end{subequations}
Using (\ref{eq:state expansion}), we see that the Schr\"odinger equations (\ref{eq:product state Schrodinger equations}) are equivalent to
\begin{subequations}\label{eq:almost product EOM}
\begin{align}
i\di_{t} \mathbf{w} & = H_{1}\mathbf{w} + \lambda (\mathbf{z}^{\dagger} V_{2} \mathbf{z}) V_{1}\mathbf{w}, \\
i\di_{t} \mathbf{z} & = H_{2}\mathbf{z} + \lambda (\mathbf{w}^{\dagger} V_{1} \mathbf{w}) V_{2}\mathbf{z}.
\end{align}
\end{subequations}
Equations (\ref{eq:almost product EOM}) represent $d_{1} + d_{2}$ complex nonlinear differential equations whose solution completely specify the quantum dynamics of the system within the context of the ``product state'' approximation.

We note that an interesting feature of the equations of motion (\ref{eq:almost product EOM}) in the product state approximation is that they are nonlinear. This may seem surprising since the QQ equations of motion (\ref{eq:QQ EOM general}) are linear. However, the reason for this nonlinearity is simple: the effective Hamiltonians for subsystems 1 and 2 are deduced from the equations of motion for the reduced density matrices, which themselves are nonlinear functions of the full quantum state. In hindsight, it appears that such a nonlinearity is inevitable, since it is unclear how to separate out the dynamics of subsystem 1 from subsystem 2 without introducing a nonlinear object like the reduced density matrix.

We can also contrast the $\rho_1$ and $\rho_2$ evolution equations (\ref{eq:product state density matrix evolution}) in the product state approximation to the well-known Lindblad master equation governing the evolution of the density matrix of a quantum system coupled to a much larger system called the ``environment'' \cite{Manzano_2020}. If we call subsystem 1 the environment, the Lindblad formalism makes several assumptions, the most important of which are:
\begin{enumerate}
\item The coupling of subsystem 2 to the environment is weak. \label{assumption:weak coupling}
\item The effects of subsystem 2 on the environment are negligible; i.e., the dynamics of the environment is unaffected by the presence of subsystem 2. (This is essentially the Born approximation.) \label{assumption:Born}
\item The system is prepared initially in a product state, which implies that it remains in a product state under the Born approximation. \label{assumption:product state}
\item The time derivative of the subsystem 2 density matrix at given time depends only on the current value of the density matrix. (This is called the Born-Markov approximation.) \label{assumption:Markovian}
\end{enumerate}
We see that the product state approximation presented above makes explicit use of assumptions \ref{assumption:weak coupling} and \ref{assumption:product state}, but does not neglect the effects of subsystem 2 on subsystem 1 (assumption \ref{assumption:Born}) and does not artificially enforce any sort of Markovian approximation (assumption \ref{assumption:Markovian}). The price to be paid is that in the product state approximation, we need to simultaneously solve for the dynamics of both systems. In contrast, the assumption that the environment is too big to be affected by subsystem 2 means that one has less degrees of freedom to solve for in the Lindblad formalism, since loss of coherence to the environment is not tracked. The same assumption also results in nonunitary linear equations of motion in the Lindblad case, as opposed to the unitary nonlinear equations we have in the product state approximation.

\subsection{Classical-Quantum (CQ) approximation}\label{sec:classical-quantum}

The classical-quantum approximation builds on the product state approximation of the previous section by making a significant assumption, namely that one subsystem behaves ``classically''; i.e. it is acceptable to replace quantum expressions associated with one subsystem with their classical analogues.  Stated another way, this approximation is based on the belief that it is safe to apply the generalized Bohr correspondence principle to one subsystem. Heuristically, one would expect this to be a valid approximation when the quantum number or energy of that subsystem is large, or when the associated wavefunction is ``sharply peaked''.

How does this work in practical terms? In this subsection, we discuss the  assumptions  required to arrive at the classical-quantum approximation in a general setting, while in appendix \ref{sec:CQ in 1D potential} we show how these assumptions can be realized when the classical subsystem represents a particle moving in a one-dimensional potential. 

In the general case, we start with the effective Hamiltonian operators in the product state approximation (\ref{eq:product state Hamiltonian operators}) and assume that subsystem 1 behaves ``classically''. The key approximation is that the expectation value of the potential $\hat{V_{1}}$ as a function of time is well approximated by the potential function $\mathcal{V}_{1}$ evaluated on a classical trajectory $(q_{1}(t),p_{1}(t))$. More specifically, we write
\begin{equation}\label{eq:classical-quantum approximation}
	\mathcal{V}_{1}(q_{1}(t),p_{1}(t)) = \langle \varphi_{1}(t)|\hat{V}_{1}|\varphi_{1}(t)\rangle [1 - \varepsilon(t)],
\end{equation}
under the assumption that $\varepsilon(t)$ is in some suitable sense ``small''. Here, $q_{1}(t)$ and $p_{1}(t)$ are the solutions of the equations of motion generated by an effective classical Hamiltonian
\begin{align}\label{eq:CQ classical effective Hamiltonian}
\mathcal{H}_{\CQ}^{\eff} & = \mathcal{H}_{1} + \lambda \mathcal{V}_{1} \langle \varphi_{2}|\hat{V}_{2}|\varphi_{2}\rangle.
\end{align}
This classical Hamiltonian is obtained by replacing the subsystem 1 quantum operators in the expression (\ref{eq:product state Hamiltonian operators 1}) for $\hat{H}_{1}^{\eff}$ by their classical counterparts. To complete the picture, we demand that the state vector for subsystem 2 evolves via the Hamiltonian operator
\begin{align}\label{eq:CQ classical effective Hamiltonian 2}
\hat{H}_{\CQ}^{\eff} = \hat{H}_{2} + \lambda \mathcal{V}_{1} \hat{V}_{2}.
\end{align}
Here, $\hat{H}_{\CQ}^{\eff}$ has been obtained by substituting (\ref{eq:classical-quantum approximation}) into the expression (\ref{eq:product state Hamiltonian operators 2}) for $\hat{H}_{2}^{\eff}$ in the product state approximation and neglecting $\varepsilon(t)$.  

We pause here to emphasize that the smallness of $|\varepsilon(t)|$ does \emph{not necessarily} follow from a $\lambda \ll 1$ approximation; that is, we expect 
$\lim_{\lambda \to 0} |\varepsilon(t)| \ne 0$.
The demand that $|\varepsilon(t)|$ is small is really a requirement that the quantum corrections to the classical dynamics of subsystem 1 are small irrespective of the size of $\lambda$, which in turn depends crucially on the nature of the initial quantum state $|\varphi_{1}\rangle$ of subsystem 1. Quantifying the conditions under which $\varepsilon(t)|$ is small is a subtle problem that we return to in \S\ref{sec:classical-quantum approx validity} below.

Explicitly, the complete equations of motion for the system in this classical-quantum approximation are
\begin{subequations}
\begin{gather}
	\label{eq:CQ classical EOMS} \di_{t}q_{1} = \{q_{1}, \mathcal{H}_{\CQ}^{\eff}\}, \quad \di_{t}p_{1} = \{p_{1}, \mathcal{H}_{\CQ}^{\eff}\}, \\
	i\di_{t} |\varphi_{2}(t)\rangle = \hat{H}_{\CQ}^{\eff}  |\varphi_{2}(t)\rangle.
\end{gather}
\end{subequations}
If we again make use of the expansion (\ref{eq:state expansion}) of the state vector into eigenstates of $\hat{H}_{2}$, we obtain
\begin{subequations}\label{eq:CQ quantum EOM}
\begin{align}
\di_{t}q_{1} & = +\di_{p_{1}}\mathcal{H}_{1} + \lambda \, (\mathbf{z}^{\dagger} V_{2} \mathbf{z}) \, \di_{p_{1}}\mathcal{V}_{1}, \label{eq:CQ Hamilton 1} \\
\di_{t}p_{1} & = -\di_{q_{1}}\mathcal{H}_{1} - \lambda \, (\mathbf{z}^{\dagger} V_{2} \mathbf{z}) \, \di_{q_{1}}\mathcal{V}_{1}, \label{eq:CQ Hamilton 2} \\
i\di_{t} \mathbf{z} & = H_{2}\mathbf{z} + \lambda \mathcal{V}_{1} V_{2}\mathbf{z}. \label{eq:CQ quantum EOM 2}
\end{align}
\end{subequations}
Equations (\ref{eq:CQ quantum EOM}) comprise 2 real and $d_{2}$ complex nonlinear ordinary differential equations that completely specify the system dynamics.

Once equations (\ref{eq:CQ quantum EOM}) have been solved, the density matrix for oscillator 2 and the expectation value of any operator $\hat{A}_{2}:\mathscr{H}_{2} \to \mathscr{H}_{2}$ is given by
\begin{equation}
	\rho_{\CQ}(t) = \mathbf{z}(t)\mathbf{z}^{\dagger}(t), \quad \langle A_{2}(t) \rangle = \text{Tr}\,[\rho_{\CQ}(t)A_{2}].
\end{equation}
This density matrix will evolve according to the von Neumann equation $i\di_{t}\rho_{\CQ} = [H_{\CQ}^{\eff},\rho_{\CQ}]$, or
\begin{equation}\label{eq:CQ density matrix evolution}
	i\di_{t}\rho_{\CQ} = [H_{2}+\lambda \mathcal{V}_{1} V_{2},\rho_{\CQ}].
\end{equation}

\subsection{Classical-Classical (CC) approximation}

The CC  approximation is the logical extension of the CQ approximation. We assume it is reasonable to treat all the system's degrees of freedom classically. The Hamiltonian is just the full classical Hamiltonian (\ref{eq:full classical Hamiltonian}), and the equations of motion are Hamilton's equations:
\begin{gather}
\nonumber \di_{t}q_{1} = \{q_{1}, \mathcal{H}\}, \quad \di_{t}p_{1} = \{p_{1}, \mathcal{H}\}, \\
\di_{t}q_{2} = \{q_{2}, \mathcal{H}\}, \quad \di_{t}p_{2} = \{p_{2}, \mathcal{H}\}.
\end{gather}
These represent four real nonlinear differential equations that describe the system dynamics.

\subsection{Classical background approximation}

One can obtain a different approximation from the CQ formalism by assuming that the backreaction of the quantum system on the classical system is negligibly small. Practically, this involves neglecting the terms proportional to $\lambda$ in the classical part of the CQ equations of motion (\ref{eq:CQ quantum EOM}):
\begin{subequations}\label{eq:CB EOM}
\begin{align}
\di_{t}q_{1} & = +\di_{p_{1}}\mathcal{H}_{1}, \\ 
\di_{t}p_{1} & = -\di_{q_{1}}\mathcal{H}_{1}, \\
i\di_{t} \mathbf{z} & = H_{2}\mathbf{z} + \lambda \mathcal{V}_{1} V_{2}\mathbf{z}. 
\end{align}
\end{subequations}
This is expected to be reasonable if the energy stored in subsystem 1 is significantly larger than the interaction energy. This is quite similar to the Born-Oppenheimer approximation discussed in \S\ref{sec:introduction}. In fact, equations (\ref{eq:CB EOM}) are really the analogues of the semiclassical Einstein equations without backreaction (\ref{eq:semiclassical system 1}) for our bipartite system (\ref{eq:full classical Hamiltonian}). However, the path to arriving to these equations via the Born-Oppenheimer approximation is distinct from what we have presented above in several subtle ways. In appendix \ref{sec:Born-Oppenheimer}, we show how to arrive at (\ref{eq:CB EOM}) using the Born-Oppenheimer approximation in the situation when each of subsystem 1 and 2 correspond to particles moving in one-dimensional potentials.

\subsection{Conservation of energy and probability via alternative Hamiltonians}

It it interesting to note that the dynamical systems governing the system's evolution in the fully quantum-quantum formalism (\ref{eq:QQ EOM general}), the product state approximation (\ref{eq:almost product EOM}), and the classical-quantum approximation (\ref{eq:CQ quantum EOM}) can each be derived directly from alternative Hamiltonians. For the quantum-quantum case, we can define an alternative Hamiltonian by the expectation value of the Hamiltonian operator:
\begin{equation}
	\mathfrak{H}_{\QQ} = \langle \psi(t) | \hat{H} | \psi(t) \rangle.
\end{equation}
Using the state expansion (\ref{eq:QQ state expansion}), we find
\begin{multline}\label{eq:QQ alternative Hamiltonian}
	\mathfrak{H}_{\QQ} = \text{Tr}( Z^{\dagger} H_{1} Z) +  \text{Tr}(ZH_{2}^{\T}Z^{\dagger} )  + \\ \lambda  \text{Tr}(V_{1} Z V_{2}^{\T} Z^{\dagger}).
\end{multline}
If we assume the Poisson brackets
\begin{equation}
	\{ z_{nm},z^{*}_{n'm'} \} = -i\delta_{nn'}\delta_{mm'},
\end{equation}
then we see that the ordinary expression for time evolution
\begin{align}
	\di_{t}z_{nm} = \{z_{nm},\mathfrak{H}_{\QQ}\},
\end{align}
is equivalent to the equations (\ref{eq:QQ EOM general}); hence, the dynamical system (\ref{eq:QQ EOM general}) is Hamiltonian.  

For the product state approximation, an alternative Hamiltonian is defined by
\begin{equation}\label{eq:PS alternative Hamiltonian}
	\mathfrak{H}_{\PS} = \mathbf{w}^{\dagger}H_{1}\mathbf{w} + \mathbf{z}^{\dagger}H_{2}\mathbf{z} + \lambda (\mathbf{w}^{\dagger} V_{1} \mathbf{w}) (\mathbf{z}^{\dagger} V_{2} \mathbf{z}),
\end{equation}
which can be obtained from (\ref{eq:QQ alternative Hamiltonian}) under the assumption that $Z \approx \mathbf{w}\mathbf{z}^{\T}$. The Poisson brackets
\begin{equation}
	\{w_{n},w_{n'}^{*}\} = -i\delta_{nn'}, \quad \{z_{m},z_{m'}^{*}\} = -i\delta_{mm'},
\end{equation}
combined with
\begin{equation}
	\di_{t}w_{n} = \{w_{n},\mathfrak{H}_{\PS}\}, \quad \di_{t}z_{m} = \{z_{m},\mathfrak{H}_{\PS}\},
\end{equation}
then reproduce (\ref{eq:almost product EOM}). Hence, (\ref{eq:almost product EOM}) is a Hamiltonian dynamical system.  

For the classical-quantum approximation, the alternative Hamiltonian is
\begin{equation}
	\mathfrak{H}_{\CQ} = \mathcal{H}_{1} + \mathbf{z}^{\dagger}H_{2}\mathbf{z} + \lambda \mathcal{V}_{1} (\mathbf{z}^{\dagger} V_{2} \mathbf{z}),
\end{equation}
which can be obtained from (\ref{eq:PS alternative Hamiltonian}) under the assumption that $\mathbf{w}^{\dagger}H_{1}\mathbf{w}\approx \mathcal{H}_{1}$ and $\mathbf{w}^{\dagger}V_{1}\mathbf{w}\approx \mathcal{V}_{1}$. The Poisson brackets
\begin{equation}
	\{z_{m},z_{m'}^{*}\} = -i\delta_{mm'},
\end{equation}
combined with
\begin{gather}\nonumber
	\di_{t}q_{1} = \{q_{1},\mathfrak{H}_{\CQ}\}, \quad \di_{t}p_{1} = \{p_{1},\mathfrak{H}_{\CQ}\}, \\ \di_{t}z_{m} = \{z_{m},\mathfrak{H}_{\CQ}\},
\end{gather}
are equivalent to (\ref{eq:CQ quantum EOM}), which implies the dynamical system is Hamiltonian. 

All three of the above alternative Hamiltonians are time-translation invariant, implying that each Hamiltonian is conserved on-shell. This in turn implies conservation of energy in solutions of the quantum-quantum, product-state approximation, and classical-quantum equations of motion. Similarly, each of the Hamiltonians are invariant under phase shifts of the quantum variables:
\begin{subequations}
\begin{align}
	Z \mapsto e^{i\phi}Z \quad & \Rightarrow \quad \mathfrak{H}_{\QQ} \mapsto \mathfrak{H}_{\QQ}, \\
	(\mathbf{w},\mathbf{z}) \mapsto (e^{i\phi_{1} }\mathbf{w},e^{i\phi_{2} }\mathbf{z}) \quad & \Rightarrow \quad \mathfrak{H}_{\PS} \mapsto \mathfrak{H}_{\PS}, \\
	\mathbf{z} \mapsto e^{i\phi}\mathbf{z} \quad & \Rightarrow \quad \mathfrak{H}_{\CQ} \mapsto \mathfrak{H}_{\CQ}.
\end{align}
\end{subequations}
Via Noether's theorem, these symmetries imply that the Frobenius norms of $Z$, $\mathbf{w}$ and $\mathbf{z}$ are conserved on-shell. This in turn implies the conservation of probability in each calculations; i.e.,
\begin{subequations}
\begin{align}
\text{quantum-quantum:} \,\, &  \di_{t} \langle \psi | \psi \rangle = 0, \\
\text{product-state:} \,\, &   \di_{t} \langle \varphi_{1} | \varphi_{1} \rangle = \di_{t} \langle \varphi_{2} | \varphi_{2} \rangle = 0,\\
\text{classical-quantum:} \,\, &  \di_{t} \langle \varphi_{2} | \varphi_{2} \rangle = 0.
\end{align}
\end{subequations}

Before moving on, we note that while energy is explicitly conserved in the QQ, product-state, CQ and CC approximations, it will not in be conserved in the classical background approximation as the influence of the classical system on the quantum system takes the form of an external source. That is, the quantum Hamiltonian contains explicit time dependence.

\section{Perturbative analysis}\label{sec:perturbative}

As noted above, the equations of motion for the reduced density matrices in the quantum-quantum case (\ref{eq:QQ density matrix evolution general}) are not closed, but when the coupling is small and if the system is in a nearly product state it is possible to write a self-contained dynamical system for $\rho_{1}$ and $\rho_{2}$. In this section, we investigate this regime more formally in the context of perturbation theory with the purpose of quantifying the regimes of validity in the product state and classical-quantum approximations.

\subsection{Interaction picture and perturbative solutions in the quantum-quantum case}\label{sec:QQ perturbative}

In order to analyze the dynamics of the system under the assumption that the coupling is small $\lambda \ll 1$, it will be useful to move from the Schr\"odinger picture to the interaction picture. Our notation is that an under-tilde denotes the interaction picture version of a given quantity. We define the interaction picture coefficient matrix $\INT{Z}$ by
\begin{equation}
	\INT{Z} = e^{iH_{1}t} Z e^{iH_{2}t}.
\end{equation}
We also define the interaction picture density matrices via
\begin{subequations}
\begin{align}
	\INT{\rho}_{1} = \INT{Z}\INT{Z}^{\dagger} & = e^{iH_{1}t} \rho_{1} e^{-iH_{1}t}, \\
	\INT{\rho}_{2} = \INT{Z}^{\T}\INT{Z}^{*} & = e^{iH_{2}t} \rho_{2} e^{-iH_{2}t}.
\end{align}
\end{subequations}
The matrix representations of generic operators $\hat{A}_{1}$ or $\hat{A}_{2}$ on $\mathscr{H}_{1}$ or $\mathscr{H}_{2}$ are, respectively,
\begin{align}
	\INT{A}_{1} & = e^{iH_{1}t} A_{1} e^{-iH_{1}t}, \quad \INT{A}_{2} = e^{iH_{2}t} A_{2} e^{-iH_{2}t},
\end{align}
which implies
\begin{subequations}
\begin{align}
	\langle \hat{A}_{1} \otimes \hat{I}_{2} \rangle = \Tr(\rho_{1}A_{1}) = \Tr(\INT{\rho}_{1}\INT{A}_{1}), \\
	\langle \hat{I}_{1} \otimes \hat{A}_{2} \rangle = \Tr(\rho_{2}A_{2}) = \Tr(\INT{\rho}_{2}\INT{A}_{2}).
\end{align}
\end{subequations}
That is, formulae for expectation values are the same in either picture. Finally, we note that $H_{1} = \INT{H}_{1}$ and $H_{2} = \INT{H}_{2}$.

Substituting the above definitions into (\ref{eq:QQ EOM general}), we find that
\begin{align}\label{eq:QQ EOM interaction}
	i \di_{t} \INT{Z} = \lambda \INT{V}_{1} \INT{Z} \INT{V}_{2}^{\T}.
\end{align}
Similarly, the evolution equations for the interaction picture reduced density matrices are
\begin{subequations}\label{eq:QQ density matrix evolution interaction}
\begin{align}
i\di_{t}\INT{\rho}_{1} & = \lambda  [\INT{V}_{1},\INT{Z} \INT{V}_{2}^{\T}\INT{Z}^{\dagger}], \\
i\di_{t}\INT{\rho}_{2} & = \lambda  [\INT{V}_{2},\INT{Z}^{\T} \INT{V}_{1}^{\T}\INT{Z}^{*}].
\end{align}
\end{subequations}

We now substitute a perturbative expansion of the coefficient matrix,
\begin{equation}\label{eq:Z expansion}
	\INT{Z} = \INT{Z}^{\zero} + \lambda \INT{Z}^{\one} + \cdots,
\end{equation}
into (\ref{eq:QQ EOM interaction}). Setting coefficients of different powers of $\lambda$ equal to zero yields
\begin{equation}
	i \di_{t} \INT{Z}^{\zero} = 0, \quad i \di_{t} \INT{Z}^{(k+1)} = \INT{V}_{1} \INT{Z}^{(k)} \INT{V}_{2}^{\T},
\end{equation}
for $k = 0,1,2\ldots$ We assume a product state solution for the zeroth order equation
\begin{equation}
	\INT{Z}^{\zero}(t) = \INT{\mathbf{u}}_{1}\INT{\mathbf{u}}_{2}^{\T},
\end{equation}
where $\INT{\mathbf{u}}_{1}$ and $\INT{\mathbf{u}}_{2}$ are constant column vectors of dimension $d_{1}$ and $d_{2}$, respectively. We also select initial conditions
\begin{equation}
	\INT{Z}^{(k)}(0) = 0, \quad k = 1,2,3\ldots
\end{equation}
This ensures that the subsystems are unentangled at $t=0$. Then, the solution to the first order equation is
\begin{equation}\label{eq:perturbative Z soln}
	\INT{Z}^{\one}(t) = -i \int\limits_{\mathcal{T}}  \INT{V}'_{1} \INT{\mathbf{u}}_{1}\INT{\mathbf{u}}_{2}^{\T}\INT{V}'^{\T}_{2},
\end{equation}
where here and below we make use of the notation
\begin{gather}\nonumber
	\int\limits_{\mathcal{T}} = \int\limits_{0}^{t} dt' , \quad \iint\limits_{\mathcal{T}^{2}} = \int\limits_{0}^{t} \!\! \int\limits_{0}^{t} dt'dt'' , \\ V_{1,2}(t') = V'_{1,2}, \quad V_{1,2}(t'') = V''_{1,2}.
\end{gather}
Since the sum (or integral) of outer products of vectors is not in general itself an outer product, we expect that neither $\INT{Z}^{\one}(t)$ or $\INT{Z}(t)$ will be expressible as an outer product for $t>0$.  That is, the system will generally be entangled for all $t>0$.

The expansion (\ref{eq:Z expansion}) implies that the density matrices are of the form
\begin{align}\label{eq:density matrix expansion}
	\INT{\rho}_{1,2}  =  \INT{\rho}_{1,2}^{\zero} + \lambda \INT{\rho}^{\one}_{1,2} + \cdots,  \quad \INT{\rho}_{1,2}^{\zero} & =  \INT{\mathbf{u}}_{1,2} \INT{\mathbf{u}}_{1,2}^{\dagger} ,   
\end{align}
with
\begin{equation}
	 \INT{\rho}^{(k)}_{1}(0) = 0 = \INT{\rho}^{(k)}_{2}(0), \quad k = 1,2,3\dots
\end{equation}
Then, (\ref{eq:QQ density matrix evolution interaction}) gives
\begin{align}\label{eq:small coupling EOMs}
	i \di_{t}\INT{\rho}_{1,2}^{\one} = \text{Tr}(\INT{\rho}_{2,1}^{\zero}\INT{V}_{2,1})[\INT{V}_{1,2},\INT{\rho}_{1,2}^{\zero}].
\end{align}
These equations allow us to obtain $\INT{\rho}_{1}^{\one}$ and $\INT{\rho}_{2}^{\one}$ via quadratures:
\begin{align}\label{eq:perturbative rho solution}
	\INT{\rho}_{1,2}^{\one}(t) = -i\int\limits_{\mathcal{T}}  \langle V'_{2,1} \rangle_{\zero} [\INT{V}'_{1,2},\INT{\mathbf{u}}_{1,2} \INT{\mathbf{u}}_{1,2}^{\dagger}].
\end{align}
Here and below, the $(0)$ subscript indicates evaluation at $\lambda = 0$; i.e. when the interaction between the subsystems is neglected.

\subsection{Scrambling time}\label{sec:validity of product state approx}

In this subsection, we estimate how long after $t=0$ the entanglement between the subsystems remains small and the product state approximation is expected to be valid. We call this timescale the ``scrambling time'' of the system. Our first step will be to derive bounds on the linear and von Neumann entropies. We begin by noting that the Frobenius norm of matrices is in general the same in the Schr\"odinger and interaction pictures by equation (\ref{eq:unitary frob}), which allows us to express the linear entropy as
\begin{equation}
	S_\lin = 1 - \frob{\INT{\rho}_{1,2}}^{2}.
\end{equation}
We write
\begin{equation}
	\INT{\rho}_{1,2} = \INT{\rho}_{1,2}^{\zero} + \lambda \, \delta\INT{\rho}_{1,2},
\end{equation}
where $\INT{\rho}_{1,2}^{\zero}$ is the leading term in the perturbative $\lambda$-expansion of $\INT{\rho}_{1,2}$. It then it follows from equations (\ref{eq:frob inequality}), and the triangle inequality that:
\begin{equation}\label{eq:linear entropy inequality 2}
	S_\lin \le 2\lambda \frob{\delta\INT{\rho}_{1,2}} ( 1+ \tfrac{1}{2} \lambda \frob{\delta\INT{\rho}_{1,2}}).
\end{equation}
This is exact, but it is useful to re-express it in a less precise form using the perturbative expansions (\ref{eq:density matrix expansion}):
\begin{align}
	S_\lin & \le 2\lambda \frob{\INT{\rho}^{\one}_{1,2}} +\mathcal{O}(\lambda^{2}).
\end{align}
Note that this actually represents two distinct upper bounds on $S_\lin$; i.e., one derived from $\frob{\INT{\rho}^{\one}_{1}}$ and another from $\frob{\INT{\rho}^{\one}_{2}}$. The two bounds can be combined as
\begin{align}\label{eq:linear entropy inequality 3}
	S_\lin & \le 2\lambda \, \text{min}(  \frob{\INT{\rho}^{\one}_{1,2}}) +\mathcal{O}(\lambda^{2}).
\end{align}

We can also derive a bound on the von Neumann entropy by noting that the singular values $\{\chi_{i}\}$ of $Z$ are the same as the singular values of $\INT{Z}$. We then write
\begin{equation}
	\INT{Z} = \INT{Z}^{\zero} + \lambda \, \delta \INT{Z}.
\end{equation}
where $\INT{Z}^{\zero} =  \INT{\mathbf{u}}_{1}\INT{\mathbf{u}}_{2}^{\T}$ is the zeroth order term in the perturbative expansion of $\INT{Z}$. The singular values of $\INT{\mathbf{u}}_{1}\INT{\mathbf{u}}_{2}^{\T}$ are trivially 
\begin{equation}
	\chi_{i}^{\zero} = \begin{cases} 1, & i = 1, \\ 0, & i = 2,3,4\ldots \end{cases}
\end{equation}
Let us now write
\begin{equation}
	\chi_{i} = \chi_{i}^{\zero} + \lambda \, \delta\chi_{i}.
\end{equation}
Substituting this into the formula (\ref{eq:entanglement entropy}) for the von Neumann entropy, we obtain
\begin{equation} \label{eq:VN entropy inequality 1}
	S_{\VN} = -2\lambda \, \delta\chi_{1} + \mathcal{O}(\lambda^{2}\ln \lambda).
\end{equation}
Now, due to a theorem by Weyl, we have
\begin{equation}
	|\delta\chi_{1}| \le \Vert \delta\INT{Z} \Vert_{2} \le \frob{ \delta\INT{Z} }.
\end{equation}
Here, $\Vert \delta\INT{Z} \Vert_{2}$ is the spectral norm of $\delta\INT{Z}$. This gives us the following bound on $S_{\VN}$:
\begin{align}\nonumber
	S_{\VN} & \le 2\lambda \frob{ \delta\INT{Z} } + \mathcal{O}(\lambda^{2}\ln \lambda) \\
	& = 2\lambda \frob{ \INT{Z}^{\one} } + \mathcal{O}(\lambda^{2}\ln \lambda), \label{eq:VN entropy inequality 2}
\end{align}
where we have made use of (\ref{eq:Z expansion}) in moving from the first to second lines. We note that in comparing the inequalities (\ref{eq:linear entropy inequality 2}) and (\ref{eq:VN entropy inequality 1}), it is useful to make note of the fact that
\begin{equation}\label{eq:entropy relative sizes}
\frob{\delta\INT{\rho}_{1,2}} \le 2\frob{\delta\INT{Z}},
\end{equation} which is straightforward to prove. This is consistent with our expectation that $S_\lin < S_{\VN}$.

Comparing the inequalities (\ref{eq:linear entropy inequality 3}) and (\ref{eq:VN entropy inequality 2}) with (\ref{eq:nearly product state conditions}) and retaining the leading order terms, we find two conditions to be satisfied for nearly product states:
\begin{equation}\label{eq:product state inequalities 1}
	\lambda \, \text{min}( \frob{\INT{\rho}^{\one}_{1,2}} ) \ll 1, \quad \lambda \frob{\INT{Z}^{\one}}  \ll \ln d.
\end{equation}
To analyze these further, we note that an explicit formula for $ \frob{\INT{Z}^{\one}}$ can be obtained from (\ref{eq:perturbative Z soln}):
\begin{align}\label{eq:Z norm}
	 \frob{\INT{Z}^{\one}}  = \left\{ \iint\limits_{\mathcal{T}^{2}} \langle V'_{1} V''_{1} \rangle \langle V'_{2} V''_{2} \rangle   \right\}^{1/2}_{\zero}.
\end{align}
Similarly, (\ref{eq:perturbative rho solution}) yields an expression for $\frob{\INT{\rho}^{\one}_{1,2}} $:
\begin{align}\label{eq:rho norm}
	 \frob{\INT{\rho}^{\one}_{1,2}}  =  \left\{ 2 \iint\limits_{\mathcal{T}^{2}} \langle V'_{2,1}\rangle \langle V''_{2,1} \rangle \text{Cov}(V'_{1,2},V''_{1,2})]    \right\}^{1/2}_{\zero}.
\end{align}
Here, $\text{Cov}(V'_{1,2},V''_{1,2})$ is the quantum covariance of $V'_{1,2}$ and $V''_{1,2}$:
\begin{equation}
	\text{Cov}(V'_{1,2},V''_{1,2}) = \tfrac{1}{2} \langle \left\{ V'_{1,2},V''_{1,2} \right\} \rangle - \langle V'_{1,2} \rangle \langle V''_{1,2} \rangle,
\end{equation}
where $\left\{ V'_{1,2},V''_{1,2} \right\} = V'_{1,2}V''_{1,2}+V''_{1,2}V'_{1,2}$ is the anti-commutator.  This can be written in an alternate form:
\begin{multline}
	 \frob{\INT{\rho}^{\one}_{1,2}}  =   \sqrt{2} \left\langle  \left[ \int\limits_{\mathcal{T}} \langle V'_{2,1} \rangle \left(V'_{1,2} - \langle V'_{1,2}\rangle I_{1,2} \right) \right]^{2} \right\rangle^{1/2}_{\zero}.
\end{multline}
The integrals (\ref{eq:Z norm}) and (\ref{eq:rho norm}) can be bounded by noting that
\begin{subequations}
\begin{align}
	|\langle V'_{1,2} \rangle| & \le \sqrt{\langle V'^{2}_{1,2} \rangle}, \\ 
	|\langle V'_{1,2} V'_{1,2} \rangle| & \le \sqrt{ \langle V'^{2}_{1,2} \rangle\langle V''^{2}_{1,2} \rangle}, \\
	|\text{Cov}(V'_{1,2},V''_{1,2})| & \le \sqrt{\text{Var}(V'_{1,2})\text{Var}(V''_{1,2})},
\end{align}
\end{subequations}
with the variance defined as usual:
\begin{equation}
	\text{Var}(V'_{1,2}) =  \langle  V'^{2}_{1,2} \rangle - \langle V'_{1,2} \rangle^{2}.
\end{equation}
We obtain
\begin{subequations}
\begin{align}
	\frob{\INT{Z}^{\one}} & \le \left\{ \int\limits_{\mathcal{T}}  \sqrt{ \langle V'^{2}_{1}  \rangle \langle V'^{2}_{2} \rangle }  \right\}_{\zero}, \\
	 \frob{\INT{\rho}^{\one}_{1,2}} & \le \left\{ \int\limits_{\mathcal{T}}  \sqrt{ 2 \langle V'^{2}_{2,1}  \rangle \left( \langle  V'^{2}_{1,2} \rangle - \langle V'_{1,2} \rangle^{2} \right) }  \right\}_{\zero}.
\end{align}
\end{subequations}
Then, we see that a sufficient condition for the inequality $S_{\VN} \sim \lambda \frob{\INT{Z}^{\one}}  \ll \ln d$ to be satisfied is
\begin{equation}\label{eq:product state inequalities 2}
	t \ll t_{\VN}, \quad t_{\VN} = \frac{\ln d}{\mathcal{E}_\text{int}(t_{\VN})},
\end{equation}
where $\mathcal{E}_\text{int}(t)$ is a measure of the average value of an upper bound on the interaction energy between the subsystems over the time interval $[0,t]$:
\begin{equation}\label{eq:avg interaction energy}
	\mathcal{E}_\text{int}(t) = \frac{\lambda}{t} \left\{ \int\limits_{\mathcal{T}}  \sqrt{ \langle V'^{2}_{1}  \rangle \langle V'^{2}_{2} \rangle }  \right\}_{\zero} = \frac{1}{t} \left\{ \int\limits_{\mathcal{T}} \sqrt{\langle H'^{2}_{\text{int}} \rangle} \right\}_{\zero},
\end{equation}
with $\hat{H}_{\text{int}} = \lambda \, \hat{V}_{1} \otimes \hat{V}_{2}$. Furthermore, a sufficient condition for $S_{\lin} \sim \lambda \, \text{min}( \frob{\INT{\rho}^{\one}_{1,2}} ) \ll 1$ to hold is
\begin{equation}\label{eq:product state inequalities 3}
	t \ll t_{\lin}, \quad t_{\lin} = \frac{1}{\mathcal{E}_\text{int}(t_{\lin}) \, \text{min}[\mathcal{N}_{1,2}(t_{\lin})] },
\end{equation}
where the functions $\mathcal{N}_{1,2}(t)$ are given by
\begin{equation}
	\mathcal{N}_{1,2}(t) =  \left\{ \frac{ \int_{\mathcal{T}}  [ \langle V'^{2}_{2,1}  \rangle \text{Var}(V'_{1,2})]^{1/2} }{\int_{\mathcal{T}}  [ \langle V'^{2}_{1}  \rangle \langle V'^{2}_{2} \rangle ]^{1/2} }  \right\}_{\zero}.
\end{equation}
It is easy to confirm that
\begin{equation}
	\mathcal{N}_{1,2}(t) \in [0,1],
\end{equation}
and that $\mathcal{N}_{1,2}(t)=0$ implies that $\text{Var}(V'_{1,2})=0$ for all $t' \in [0,t]$. That is, $\mathcal{N}_{1,2}$ essentially measures the degree to which the quantum probability distributions for $V_{1,2}$ are peaked at zero coupling; i.e. the degree of coherence of the state of subsystem 1 or 2 with respect to the $V_{1}$ or $V_{2}$ observable when $\lambda=0$, respectively.  

To summarize, the main results of this subsection are (\ref{eq:product state inequalities 2}) and (\ref{eq:product state inequalities 3}), which represent two alternate criterion for the validity of the product state approximation. They both bound the time interval after $t=0$ for which the approximation is expected to hold; i.e, the scrambling time of the system. The inequality derived from the von Neumann entropy (\ref{eq:product state inequalities 2}) is ill-defined in the infinite dimensional case. The inequality derived from the linear entropy suggests that the temporal regime of validity of the product state approximation will be longer when probability distribution for $V_{1}$ or $V_{2}$ are sharply peaked; i.e., when subsystem 1 or 2 are described by a coherent state at $t=0$.

\subsection{Discrepancy between the quantum-quantum and classical-quantum calculations}\label{sec:classical-quantum approx validity}

We now turn our attention to the deviations between the quantum-quantum and classical-quantum calculations in the context of small-$\lambda$ perturbation theory. The classical-quantum equations of motion that we will attempt to solve perturbatively are (\ref{eq:CQ quantum EOM}) and (\ref{eq:CQ density matrix evolution}). As in the previous subsection, we will find it useful to work in the interaction picture for the quantum variables:
\begin{equation}
	\INT{\mathbf{z}}=e^{iH_{2}t}\mathbf{z}, \quad \INT{\rho}_{\CQ} = e^{iH_{2}t} \rho_{\CQ}  e^{-iH_{2}t}.
\end{equation}
Furthermore, let us define
\begin{align}
	\mathbf{Q} =  \begin{pmatrix*}[c]
	q_{1} \\ p_{1}
	\end{pmatrix*} , \quad M = \begin{pmatrix*}[c]
	0 & 1 \\ -1 & 0
	\end{pmatrix*}.
\end{align}
In terms of these, the equations of motion are
\begin{subequations}
\begin{align}
\di_{t} \mathbf{Q} & = M [ \text{grad}_{\mathcal{H}_{1}}(\mathbf{Q}) + \lambda \, (\INT{\mathbf{z}}^{\dagger} \INT{V}_{2} \INT{\mathbf{z}}) \, \text{grad}_{\mathcal{V}_{1}}(\mathbf{Q}) ],\\
i\di_{t} \INT{\mathbf{z}} & = \lambda \mathcal{V}_{1} \INT{V}_{2}\INT{\mathbf{z}}, \\
i\di_{t}\INT{\rho}_{\CQ} & = \lambda \mathcal{V}_{1} [\INT{V}_{2},\INT{\rho}_{\CQ}],
\end{align}
\end{subequations}
where $\text{grad}_{\mathcal{H}_{1}}(\mathbf{Q})$ and $\text{grad}_{\mathcal{V}_{1}}(\mathbf{Q})$ are the gradients of $\mathcal{H}_{1}$ and $\mathcal{V}_{1}$ evaluated at $\mathbf{Q}$, respectively. We assume the following perturbative expansion:
\begin{subequations}
\begin{align}
	\mathbf{Q} & =  \mathbf{Q}^{\zero}  + \lambda \mathbf{Q}^{\one} + \cdots \\
	\INT{\mathbf{z}} & =  \INT{\mathbf{z}}^{\zero} +\lambda \INT{\mathbf{z}}^{\one}  + \cdots \\
	\INT{\rho}_{\CQ} & =  \INT{\rho}_{\CQ}^{\zero}  + \lambda \INT{\rho}_{\CQ}^{\one} + \cdots
\end{align}
\end{subequations}
The zeroth order equations of motion are
\begin{gather}
	\nonumber \di_{t} \mathbf{Q}^{\zero} = M \, \text{grad}_{\mathcal{H}_{1}}(\mathbf{Q}^{\zero}) , \\
	\di_{t}\INT{\mathbf{z}}^{\zero} = 0, \quad  \di_{t}\INT{\rho}_{\CQ}^{\zero} = 0,\label{eq:CQ zeroth order}
\end{gather}
The first equation in (\ref{eq:CQ zeroth order}) states that $\mathbf{Q}^{\zero}$ is just the classical phase space trajectory of subsystem 1 when subsystem 2 is completely neglected. This equation may or may not be easy to solve depending on the exact form of $\mathcal{H}_{1}$. Conversely, the last two equations are always easy to solve: Assuming that the density matrices for subsystem 2 in the QQ and CQ calculations coincide to zeroth order in $\lambda$,
\begin{equation}
	\INT{\rho}_{\CQ}^{\zero} = \INT{\rho}_{2}^{\zero}, 
\end{equation}
we obtain
\begin{equation}
	\INT{\mathbf{z}}^{\zero} = \INT{\mathbf{u}}_{2}, \quad \INT{\rho}_{\CQ}^{\zero} = \INT{\mathbf{u}}_{2}\INT{\mathbf{u}}_{2}^{\dagger},
\end{equation}
with $\INT{\mathbf{u}}_{2}$ being the same constant column vector appearing in \S\ref{sec:QQ perturbative}.

The first order evolution equations for the classical degrees of freedom are of the form
\begin{gather}\nonumber
	\di_{t} \mathbf{Q}^{\one} = M [ \text{Hess}_{\mathcal{H}_{1}}\!(\mathbf{Q}^{\zero}) \mathbf{Q}^{\one} +  \lambda \langle V_{2} \rangle_{\zero} \, \text{grad}_{\mathcal{V}_{1}}\!(\mathbf{Q}^{\zero}) ] , \\
i\di_{t} \INT{\mathbf{z}}^{\one} = \lambda \mathcal{V}_{1}^{\zero} \INT{V}_{2}\INT{\mathbf{u}}_{2} , \quad
i\di_{t}\INT{\rho}_{\CQ}^{\one} = \lambda \mathcal{V}_{1}^{\zero} [\INT{V}_{2}, \INT{\mathbf{u}}_{2}\INT{\mathbf{u}}_{2}^{\dagger}],	
\end{gather}
where
\begin{equation}
	\mathcal{V}_{1}^{\zero} = \mathcal{V}_{1}(\mathbf{Q}^{\zero}),
\end{equation}
and $\text{Hess}_{\mathcal{H}_{1}}(\mathbf{Q}^{\zero})$ is the Hessian matrix of $\mathcal{H}_{1}$ evaluated at $\mathbf{Q}^{\zero}$.  We can easily solve for the first order density matrix $\INT{\rho}_{\CQ}^{\one}$:
\begin{align}
	\INT{\rho}_{\CQ}^{\one}(t) = -i\int\limits_{\mathcal{T}} \mathcal{V}'^{\zero}_{1} [\INT{V}'_{2},\INT{\mathbf{u}}_{2} \INT{\mathbf{u}}_{2}^{\dagger}].
\end{align}
This is very similar in form to the first order density matrices in the QQ calculation (\ref{eq:perturbative rho solution}). In fact, the difference between the density matrices for subsystem 2 in the QQ and CQ calculations is simply
\begin{align}\nonumber
	 \delta \INT{\rho}(t) & = \INT{\rho}_{2}(t) - \INT{\rho}_{\CQ}(t) \\   & = -i \lambda \int\limits_{\mathcal{T}} \left\{ \langle V'_{1} \rangle- \mathcal{V}'_{1}\right\}_{0} [\INT{V}'_{2},\INT{\mathbf{u}}_{2} \INT{\mathbf{u}}_{2}^{\dagger}] + \mathcal{O}(\lambda^{2}).
\end{align}
Interestingly, we see that the leading order contribution to $\delta \INT{\rho}(t)$ is small if $\langle V'_{1} \rangle \approx \mathcal{V}'_{1}$ when $\lambda = 0$. That is, the difference between the QQ and CQ density matrices for subsystem 2 will be minimized if the discrepancy between the expectation value of $\hat{V}_{1}$ and its classical value is small at zero coupling.

Now, we define the relative error between observables in the QQ and CQ schemes. As above, suppose $\hat{A}_{1} : \mathscr{H_{1}} \to \mathscr{H_{1}}$ and $\hat{A}_{2} : \mathscr{H_{2}} \to \mathscr{H_{2}}$ are generic operators corresponding to the classical phase space functions $\mathcal{A}_{1} = \mathcal{A}_{1}(q_{1},p_{1})$ and $\mathcal{A}_{2} = \mathcal{A}_{2}(q_{2},p_{2})$, respectively. Then, the relative error in the observed value of $\hat{A}_{1}$ is defined as
\begin{equation}
	\Delta A_{1} = \frac{\langle A_{1}\rangle_{\QQ} - \mathcal{A}_{1,\CQ}}{\langle A_{1}\rangle_{\QQ}},
\end{equation}
where $\langle A_{1}\rangle_{\QQ}$ is evaluated in the QQ scheme and $\mathcal{A}_{1,\CQ}$ is evaluated in the CQ scheme. The relative error in $\hat{A}_{2}$ is:
\begin{equation}
	\Delta A_{2} = \frac{\langle A_{2}\rangle_{\QQ} - \langle A_{2}\rangle_{\CQ} }{\langle A_{2}\rangle_{\QQ}}.
\end{equation}
Using the above perturbative expansions, we find
\begin{equation}
	\Delta A_{1} = \left\{ \frac{\langle A_{1}\rangle - \mathcal{A}_{1}}{\langle A_{1}\rangle} \right\}_{\zero} + \mathcal{O}(\lambda),
\end{equation}
and
\begin{equation}\label{eq:CQ error in subsystem 2 observables}
	\Delta A_{2} = -i\lambda \left\{ \frac{ \int_{\mathcal{T}}  \langle (V'_{1} \rangle- \mathcal{V}'_{1}) \langle[A_{2}',V'_{2}] \rangle  }{\langle A_{2}\rangle} \right\}_{\zero} + \mathcal{O}(\lambda^{2}).
\end{equation}
An important point is that $\Delta A_{1}$ does not vanish in the $\lambda \rightarrow 0$ limit. That is, even at zero coupling there will be a discrepancy between the QQ and CQ predictions for the $\hat{A}_{1}$ observable. This is directly due to the fact that subsystem 1 is treated quantum mechanically in the QQ case and classically in the CQ case.

In equation (\ref{eq:classical-quantum approximation}), we defined a quantity $\varepsilon$ that quantifies the magnitude of the approximation that led from the nearly product state to classical-quantum formalisms.  It is fairly easy to confirm that $\varepsilon = \Delta V_{1} + \mathcal{O}(\lambda)$, or, equivalently
\begin{equation}\label{eq:epsilon def}
	\varepsilon = \varepsilon^{\zero} + \mathcal{O}(\lambda), \quad   \varepsilon^{\zero}= \left\{ \frac{\langle V_{1}\rangle - \mathcal{V}_{1}}{\langle V_{1}\rangle} \right\}_{\zero}.
\end{equation}
If $\lambda \ll 1$, it would appear that a necessary condition for the validity of the CQ approximation is that $\varepsilon^{\zero}$ is small.

We conclude this section by noting that both equations (\ref{eq:CQ error in subsystem 2 observables}) and (\ref{eq:epsilon def}) imply that the CQ approximation gets better the smaller the difference between $\langle V_{1} \rangle_{0}$ and $\mathcal{V}_{1}^{\zero}$ becomes. This implies that for the CQ approximation to be useful, the initial quantum state of subsystem 1 has to be selected such that the expectation value of $\hat{V}_{1}$ is close to its classical value. We can state this another way by recalling that the generalized Bohr correspondence principle states that, in some limit, there should exist quantum states where expectation values match classical trajectories.  Hence, it appears that a necessary condition for the validity of the CQ approximation is the imposition of initial data consistent with the Bohr correspondence principle. Actually finding such data will depend very much on the nature of the unperturbed Hamiltonian $\hat{H}_{1}$.  In \S\ref{sec:nonlinearly coupled oscillators}, we concentrate on the case where each subsystem is a simple harmonic oscillator and it is relatively straightforward to find states consistent with $\langle V_{1} \rangle_{0} \approx \mathcal{V}_{1}^{\zero}$. In appendix \S\ref{sec:CQ in 1D potential}, we demonstrate how $|\varepsilon|\ll 1$ can be realized for more general systems where subsystem 1 corresponds to a particle moving in a one-dimensional potential with a sharply peaked wavefunction.

\section{Example: nonlinearly coupled oscillators}\label{sec:nonlinearly coupled oscillators}

To illustrate the various computational approaches described above in action, let us focus on a simple model in this section: two simple harmonic oscillators with a nonlinear coupling. We will compare numeric solutions to the equations of motion in the quantum-quantum, classical-quantum, and classical-classical calculation methods. Our motivation is to investigate when the CQ and CC schemes can yield reasonable approximations to the QQ dynamics. We therefore concentrate on quantum initial data that is expected to yield the most ``classical'' evolution in the absence of coupling; i.e., we will use coherent state initial data for the oscillators whenever applicable.

\subsection{Hamiltonian and operators}

We consider a system of two simple harmonic oscillators with nonlinear coupling. The oscillators are described by canonical position and momentum variables $(q_{1},p_{1})$ and $(q_{2},p_{2})$, respectively.  We can define alternative complex coordinates, as usual, by
\begin{equation}
	a = \frac{q_{1}+ip_{1}}{\sqrt{2}}, \quad b = \frac{q_{2}+ip_{2}}{\sqrt{2}}.
\end{equation}
The relevant nonzero Poisson brackets are
\begin{equation}
	\{q_{1},p_{1}\}=\{q_{2},p_{2}\} = 1, \quad \{a,a^{*}\}=\{b,b^{*}\} = -i.
\end{equation}
All the above variables are assumed to be dimensionless.

The classical Hamiltonian is
\begin{align}\nonumber
	\mathcal{H} &= \tfrac{1}{2} \Omega_{1}(q_{1}^{2}+p_{1}^{2}) +  \tfrac{1}{2} \Omega_{2} (q_{2}^{2}+p_{2}^{2})+ \lambda \bar{\Omega} q_{1}^{\nu} q_{2}^{\nu} \\
	   & = \Omega_{1} aa^{*} + \Omega_{2} bb^{*} + 2^{-\nu} \lambda \bar{\Omega} (a+a^{*})^{\nu}(b+b^{*})^{\nu}. 
\end{align}
Here, $\Omega_{1}>0$ and $\Omega_{2}>0$ are the natural frequencies of each oscillator and their arithmetic mean is
\begin{equation}
	\bar{\Omega} = \tfrac{1}{2}(\Omega_{1}+\Omega_{2}).
\end{equation}
As before, $\lambda$ is a dimensionless coupling parameter, and $\nu = 1, 2, 3,\ldots$ is a positive integer. Note that if $\nu$ is odd, the Hamiltonian is not bounded from below.

For this example, we identify the various components of the classical Hamiltonian defined in section \ref{sec:general Hamiltonian} as follows:
\begin{align}\nonumber
	\mathcal{H}_{1} & =  \Omega_{1} aa^{*}, & \mathcal{V}_{1} & = 2^{-\nu/2} \bar{\Omega}^{1/2} (a+a^{*})^{\nu}, \\
	\mathcal{H}_{2} & =  \Omega_{2} bb^{*}, & \mathcal{V}_{2} & = 2^{-\nu/2} \bar{\Omega}^{1/2} (b+b^{*})^{\nu}.
\end{align}
with the corresponding quantum operators
\begin{align}\nonumber
	\hat{H}_{1} & = \Omega_{1} (\hat{a}^{\dagger}\hat{a} + \tfrac{1}{2}), & \hat{V}_{1} & = 2^{-\nu/2} \bar{\Omega}^{1/2} (\hat{a}+\hat{a}^{\dagger})^{\nu}, \\
	\hat{H}_{2} & = \Omega_{2} (\hat{b}^{\dagger}\hat{b} + \tfrac{1}{2}),  & \hat{V}_{2} & = 2^{-\nu/2} \bar{\Omega}^{1/2} (\hat{b}+\hat{b}^{\dagger})^{\nu}.\label{eq:SHO operators}
\end{align}
with
\begin{equation}
	[\hat{a},\hat{a}^{\dagger}] = [\hat{b},\hat{b}^{\dagger}] = 1.
\end{equation}
The solution of the eigenvalue problems for $\hat{H}_{1}$ and $\hat{H}_{2}$ are, of course, easy to solve:
\begin{subequations}
\begin{align}
	\hat{H_{1}}|n\rangle & = \Omega_{1}(n+\tfrac{1}{2}) |n\rangle, \\
	\hat{H_{2}}|m\rangle & = \Omega_{2}(m+\tfrac{1}{2}) |m\rangle,
\end{align}
\end{subequations}
with $n,m=0,1,2\ldots$  In these bases, matrix representations of the operators (\ref{eq:SHO operators}) are
\begin{align}\nonumber
	H_{1} & = \Omega_{1} (N + \tfrac{1}{2}I), & V_{1} & = \bar{\Omega}^{1/2} Q^{\nu}, \\
	H_{2} & = \Omega_{2} (N + \tfrac{1}{2}I),  & V_{2} & = \bar{\Omega}^{1/2} Q^{\nu}, \label{eq:SHO matrices}
\end{align}
where $Q$ and $N$ are matrix representations of the position and number operators
\begin{gather}
	Q  = \frac{1}{\sqrt{2}}\begin{pmatrix*}[c]
	 0 & \sqrt{1} &  & &  \\
	 \sqrt{1} & 0 & \sqrt{2} & & \\
	  & \sqrt{2} & 0& & \\
	  & & & \ddots
	\end{pmatrix*},  \,\,\,
	N = \begin{pmatrix*}[c]
	 0 & &  & &  \\
	 & 1 &  & & \\
	  &  & 2 &  & \\
	  & &  & \ddots
	\end{pmatrix*},
\end{gather}
and $I$ is the identity matrix. We will also require the matrix representation of the momentum operator:
\begin{gather}
	P  =  \frac{1}{\sqrt{2}i} \begin{pmatrix*}[c]
	 0 & +\sqrt{1} &  & &  \\
	 -\sqrt{1} & 0 & +\sqrt{2} & & \\
	  & -\sqrt{2} & 0& & \\
	  & & & \ddots
	\end{pmatrix*}.
\end{gather}

\subsection{Equations of motion and initial data}

\subsubsection{Dimensionless variables}

In order to simplify the formulae below, we will normalize all quantities by the mean frequency of the free oscillators $\bar{\Omega} = \tfrac{1}{2}(\Omega_{1}+\Omega_{2})$. A useful dimensionless time variable is then
\begin{equation}
	\tau = \bar{\Omega}t = \tfrac{1}{2}(\Omega_{1}+\Omega_{2})t.
\end{equation}
We define an oscillator asymmetry parameter $\sigma \in (-1,1)$ by
\begin{equation}
	\sigma = \frac{\Omega_{1}-\Omega_{2}}{\Omega_{1}+\Omega_{2}}, \quad \Omega_{1} = (1+\sigma)\bar\Omega, \quad \Omega_{2} = (1-\sigma)\bar\Omega.
\end{equation}
In the plots below, all energies are measured in units of $\bar\Omega$.

\subsubsection{Quantum-quantum case}

In dimensionless form, the dynamical system governing the system's evolution (\ref{eq:QQ EOM general}) in the quantum-quantum case is explicitly:
\begin{multline}\label{eq:QQ EOM}
	i \dot{Z} = (1+\sigma)(N + \tfrac{1}{2}I)Z  \\ +  (1-\sigma)Z(N + \tfrac{1}{2}I) + \lambda Q^{\nu} Z Q^{\nu},
\end{multline}
where $\dot{Z} = dZ/d\tau$. For all simulations, we will assume that at $\tau=0$ the system is prepared in a product of coherent states for each oscillator. Then,
\begin{equation}
	z_{nm}(0) = \exp\left( -\frac{|\alpha|^{2}+|\beta|^{2}}{2} \right) \frac{\alpha^{n}\beta^{m}}{\sqrt{n!m!}},
\end{equation}
where $\alpha$ and $\beta$ are dimensionless complex parameters. For convenience, we parametrize these as
\begin{equation}\label{eq:QQ alpha and beta}
	\alpha = \sqrt{ \frac{\zeta_{1}}{1+\sigma }} e^{i\phi_{1}}, \quad \beta = \sqrt{ \frac{\zeta_{2}}{1-\sigma }} e^{i\phi_{2}},
\end{equation}
where $\zeta_{1}$ and $\zeta_{2}$ are nonnegative parameters, and $\phi_{1}$ and $\phi_{2}$ are angles.  Unless stated otherwise, we will choose
\begin{equation}
	\phi_{1} = \phi_{2} = \pi/2 \quad \Rightarrow \quad \langle q_{1}(0) \rangle =  \langle q_{2}(0) \rangle = 0.
\end{equation}
This choice is expected to minimize the initial value of the interaction energy.

Once we have a solution for $Z(\tau)$ we can calculate the values of various observables and other quantities of interest by using the formulae in Table \ref{tab:observables}. Note that in this table, the oscillator energies are defined as
\begin{align}
	E_{1} = \langle \hat{H}_{1} \rangle, \quad E_{2} = \langle \hat{H}_{2}\rangle , \quad E_\text{int} =  \lambda \langle \hat{V}_{1} \otimes  \hat{V}_{2} \rangle.
\end{align}
From the table, we can see that $\zeta_{1}$ and $\zeta_{2}$ are equal to the initial energies of oscillator 1 and 2 above the ground state in units of $\bar\Omega$.  Once can confirm by explicit calculation that
\begin{equation}\label{eq:conservation of energy}
	\di_{\tau} (E_{1}+E_{2}+E_\text{int}) = 0.
\end{equation}
as expected.
\newlength{\width}
\width.2\textwidth
\begin{table*}
\begin{tabular}{p{1.1\width}p{1.105\width}p{\width}p{0.70\width}p{0.85\width}}\hline
 & quantum-quantum & classical-quantum & classical-classical & initial value \\ \hline\hline
O1 position $q_{1}$ & $\text{Tr}(Z Z^{\dagger} Q)$ & $\sqrt{2} \, \text{Re} \, a$ & $\sqrt{2} \, \text{Re} \, a$ & $\sqrt{2\zeta_{1}/(1+\sigma)} \cos \phi_{1}$ \\
O2 position $q_{2}$ & $\text{Tr}(Z^{\T} Z^{*}  Q)$ & $\text{Tr}\,(\mathbf{z}\mathbf{z}^{\dagger}  Q)$ & $\sqrt{2} \, \text{Re} \, b$ & $\sqrt{2\zeta_{2}/(1-\sigma)} \cos \phi_{2}$ \\
O1 momentum $p_{1}$ & $\text{Tr}(Z Z^{\dagger} P)$ & $\sqrt{2} \, \text{Im} \, a$ & $\sqrt{2} \, \text{Im} \, a$ & $\sqrt{2\zeta_{1}/(1+\sigma)} \sin \phi_{1}$ \\
O2 momentum $p_{2}$ & $\text{Tr}(Z^{\T} Z^{*}  P)$ & $\text{Tr}\,(\mathbf{z}\mathbf{z}^{\dagger} P)$ & $\sqrt{2} \, \text{Im} \, b$ & $\sqrt{2\zeta_{2}/(1-\sigma)} \sin \phi_{2}$ \\
O1 occupation number $\langle n \rangle$ & $\text{Tr}(Z Z^{\dagger}  N)$ & n/a  & n/a & $\zeta_{1}/(1+\sigma)$ \\ 
O2 occupation number $\langle m \rangle$ & $\text{Tr}(Z^{\T} Z^{*}  N)$ & $\text{Tr}(\mathbf{z}\mathbf{z}^{\dagger} N)$ & n/a & $\zeta_{2}/(1-\sigma)$ \\ 
$\text{Var}(n) = \langle n^{2} \rangle - \langle n \rangle^{2}$ & $\text{Tr}(Z Z^{\dagger} N^{2}) - \text{Tr}^{2}(Z Z^{\dagger} N)  $  & n/a & n/a & $\zeta_{1}/(1+\sigma)$ \\ 
$\text{Var}(m) = \langle m^{2} \rangle - \langle m \rangle^{2}$ & $\text{Tr}(Z^{\T} Z^{*}  N^{2}) - \text{Tr}^{2}(Z^{\T} Z^{*}  N)  $ & $\text{Tr}(\mathbf{z}\mathbf{z}^{\dagger} N^{2}) -\text{Tr}^{2}(\mathbf{z}\mathbf{z}^{\dagger} N)$ & n/a & $\zeta_{2}/(1-\sigma)$ \\ 
O1 energy $E_{1}/\bar{\Omega}$ & $(1+\sigma)[\text{Tr}(Z Z^{\dagger}N)+\tfrac{1}{2}]$ & $(1+\sigma)aa^{*}$ & $(1+\sigma)aa^{*}$ & $\zeta_{1} +$ zero-point energy \\
O2 energy $E_{2}/\bar{\Omega}$ & $(1-\sigma)[\text{Tr}(Z^{\T} Z^{*}N)+\tfrac{1}{2}]$ & $(1-\sigma)[\text{Tr}(\mathbf{z}\mathbf{z}^{\dagger} N)+\tfrac{1}{2}]$ & $(1-\sigma)bb^{*}$ & $\zeta_{2} +$ zero-point energy \\
interaction energy $E_\text{int}/\bar{\Omega}$ & $\lambda \, \text{Tr}(Z^{\dagger} Q^{\nu} Z Q^{\nu})$ & $\lambda (\sqrt{2}\,\text{Re}\,a)^{\nu}  \text{Tr}(\mathbf{z}\mathbf{z}^{\dagger}  Q^{\nu})$ & $2^{\nu} \lambda (\text{Re}\,a)^{\nu} (\text{Re}\,b)^{\nu} $& \\ 
O1 Wigner distribution & $\text{Tr}\,[Z Z^{\dagger} \mathcal{W}(q_{1},p_{1})]$ & n/a & n/a & \\ 
O2 Wigner distribution &  $\text{Tr}\,[Z^{\T} Z^{*} \mathcal{W}(q_{2},p_{2})]$ & $\text{Tr}\,[\mathbf{z}\mathbf{z}^{\dagger}\mathcal{W}(q_{2},p_{2})]$ & n/a & \\ 
\hline
\end{tabular}
\caption{Formulae for various observables and other quantities in the quantum-quantum, classical-quantum, and classical-classical schemes for nonlinearly coupled oscillators. Here, ``O1'' stands for ``oscillator 1'' and ``O2'' stands for ``oscillator 2''. Note that the ``zero-point'' energy for each oscillator is the energy of the ground state when the oscillator's evolution is treated quantum mechanically, and zero when the oscillator dynamics are classical. The Wigner matrix $\mathcal{W}(q,p)$ is defined in \S\ref{sec:Wigner}. Expressions for the initial values of the interaction energy and Wigner distributions are complicated and hence omitted from this table.}\label{tab:observables}
\end{table*}

\subsubsection{Classical-quantum case}

In dimensionless form, the equations of motion for the classical-quantum case read
\begin{subequations}\label{eq:semi-classical EOMs}
\begin{align}
	i \dot{a} & = (1+\sigma) a + \tfrac{1}{\sqrt{2}}  \lambda\nu (\sqrt{2} \, \text{Re}\,a)^{\nu-1} \mathbf{z}^{\dagger} Q^{\nu} \mathbf{z}, \\
	i \dot{\mathbf{z}} & = (1-\sigma) (N + \tfrac{1}{2} I ) \mathbf{z} + \lambda (\sqrt{2} \, \text{Re}\,a)^{\nu} Q^{\nu} \mathbf{z}.
\end{align}
\end{subequations}
We parametrize the initial value of $a(\tau)$ as follows:
\begin{equation}
	a(0) = a_{0}e^{i\phi_{1}}, \quad a_{0} = \sqrt{\frac{\zeta_{1}}{1+\sigma}},
\end{equation}
where $\zeta_{1}$ is a nonnegative real parameter. As for the quantum-quantum case above, we take oscillator 2 to be in a coherent state:
\begin{equation}
	z_{m}(0) = \exp\left( -\frac{|\beta|^{2}}{2} \right) \frac{\beta^{m}}{\sqrt{m!}}, \quad \beta = \sqrt{ \frac{\zeta_{2}}{1-\sigma }} e^{i\phi_{2}},
\end{equation}
where $\zeta_{2}$ is a nonnegative real parameter. Formulae for observables in the case are shown in Table \ref{tab:observables}. These are similar to the quantum-quantum calculation described above, but not identical.  One notable difference is the formulae for the energies:
\begin{align}
	E_{1} = \mathcal{H}_{1}, \quad E_{2} = \langle \hat{H}_{2}\rangle , \quad E_\text{int} =  \lambda \mathcal{V}_{1} \langle \hat{V}_{2} \rangle.
\end{align}
As above, we have conservation of the total energy (\ref{eq:conservation of energy}). From Table \ref{tab:observables}, we see that $\zeta_{1}$ is the dimensionless initial energy of oscillator 1, and $\zeta_{2}$ is the dimensionless initial energy of oscillator 2 above the ground state.

\subsubsection{Classical-classical case}

For the classical-classical case, the equations of motion in dimensionless form read
\begin{subequations}\label{eq:classical EOMs}
\begin{align}
	i \dot{a} & = (1+\sigma)a + 2^{\nu-1} \nu  \lambda (\text{Re}\,a)^{\nu-1} (\text{Re}\,b)^{\nu} , \\
	i \dot{b} & = (1-\sigma)b + 2^{\nu-1} \nu  \lambda (\text{Re}\,a)^{\nu} (\text{Re}\,b)^{\nu-1} .
\end{align}
\end{subequations}
We parameterize initial data as
\begin{equation}
	a(0) = \sqrt{\frac{\zeta_{1}}{1+\sigma}}e^{i\phi_{1}}, \quad b(0) = \sqrt{\frac{\zeta_{2}}{1-\sigma}}e^{i\phi_{2}}.
\end{equation}
In this case, the energies are simply
\begin{align}
	E_{1} = \mathcal{H}_{1}, \quad E_{2} = \mathcal{H}_{2} , \quad E_\text{int} =  \lambda \mathcal{V}_{1} \mathcal{V}_{2},
\end{align}
and we obviously have total energy conservation (\ref{eq:conservation of energy}).  The initial values of the normalized energies are
\begin{equation}
	E_{1}(0) = \bar{\Omega}\zeta_{1}, \quad E_{2}(0) = \bar{\Omega}\zeta_{2}.
\end{equation}

\subsection{Sample Simulations}

In order to obtain approximate numeric solutions of the relevant equations of motion for the quantum-quantum (\ref{eq:QQ EOM}) and classical-quantum (\ref{eq:semi-classical EOMs}) schemes, it is necessary to truncate the infinite dimensional matrices and vectors. We retain the first $(n_\text{max}+1)$ entries of vectors and the upper-left $(n_\text{max}+1) \times (n_\text{max}+1)$ submatrix of all matrices. We would expect such an approximation to be valid if the statistical distribution of occupation numbers of each oscillator remains below $n_\text{max}$ during simulations. For example, in the quantum-quantum case this condition would be
\begin{equation}
	\text{max} \left[ \langle n \rangle + \sqrt{\text{Var}(n)} , \langle m \rangle + \sqrt{\text{Var}(m)}  \right] < n_\text{max}
\end{equation}
Heuristically, simulation results appear to be insensitive to the value of $n_\text{max}$ if the right-hand side of the above inequality is $\gtrsim 1.5$ times larger than the left-hand side.  
\begin{figure*}
\subfigure[\,\,Position of oscillator 2 as a function of time.]%
{\includegraphics[width=0.32\textwidth]{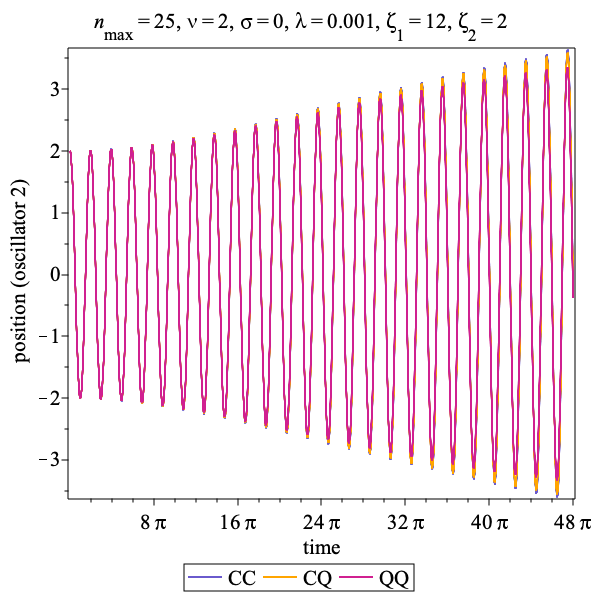} \hfill
\includegraphics[width=0.32\textwidth]{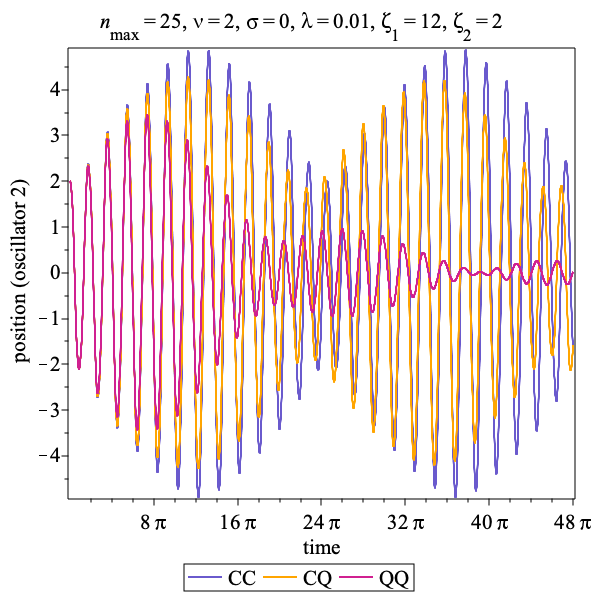} \hfill
\includegraphics[width=0.32\textwidth]{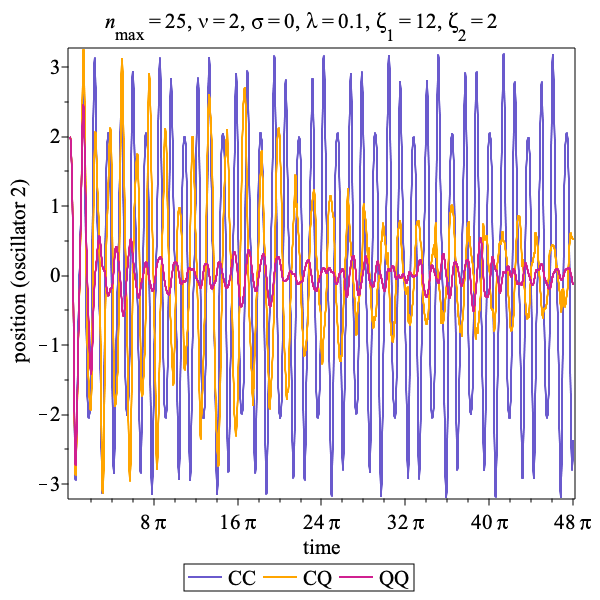}}\\
\subfigure[\,\,Distribution of energy as a function of time]
{\begin{minipage}{\textwidth}
\includegraphics[width=0.32\textwidth]{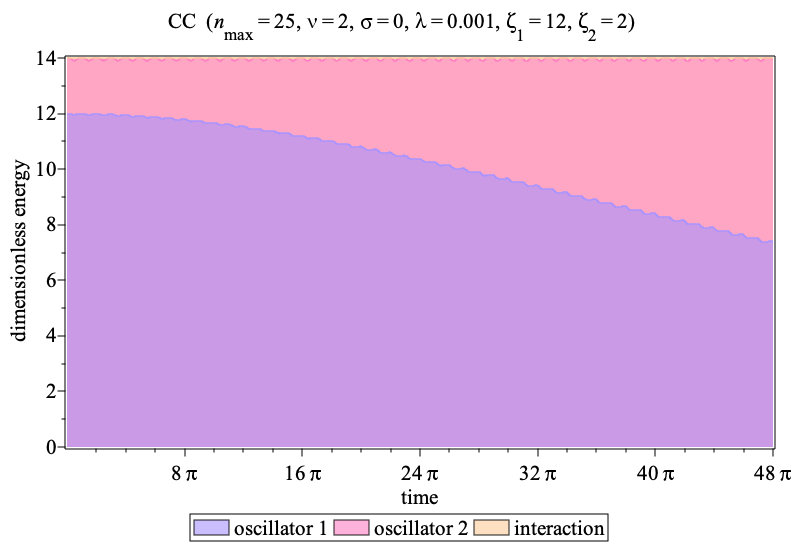} \hfill
\includegraphics[width=0.32\textwidth]{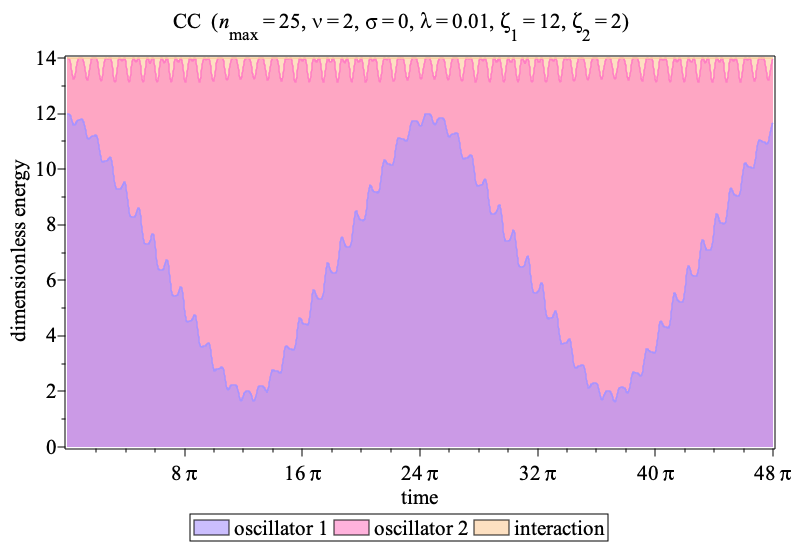} \hfill
\includegraphics[width=0.32\textwidth]{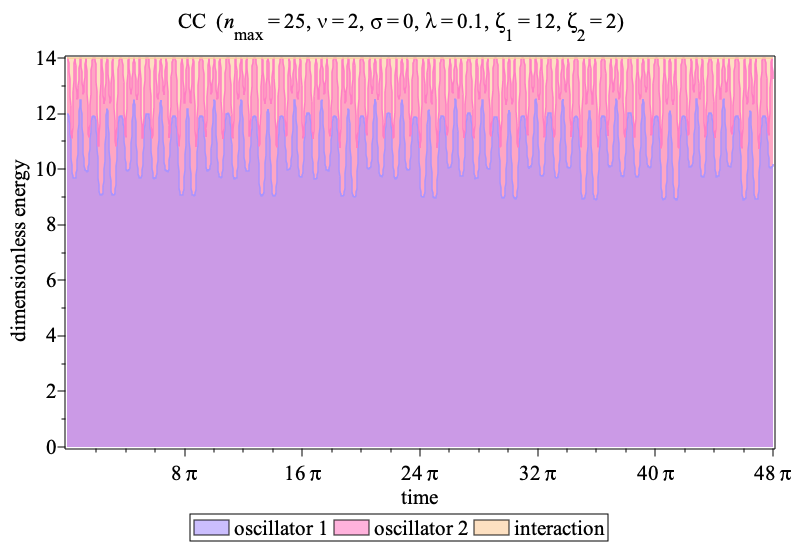} \hfill
\includegraphics[width=0.32\textwidth]{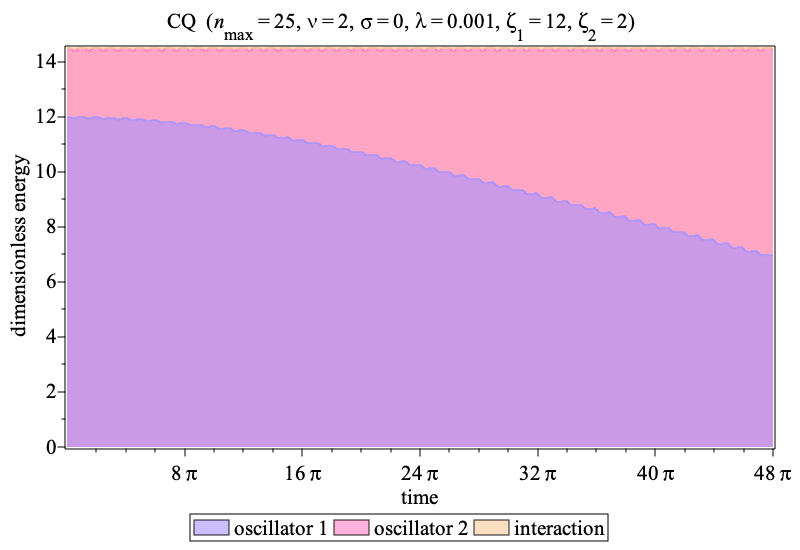} \hfill
\includegraphics[width=0.32\textwidth]{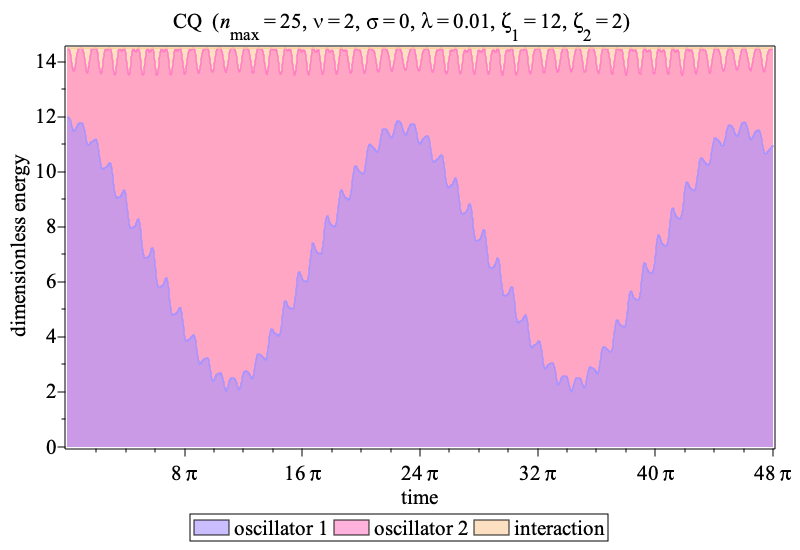} \hfill
\includegraphics[width=0.32\textwidth]{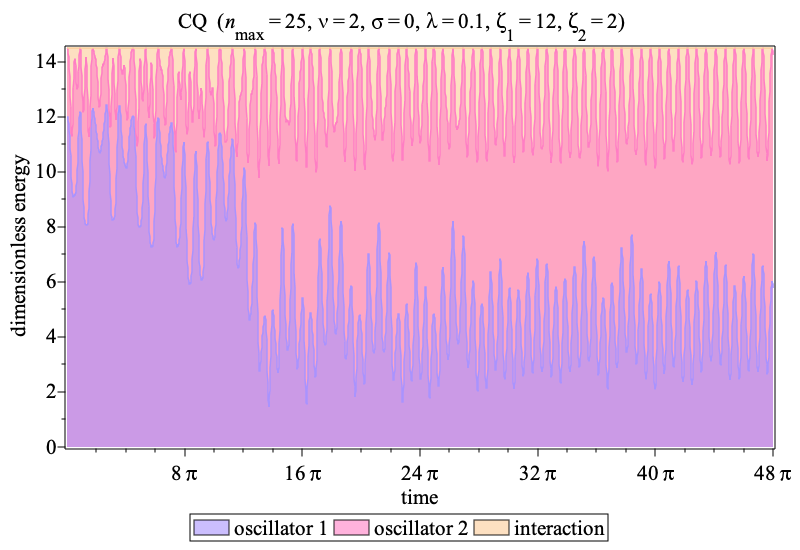} \hfill
\includegraphics[width=0.32\textwidth]{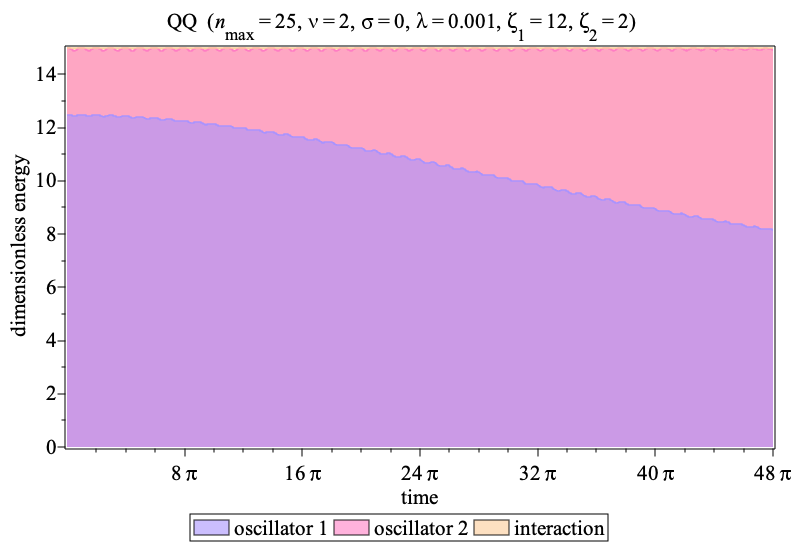} \hfill
\includegraphics[width=0.32\textwidth]{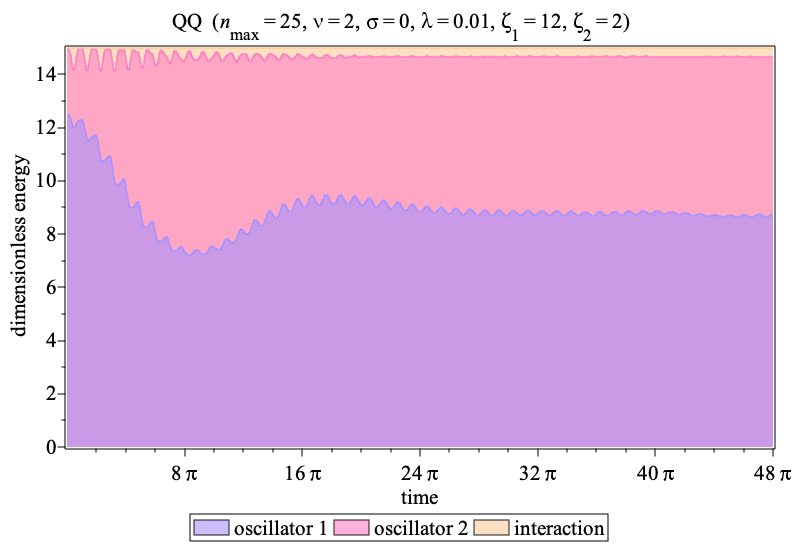} \hfill
\includegraphics[width=0.32\textwidth]{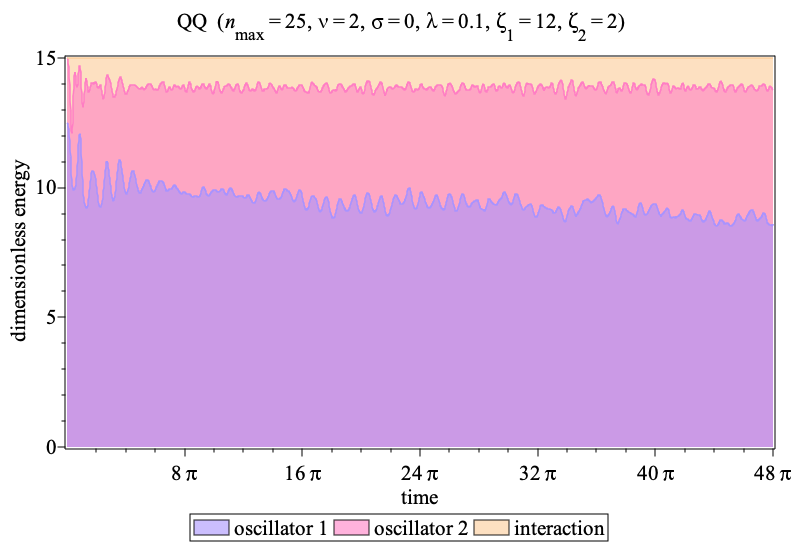} \hfill
\end{minipage}}
\caption{Typical numerical simulation results for the classical-classical (CC), classical-quantum (CQ) and quantum-quantum (QQ) schemes for $\nu=2$; i.e., $\mathcal{H}_\text{int} = \lambda \bar\Omega q_{1}^{2}q_{2}^{2}$.  Results on the left are for small coupling ($\lambda = 0.001$), in the middle are for moderate coupling ($\lambda=0.01$) and on the right are for relatively high coupling ($\lambda = 0.1$); all other parameters are the same for each panel.  In each simulation, the initial energy of oscillator 1 is greater than that of oscillator 2 ($\zeta_{1}>\zeta_{2}$), but each oscillator has identical frequencies ($\sigma=0$).  We see very good qualitative agreement between each of the calculation schemes for $\lambda = 0.001$, but the agreement is poor for $\lambda = 0.1$. Also, it is apparent that the error in the CC and CQ schemes increases with time.}\label{fig:sample results 1}
\end{figure*}
\begin{figure*}
\includegraphics[width=0.32\textwidth]{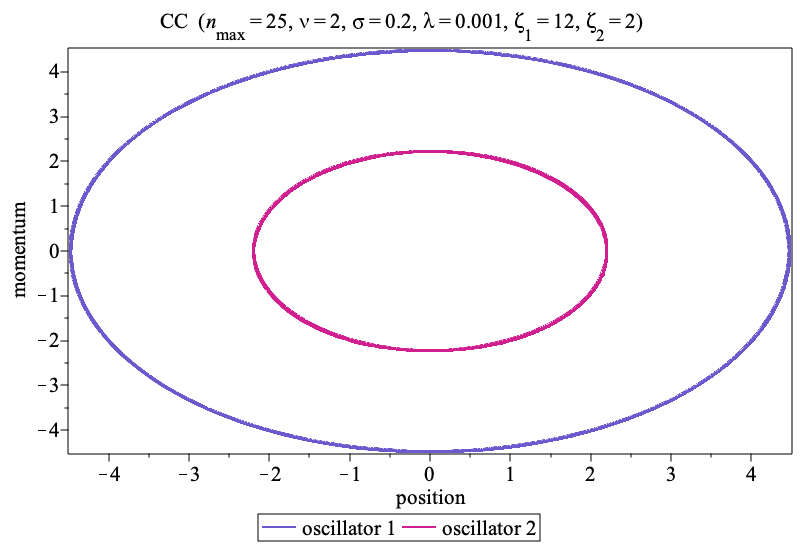} \hfill
\includegraphics[width=0.32\textwidth]{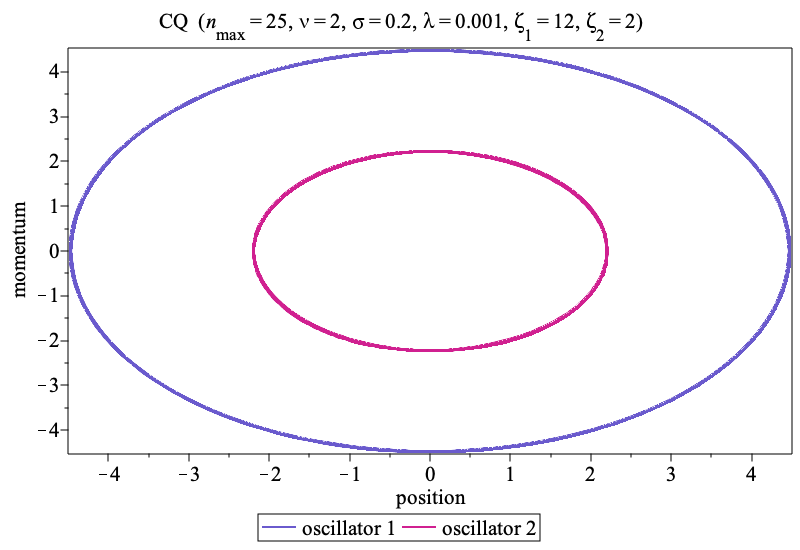} \hfill
\includegraphics[width=0.32\textwidth]{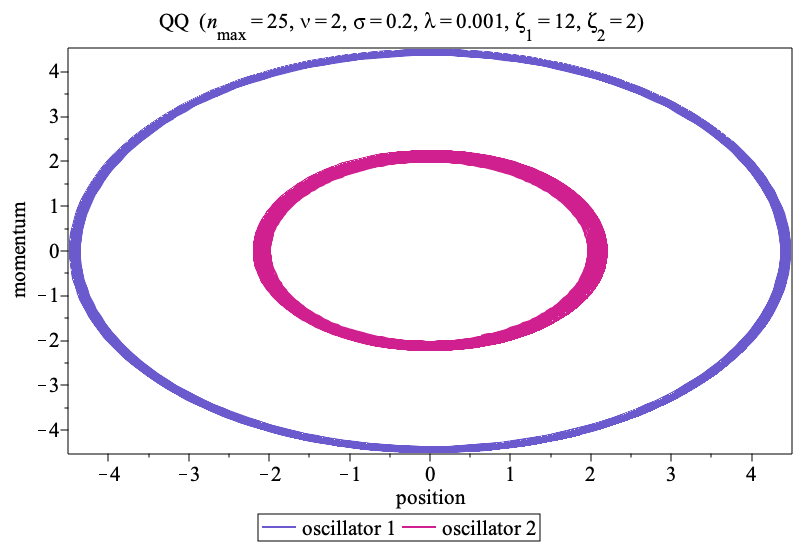} \\
\includegraphics[width=0.32\textwidth]{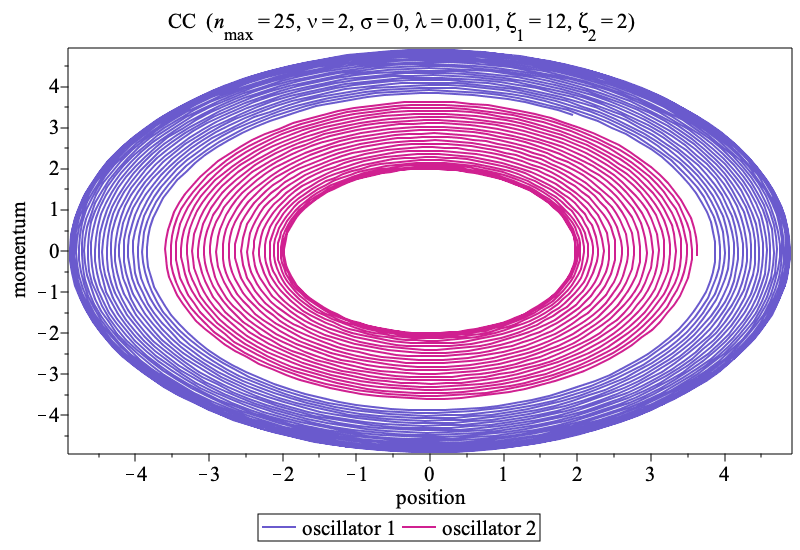} \hfill
\includegraphics[width=0.32\textwidth]{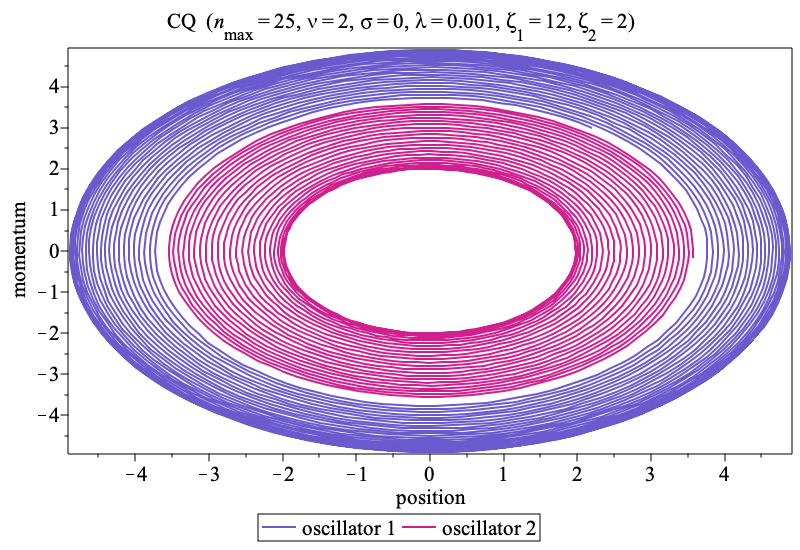} \hfill
\includegraphics[width=0.32\textwidth]{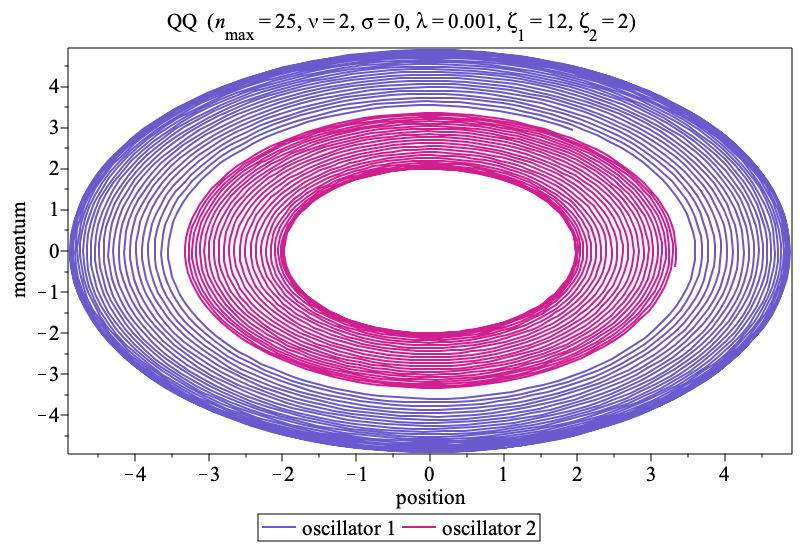} \\
\includegraphics[width=0.32\textwidth]{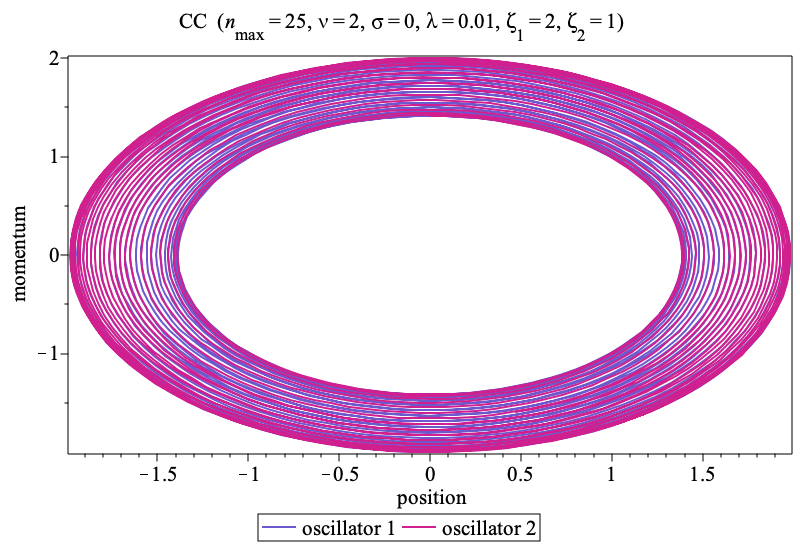} \hfill
\includegraphics[width=0.32\textwidth]{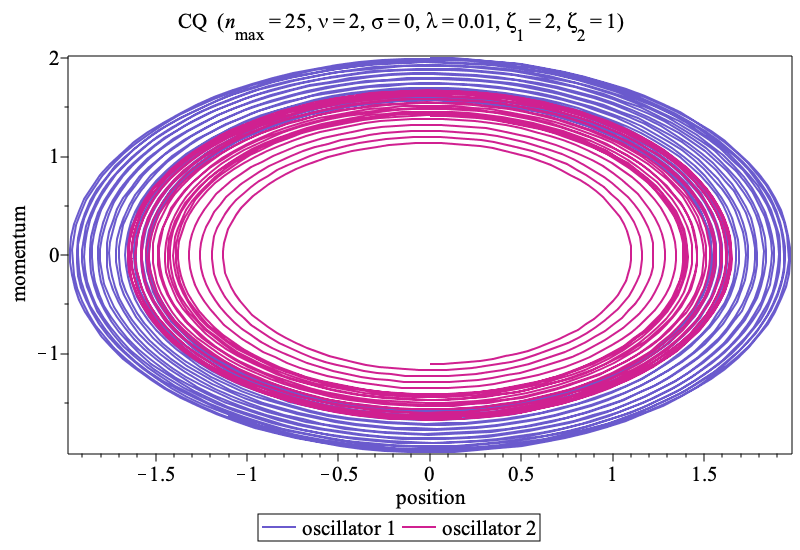} \hfill
\includegraphics[width=0.32\textwidth]{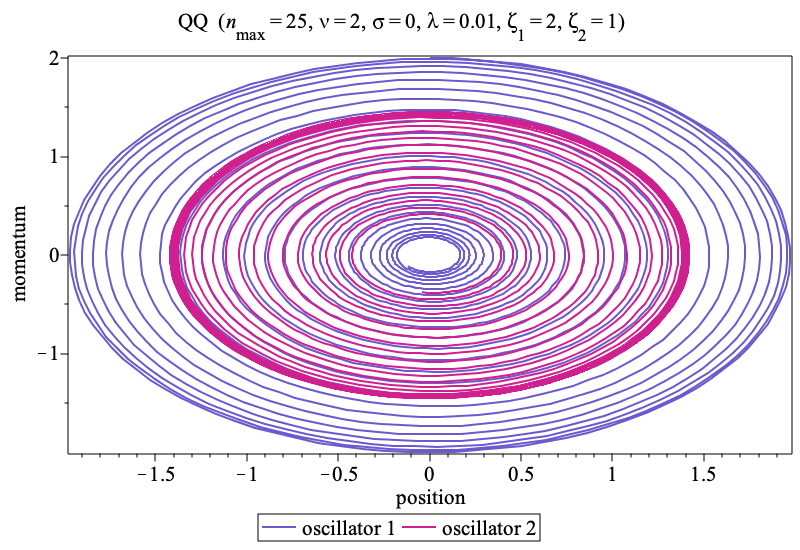} \\
\includegraphics[width=0.32\textwidth]{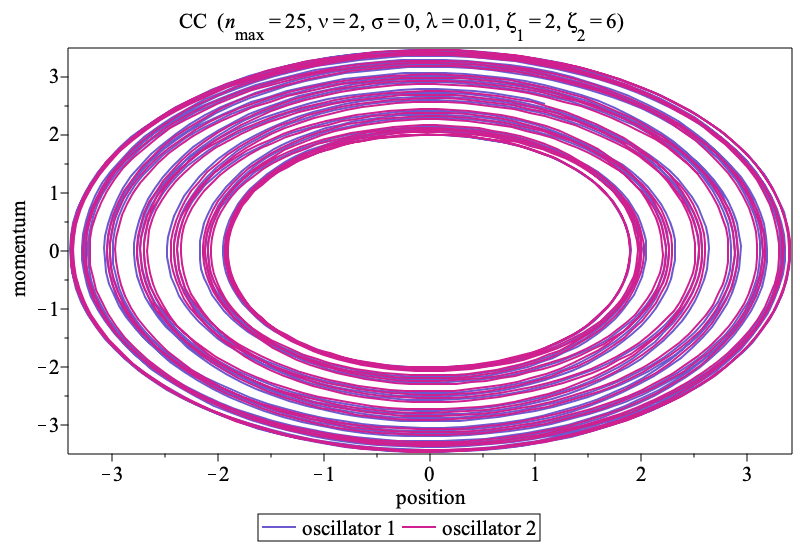} \hfill
\includegraphics[width=0.32\textwidth]{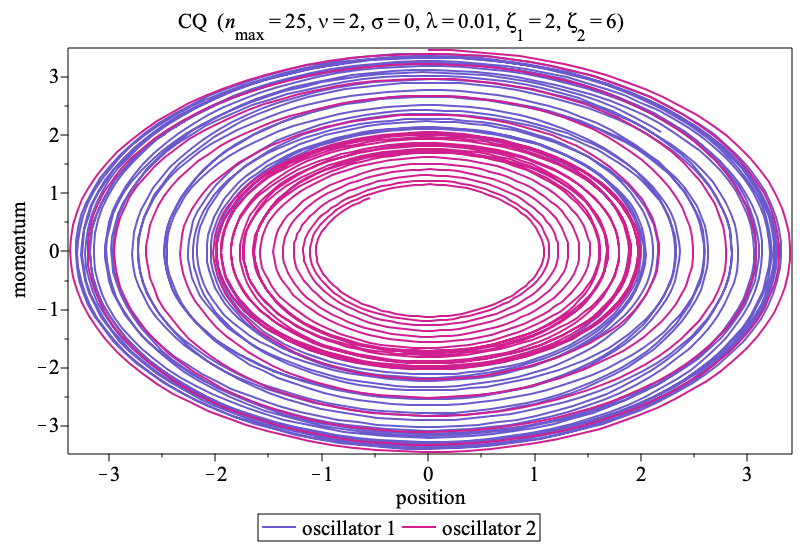} \hfill
\includegraphics[width=0.32\textwidth]{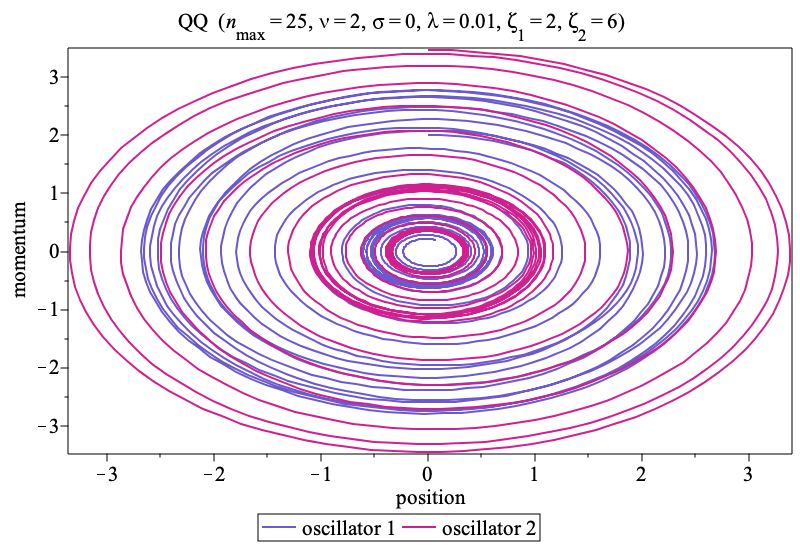} \\
\includegraphics[width=0.32\textwidth]{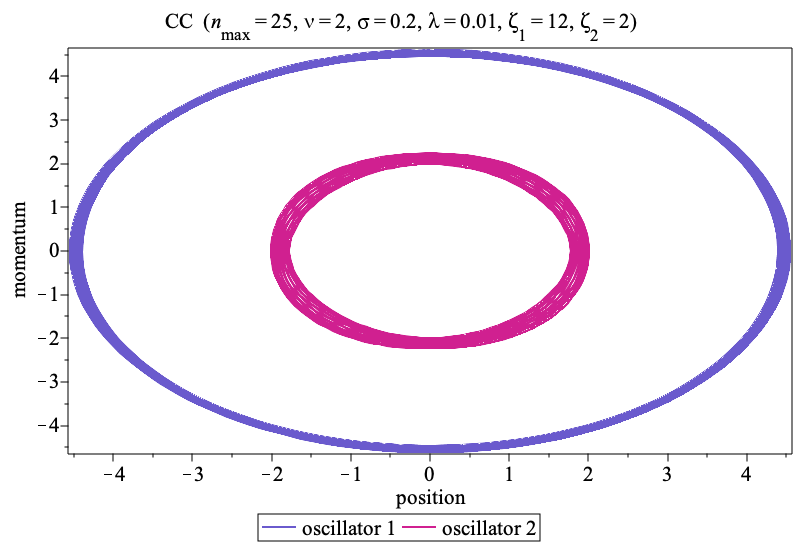} \hfill
\includegraphics[width=0.32\textwidth]{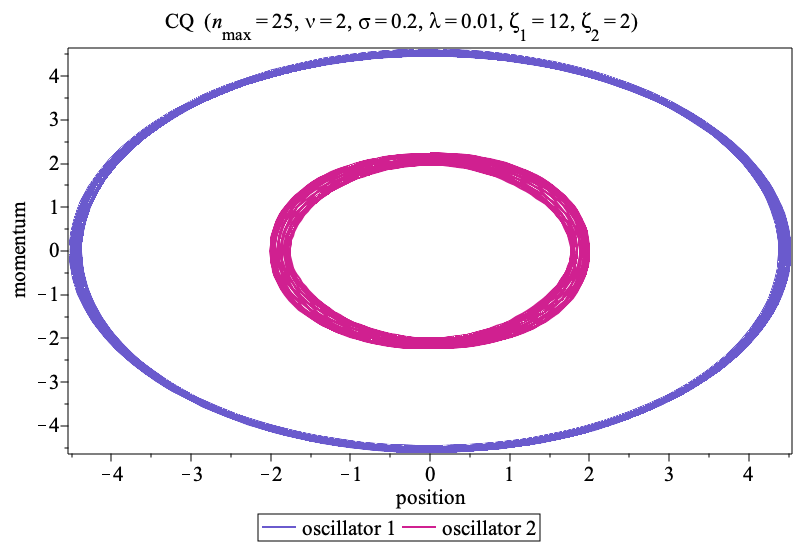} \hfill
\includegraphics[width=0.32\textwidth]{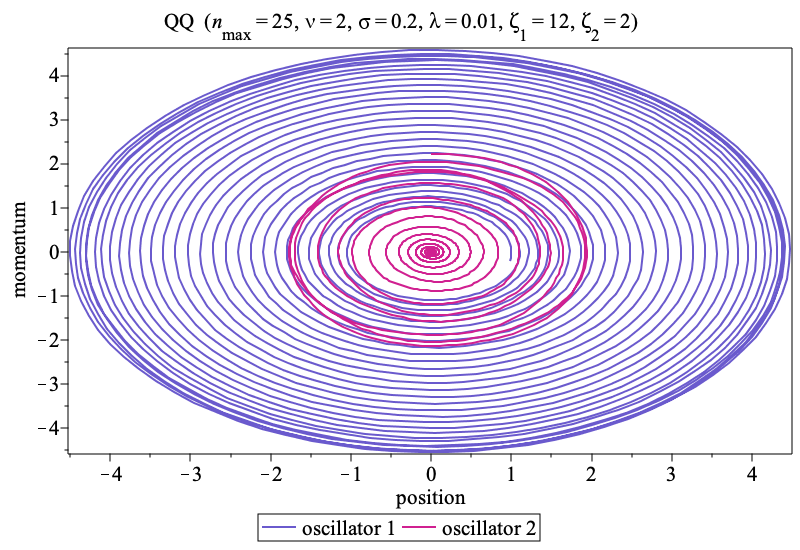}
\caption{Classical-classical (CC), classical-quantum (CQ) and quantum-quantum (QQ) phase portraits for oscillators 1 and 2 for a number of different parameter choices. We take $\nu = 2$ for all simulations.  Each row shows results using the same parameters but a different simulation method, while each column shows results for the same calculation scheme, but with different parameters. There is better agreement between all schemes for smaller values of $\lambda$. Interestingly, it is possible for the CC and CQ schemes to gives visually similar results at higher $\lambda$ that are very different than the QQ calculation, as in the last row.}\label{fig:sample results 2}
\end{figure*}

In figures \ref{fig:sample results 1} and \ref{fig:sample results 2} we show the results of some typical numeric simulations for the $\nu=2$ case. In figure \ref{fig:sample results 1}, we show simulation outputs for the trajectory of oscillator 2 and the partition of energy in the system as functions of time. In figure \ref{fig:sample results 2}, we show the phase portraits of each oscillator for a number of different parameter combinations. The main qualitative solution to be drawn is that for small coupling, the results of the CC, CQ and QQ schemes match up reasonably well, but for higher coupling discrepancies become apparent.

\begin{figure*}
\includegraphics[width=0.24\textwidth]{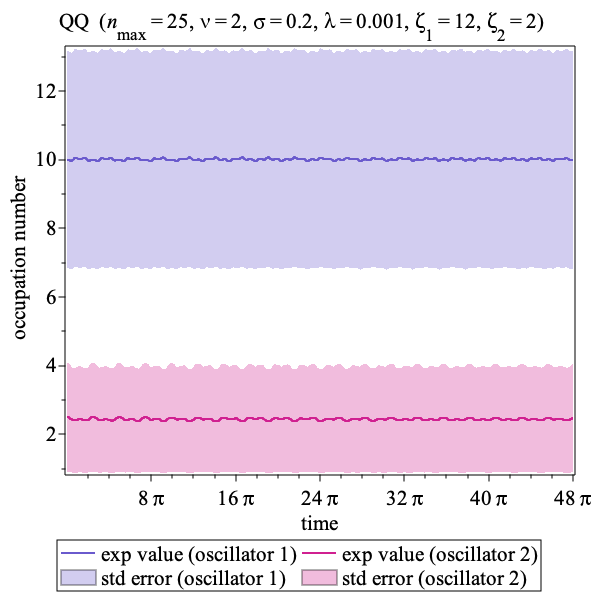} \hfill
\includegraphics[width=0.24\textwidth]{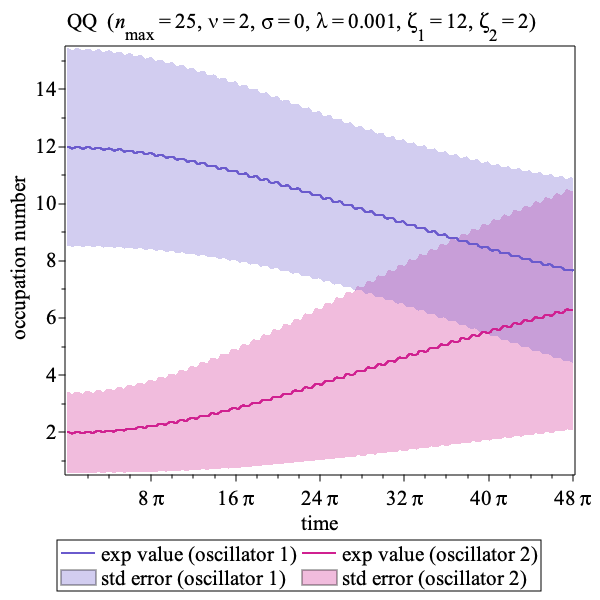} \hfill
\includegraphics[width=0.24\textwidth]{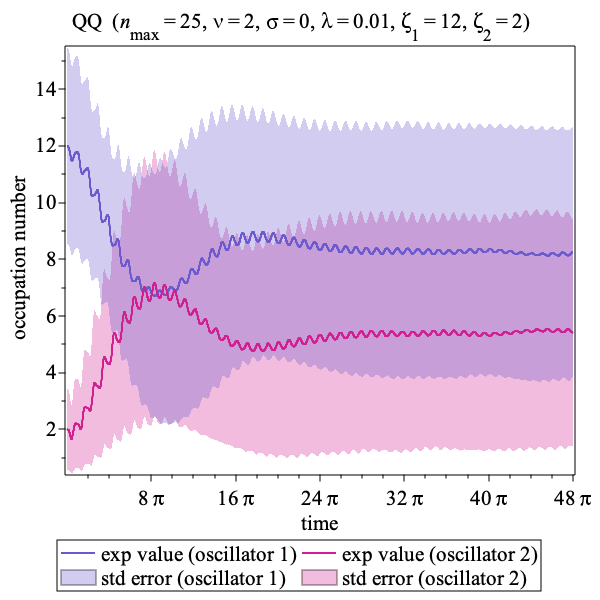} \hfill
\includegraphics[width=0.24\textwidth]{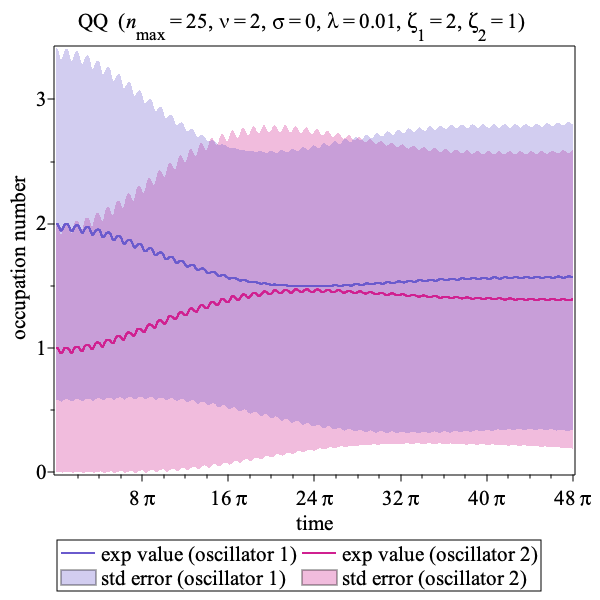} \hfill
\includegraphics[width=0.24\textwidth]{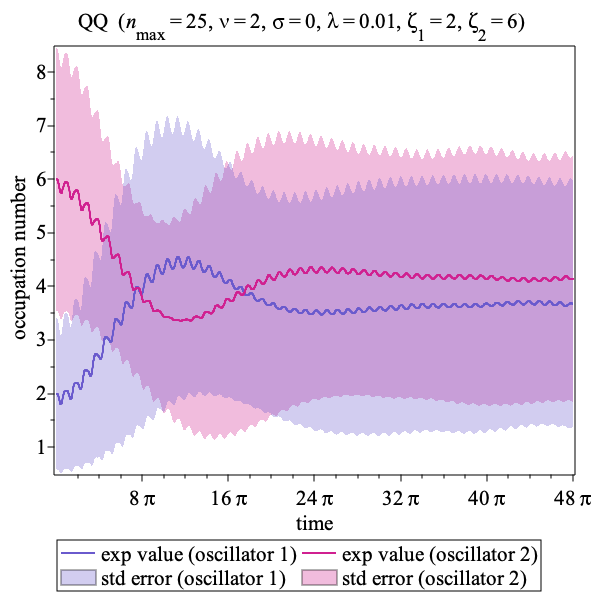} \hfill
\includegraphics[width=0.24\textwidth]{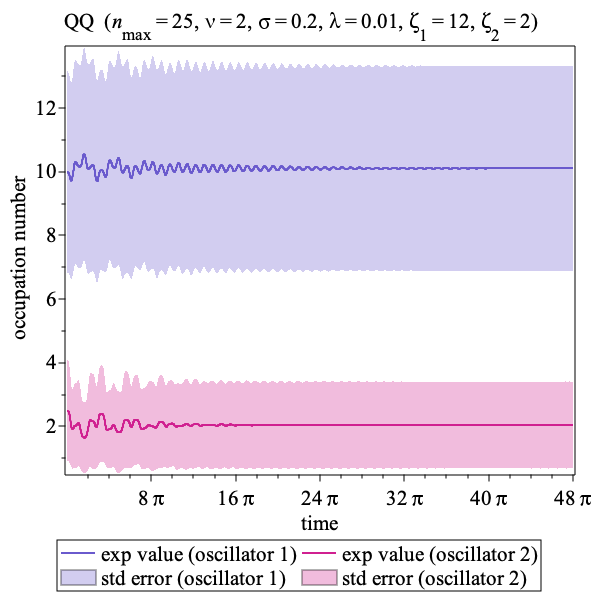} \hfill
\includegraphics[width=0.24\textwidth]{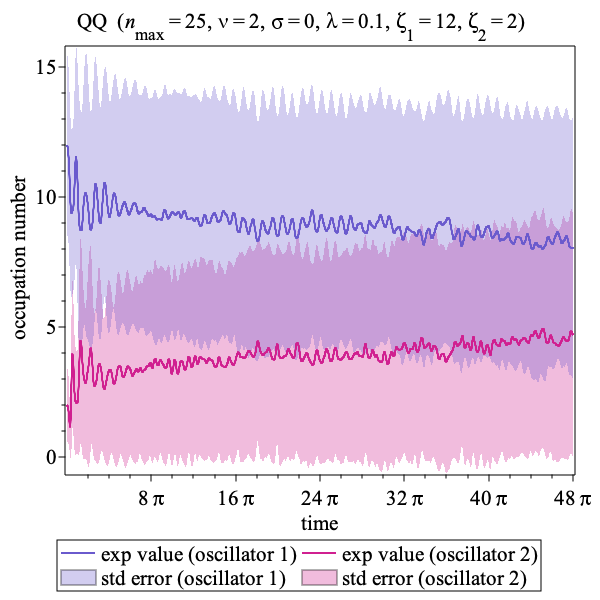} \hfill
\includegraphics[width=0.24\textwidth]{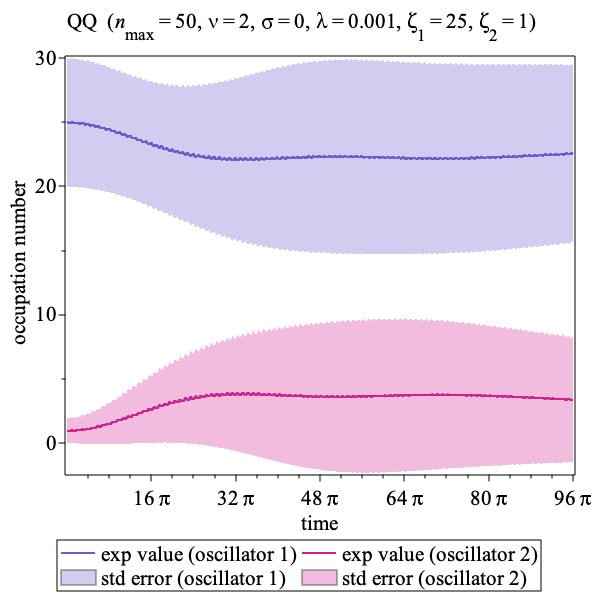} \hfill
\caption{Expectation value of occupation numbers of oscillator 1 and 2 in the quantum-quantum calculation for various choices of parameters. The shaded regions indicate the quantum uncertainty in the expected values as quantified by the standard error (i.e.\ the square root of the variances of $n$ and $m$, respectively). In each case the top of the uncertainty bands lie well below our selected value of $n_\text{max}$, in indicating that our finite truncation of each oscillator basis should be a valid approximation.}\label{fig:occupation numbers}
\end{figure*}

As mentioned above, the simulation results shown in figures \ref{fig:sample results 1} and \ref{fig:sample results 2} are only trustable if the probability distribution for the occupation numbers of each oscillator lie well below the cutoff $n_\text{max}$. In figure \ref{fig:occupation numbers}, we show the average occupation number and its statistical ``standard error'' for each oscillator for a number of different simulations. The standard error around the expectation value of the occupation number of oscillator 1 is the interval
\begin{equation}
	 \left[ \langle n \rangle - \sqrt{\text{Var}(n)},  \langle n \rangle + \sqrt{\text{Var}(n)} \right],
\end{equation}
with a similar expression for oscillator 2. We can see in this figure that the uncertainty bands lie well below our selected cutoffs, so we can be confident in the simulation results.

\subsection{Quantifying error in the CC and CQ approximations}\label{sec:quantifying CC and CQ errors}

The results in figures \ref{fig:sample results 1} and \ref{fig:sample results 2} strongly suggest that the CC and CQ approximation will give results close to the QQ calculation if $\lambda$ is small and if $t$ is not too large. In this section, we attempt to quantify these observations by introducing definitions for the discrepancy between the different calculation methods as well as the relative error in the CC and CQ schemes.

Before defining the discrepancy and relative error, it is useful to revisit the scrambling time $t_\lin$ introduced in \S\ref{sec:validity of product state approx}. Recall that this was the time at which we would estimate that the linear entropy would significantly deviate from zero for systems with infinite dimensional Hilbert spaces (such as the nonlinearly coupled oscillators of this section). Hence, we expect the product state approximation to break down at $t = t_\lin$, implying that the CC and CQ approximations may cease to be be valid. Now, the scrambling time $t_\lin$ is a function of the average interaction energy $\mathcal{E}_\text{int}$ and the degree of coherence of each subsystem $\mathcal{N}_{1,2}$. Given that at time $t=0$, each oscillator is prepared in a coherent state, we can use the results presented in Appendix \ref{sec:coherent} to explicitly write down formulae for $\mathcal{E}_\text{int}$ and $\mathcal{N}_{1,2}$ in terms of integrals of elementary functions. For example, if $\nu = 2$, we find that 
\begin{gather}\nonumber
\mathcal{E}_\text{int}(t) = \frac {\lambda \bar{\Omega}}{t}\int_{0}^{t}  \left[ (q'^{4}_{{1}}+3 q'^{2}_{{1}} +\tfrac{3}{4})(q'^{4}_{{2}} +3 q'^{2}_{{2}} +\tfrac{3}{4})\right]^{1/2} dt' , \\ 
\mathcal{N}_{1,2}(t) = \frac{\int_{0}^{t}  \left[ (q'^{2}_{{1,2}}+\tfrac{1}{4})(q'^{4}_{{2,1}} +3 q'^{2}_{{2,1}} + \tfrac{3}{4})\right]^{1/2} dt' }
{\int_{0}^{t}   \left[ (q'^{4}_{{1}}+3 q'^{2}_{{1}} +\tfrac{3}{4})(q'^{4}_{{2}} +3 q'^{2}_{{2}} +\tfrac{3}{4})\right]^{1/2} dt' },\label{eq:explicit integrals for Eint and N}
\end{gather}
where $q'_{1} = q_{1}(t')$ and $q'_{2} = q_{2}(t')$ are the classical solutions for $q_{1}$ and $q_{2}$ evaluated at time $t'$ and at zero coupling:
\begin{subequations}
\begin{align}
	q_{1}' = q_{1}(t') & = \sqrt{2} |\alpha| \cos (\Omega_{1} t' - \phi_{1}), \\ q_{2}' = q_{2}(t') & = \sqrt{2}|\beta| \cos (\Omega_{2} t' - \phi_{2}).
\end{align}
\end{subequations}
Here, $\alpha$ and $\beta$ are related to other parameters by equation (\ref{eq:QQ alpha and beta}). It does not appear to be possible to evaluate the integrals in (\ref{eq:explicit integrals for Eint and N}) analytically, but we can easily calculate the scrambling time numerically by solving
\begin{equation}\label{eq:t lin numerical def}
	1 = t_{\lin} \mathcal{E}_\text{int}(t_{\lin}) \min[\mathcal{N}_{1,2}(t_{\lin})]
\end{equation}
for $t_{\lin}$. We can derive a very crude estimate of $t_{\lin}$ for the general $\nu$ case by assuming that $|\alpha| \gg 1$ and $|\beta| \gg 1$, we obtain
\begin{equation}
	\mathcal{E_\text{int}} \sim \lambda\bar\Omega |\alpha|^{\nu} |\beta|^{\nu}, \quad \mathcal{N}_{1} \sim |\alpha|^{-1}, \quad \mathcal{N}_{2} \sim |\beta|^{-1}.
\end{equation}
which leads to the following rough estimate for the scrambling time
\begin{equation}\label{eq:scrambling time estimate}
	\bar{\Omega} \, t_{\lin} = \tau_{\lin} \sim \frac{\text{max}(|\alpha|,|\beta|)}{\lambda  |\alpha|^{\nu} |\beta|^{\nu}}.
\end{equation}

\begin{figure*}
\includegraphics[width=0.24\textwidth]{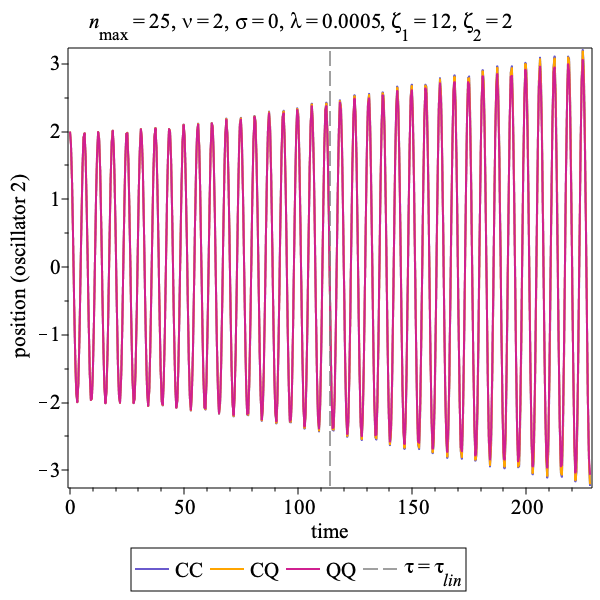} \hfill
\includegraphics[width=0.24\textwidth]{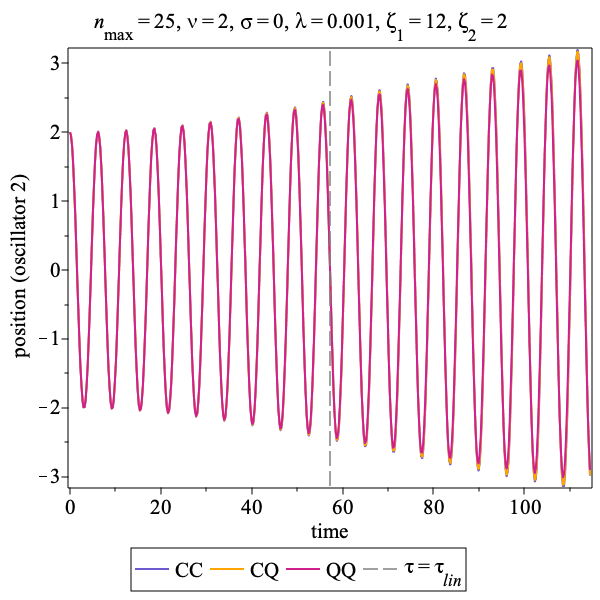} \hfill
\includegraphics[width=0.24\textwidth]{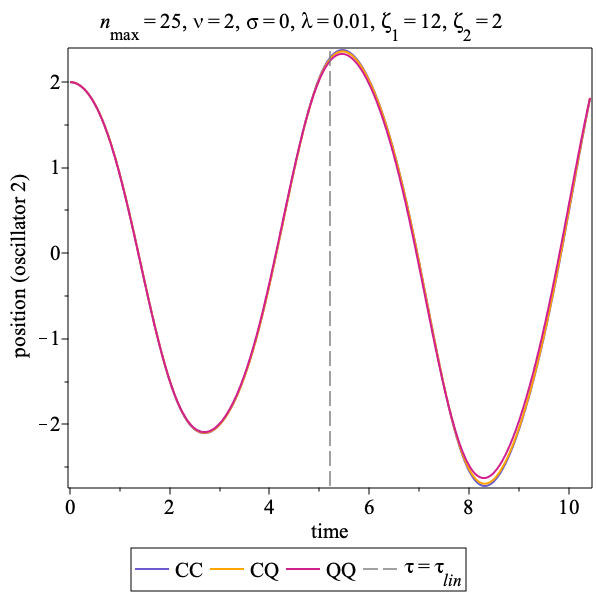} \hfill
\includegraphics[width=0.24\textwidth]{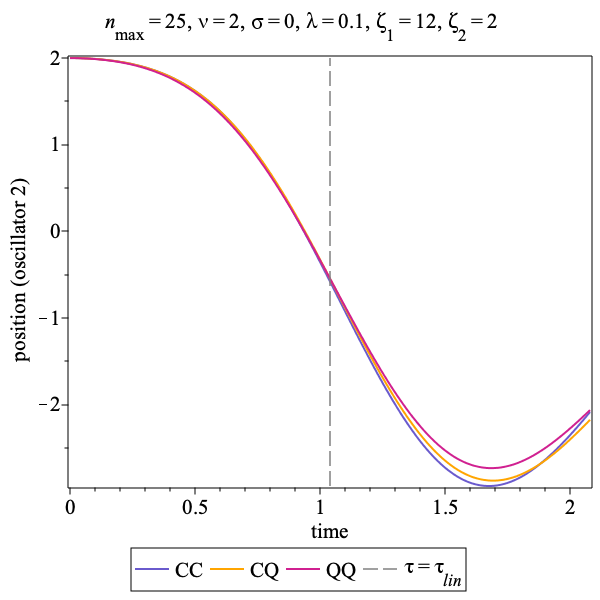} \hfill
\includegraphics[width=0.24\textwidth]{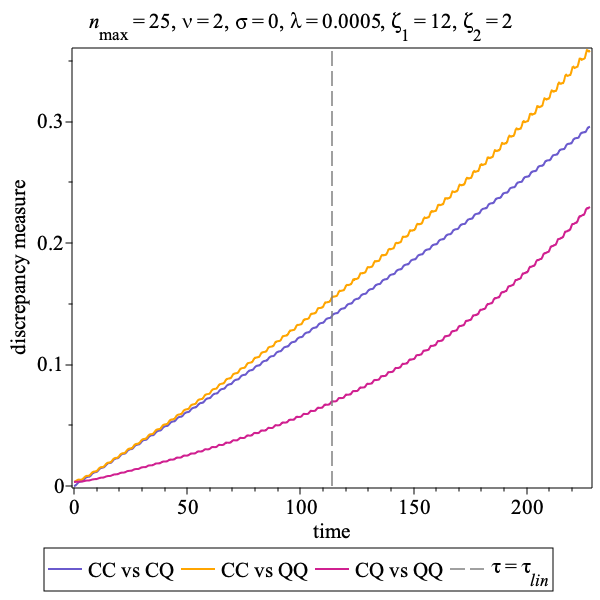} \hfill
\includegraphics[width=0.24\textwidth]{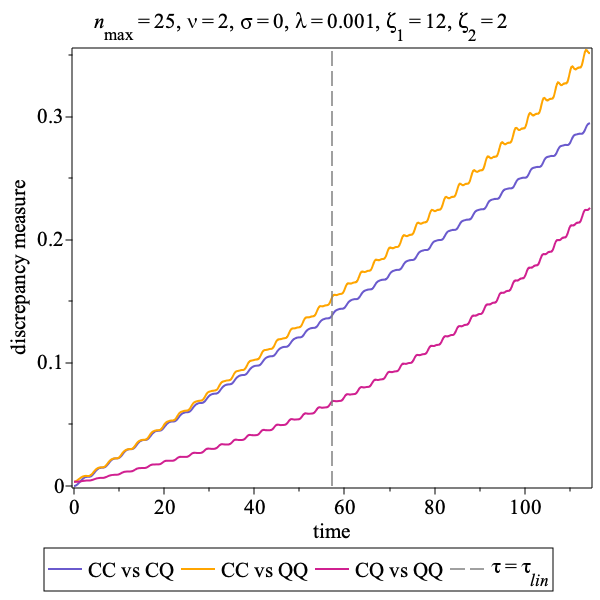} \hfill
\includegraphics[width=0.24\textwidth]{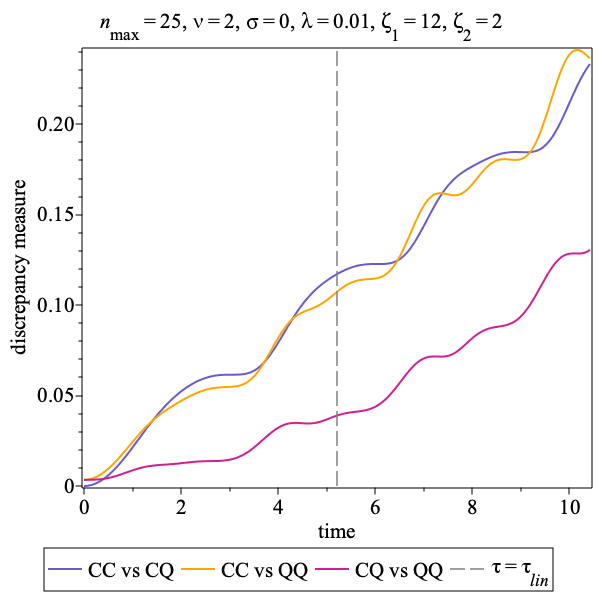} \hfill
\includegraphics[width=0.24\textwidth]{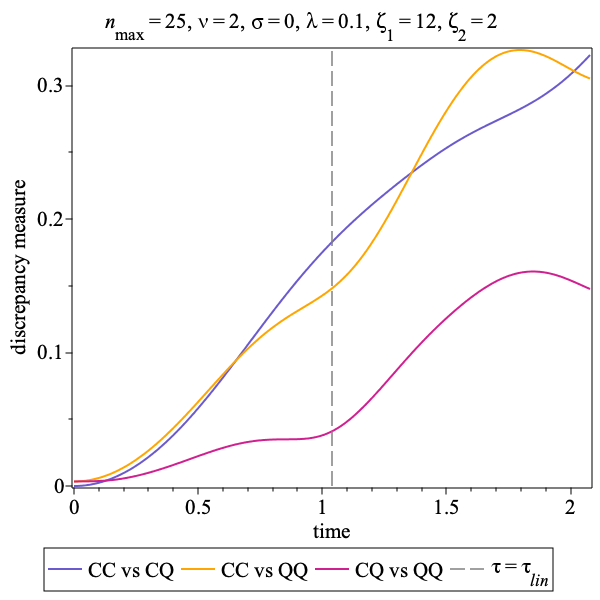} \hfill
\includegraphics[width=0.24\textwidth]{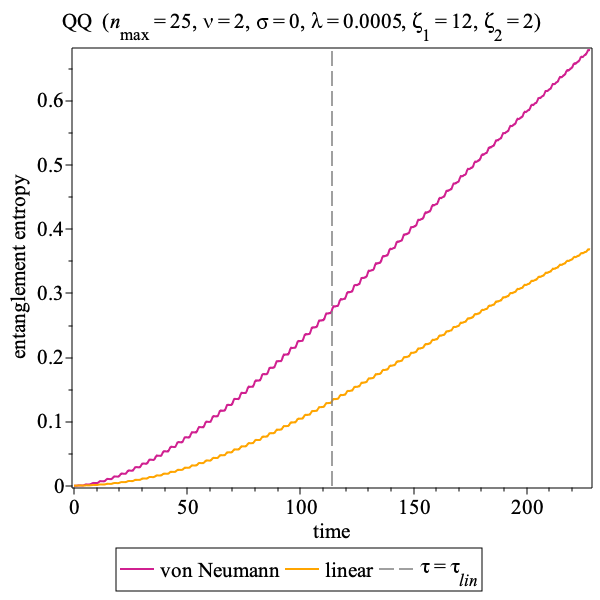} \hfill
\includegraphics[width=0.24\textwidth]{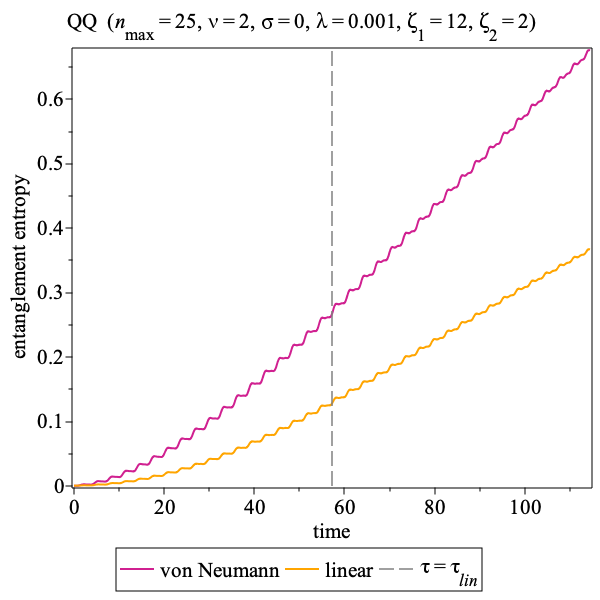} \hfill
\includegraphics[width=0.24\textwidth]{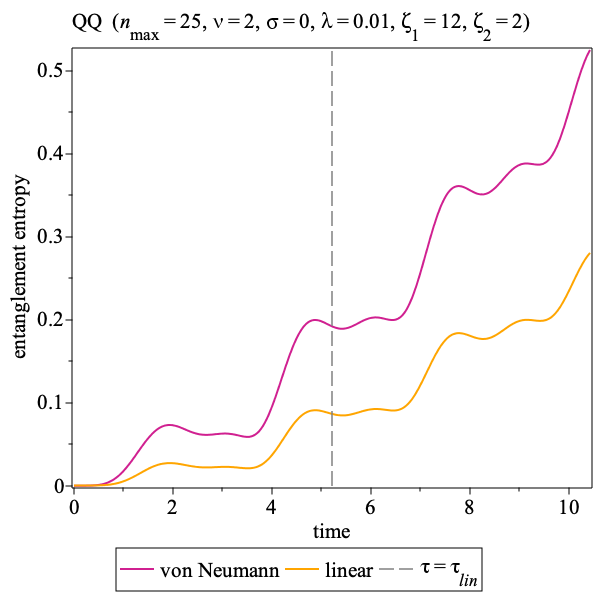} \hfill
\includegraphics[width=0.24\textwidth]{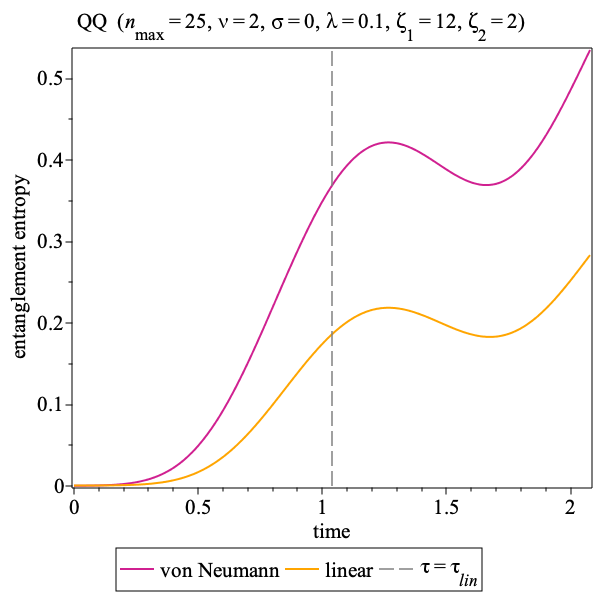} \hfill
\caption{Numeric solutions in the the CC, CQ and QQ schemes over time intervals $\tau \in [0,2\tau_{\lin}]$ with $\nu = 2$. We expect the product state approximation to break down around $\tau = \tau_{\lin}$.  All plots within a given column have the same $\lambda$, and $\lambda$ increases from left to right. We see in the top row that visual discrepancies between the various scheme are only visible for $\tau \gtrsim \tau_{\lin}$. In the middle row, we show the discrepancy between each set of simulation results as defined by the Euclidean distance in phases space as a function of time. The discrepancy between the CQ and QQ schemes is always lower that the CC vs CQ and CC vs QQ discrepancies, indicating that the CQ scheme is the better approximation over these timescales. Finally, in the third row, we show the behaviour of the von-Neumann and linear entropies in the QQ simulations. As expected, we have $S_{\VN}>S_{\lin}$, but otherwise the two entropy curves are remarkably similar.}\label{fig:discrepancy}
\end{figure*}

In figure \ref{fig:discrepancy}, we plot numeric solutions in the CC, CQ and QQ schemes over time intervals $\tau \in [0,2\tau_{\lin}]$, where we calculate $\tau_{\lin}$ numerically from (\ref{eq:t lin numerical def}). We also plot the ``discrepancy'' between each pair of calculation methods, which is defined as the Euclidean distance in phase space between the expectation value of the system's trajectory in each scheme.  More specifically, let us define phase space position vectors for each scheme:
\begin{subequations}
\begin{align}
	\mathbf{X}_{\CC} & \! = \! [q_{1}^{\CC},p_{1}^{\CC},q_{2}^{\CC},p_{2}^{\CC}]^{\T}, \\ 
	\mathbf{X}_{\CQ} & \! = \! [q_{1}^{\CQ},p_{1}^{\CQ},\langle q_{2}^{\CQ}\rangle,\langle p_{2}^{\CQ} \rangle]^{\T}, \\ 
	\mathbf{X}_{\QQ} & \! = \! [ \langle q_{1}^{\QQ} \rangle , \langle p_{1}^{\QQ} \rangle , \langle q_{2}^{\QQ} \rangle , \langle p_{2}^{\QQ} \rangle ]^{\T}.
\end{align}
\end{subequations}
Then, the discrepancy between each pair of schemes at time $\tau$ is
\begin{subequations}
\begin{align}
	\text{CC vs CQ discrepancy} & = \frob{\mathbf{X}_{\CC}-\mathbf{X}_{\CQ}}, \\   
	\text{CC vs QQ discrepancy} & = \frob{\mathbf{X}_{\CC}-\mathbf{X}_{\QQ}},\\   
	\text{CQ vs QQ discrepancy} & = \frob{\mathbf{X}_{\CQ}-\mathbf{X}_{\QQ}};
\end{align}
\end{subequations}
where the Frobenius norm reduces to the usual vector norm in $\mathbb{R}^{4}$. Each of these discrepancy metrics appear to grow with characteristic timescale set by the scrambling time, which is intuitively expected. We see that the discrepancy between the CQ and QQ schemes is generally lowest, implying that the CQ approximation is a good match to the QQ results over the timescales simulated. We also show the behaviour of the linear and von Neumann entropies as a function of time in figure \ref{fig:discrepancy}. Again, it appears that the scrambling time is the appropriate timescale for the growth of entanglement entropy, with $S_{\lin}$ is of order $0.1$ at $\tau = \tau_{\lin}$.\footnote{Interestingly, it looks like $S_{\VN}\sim 2 S_{\lin}$ from these simulations that over the interval $\tau \in [0,2\tau_{\lin}]$, which might have been guessed from equations (\ref{eq:linear entropy inequality 3}), (\ref{eq:VN entropy inequality 2}), and (\ref{eq:entropy relative sizes}).}

\begin{figure}
\includegraphics[width=0.9\columnwidth]{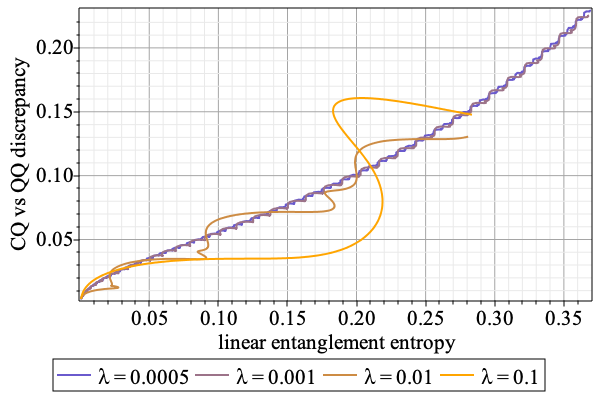} 
\caption{Discrepancy between CQ and QQ simulation results versus linear entanglement entropy for the simulations shown in \ref{fig:discrepancy}. The general trend that higher entropy is associated with higher discrepancies is readily apparent, but the correlation is much tighter for smaller coupling.}\label{fig:error vs entropy}
\end{figure}

For the simulations shown in figure \ref{fig:discrepancy}, we can plot the discrepancy between the CQ and QQ approximations versus the linear entropy; this is shown in figure \ref{fig:error vs entropy}.  This plot matches the general expectation that as the entanglement entropy of the system increases, the agreement between the CQ and QQ approximation degrades.

\begin{figure}
\includegraphics[width=0.8\columnwidth]{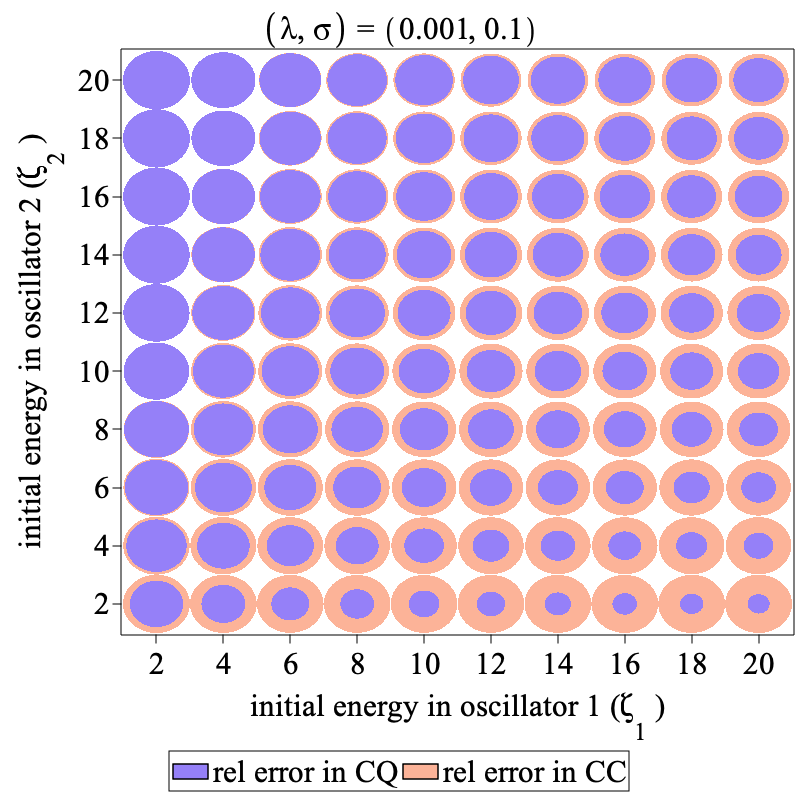} \hfill
\includegraphics[width=0.8\columnwidth]{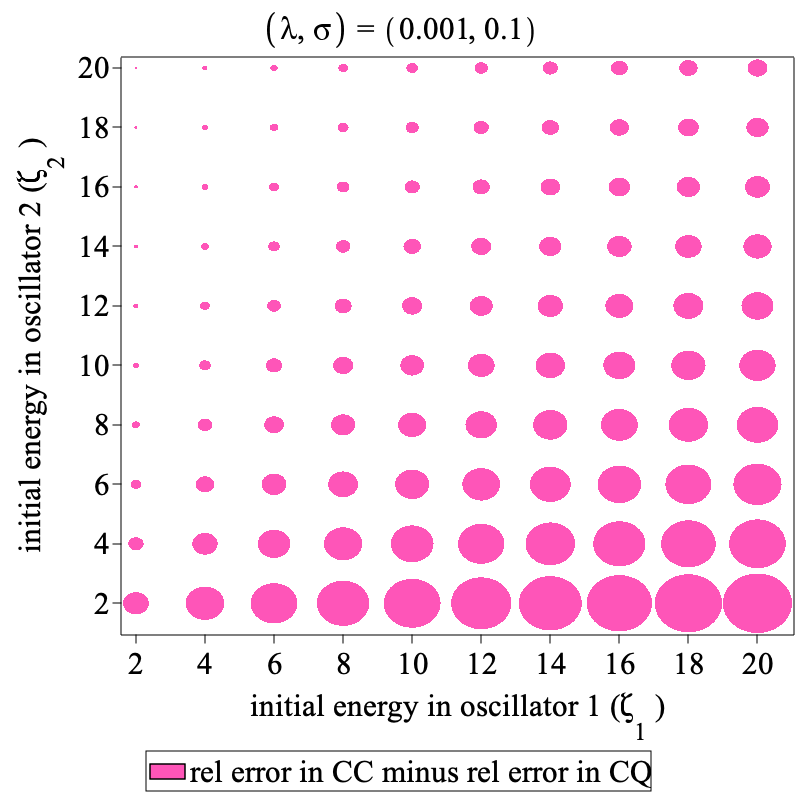}
\caption{Bubble plots comparing the relative errors in the CC and CQ approximations for $\nu =2$ simulations over the time interval $\tau \in [0,6\pi]$ as a function of the oscillator initial energies $(\zeta_{1},\zeta_{2})$. All results assume $\nu = 2$ and use $n_\text{max} = 35$. The areas of each bubble is proportional to the quantity it represents. In the top panel we see that the relative error in the CC approximation is bigger than that of the CQ approximation for all parameters simulated. In the bottom panel, we see that the accuracy advantage in the CQ approximation is largest when $\zeta_{1}$ is big and $\zeta_{2}$ is small; i.e., when the initial energy of oscillator 1 is high and that of oscillator 2 is low.}\label{fig:error bubble plots}
\end{figure}

In addition to condition that $\tau \ll \tau_{\lin}$, in order for the classical-quantum approximation to be valid when $\lambda \ll 1$ we also require that $\varepsilon^{\zero}$ is ``small''. Making use of the definition (\ref{eq:epsilon def}) and the formulae of Appendix \ref{sec:coherent}, we can again write down explicit formulae for $\varepsilon^{\zero}$.  For example, for $\nu = 2$ we get
\begin{equation}
	\varepsilon^{\zero}(t) = \frac{1}{4|\alpha|^{2} \cos^{2} (\Omega_{1} t - \phi_{1}) +1},
\end{equation}
We see that in the $|\alpha| \to \infty$ limit, this function goes to zero except for the discrete times satisfying $\Omega_{1} t - \phi_{1} = \pm\pi/2,\pm 3\pi/2 \ldots$ From this, it seems intuitively reasonable that the CQ approximation will become better and better (in some relative sense) as $\zeta_{1} \propto |\alpha|^{2}$ becomes larger and larger. In order to test this intuition, we can define the relative error in the CC and CQ schemes over a given time interval $\tau \in [0,T]$ as the $L_{2}$ norms of their discrepancy with the QQ results normalized by the $L_{2}$ norm of the QQ phase space trajectory.  Explicitly, if 
\begin{equation}
	\langle\!\langle \mathbf{X} \rangle\!\rangle_{T} = \int_{0}^{T} d\tau \,\frob{\mathbf{X}},
\end{equation}
then
\begin{subequations}
\begin{align}
	\epsilon^{\CC}_{T} & = \frac{ \langle\!\langle\mathbf{X}_{\CC}-\mathbf{X}_{\QQ}\rangle\!\rangle_{T}}{ \langle\!\langle\mathbf{X}_{\QQ}\rangle\!\rangle_{T}}, \\ \epsilon^{\CQ}_{T} & = \frac{\langle\!\langle\mathbf{X}_{\CQ}-\mathbf{X}_{\QQ}\rangle\!\rangle_{T}}{ \langle\!\langle\mathbf{X}_{\QQ}\rangle\!\rangle_{T}},
\end{align}
\end{subequations}
respectively. In figure \ref{fig:error bubble plots}, we show the relative size of these errors in the $(\zeta_{1},\zeta_{2})$ plane for a particular choice of parameters. In this plot, it can be clearly seen that the CQ approximation performs much better than the CC approximation when $\zeta_{1} > \zeta_{2}$, which supports our intuition that the CQ approximation should be appropriate when oscillator 1 has high energy.

\begin{figure*}
\includegraphics[width=0.24\textwidth]{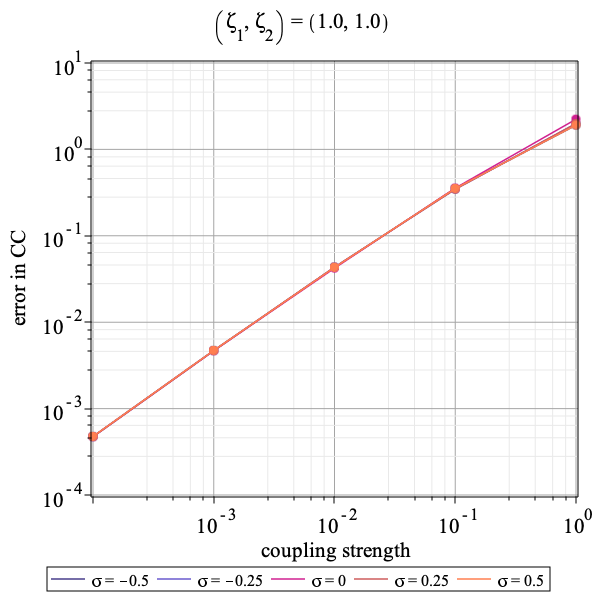} \hfill
\includegraphics[width=0.24\textwidth]{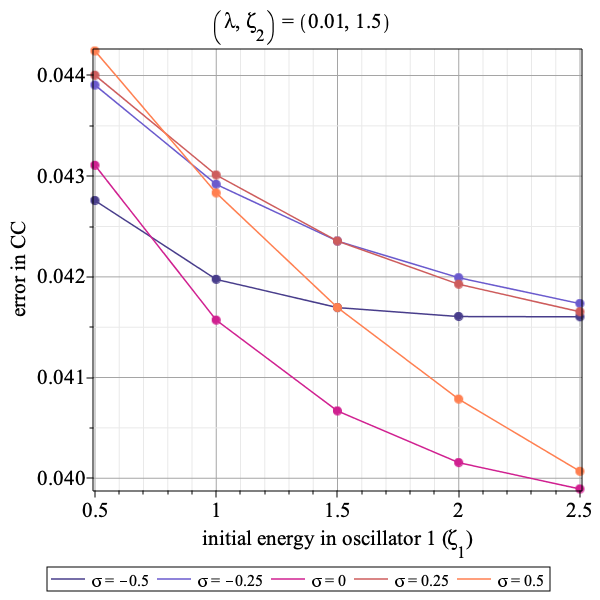} \hfill
\includegraphics[width=0.24\textwidth]{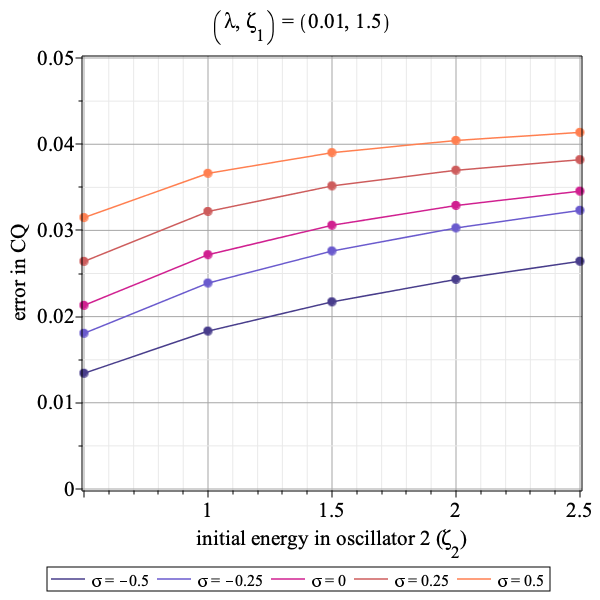} \hfill
\includegraphics[width=0.24\textwidth]{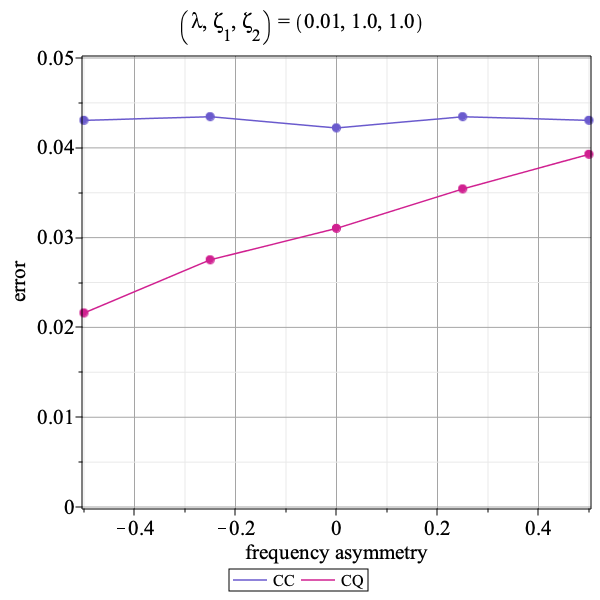} \\
\includegraphics[width=0.24\textwidth]{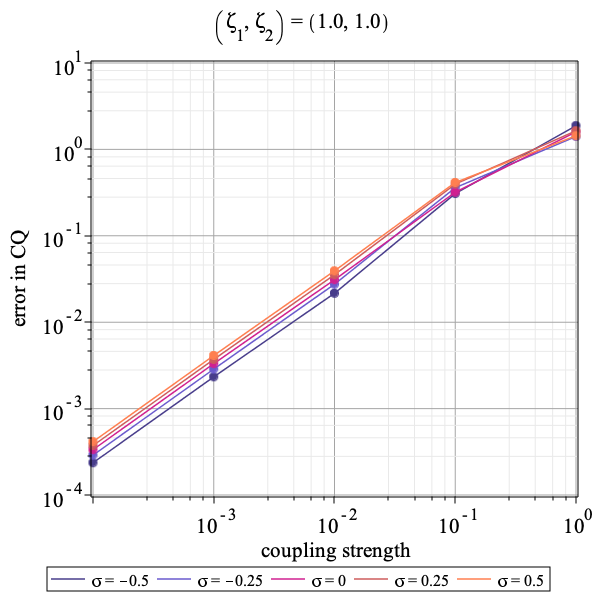} \hfill
\includegraphics[width=0.24\textwidth]{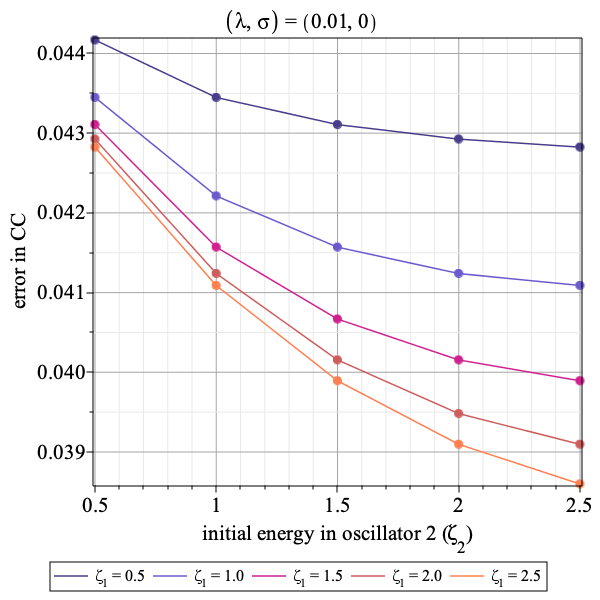} \hfill
\includegraphics[width=0.24\textwidth]{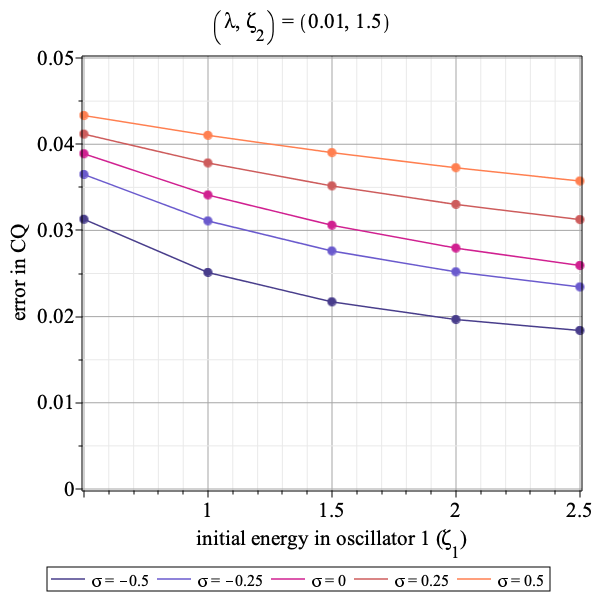} \hfill
\includegraphics[width=0.24\textwidth]{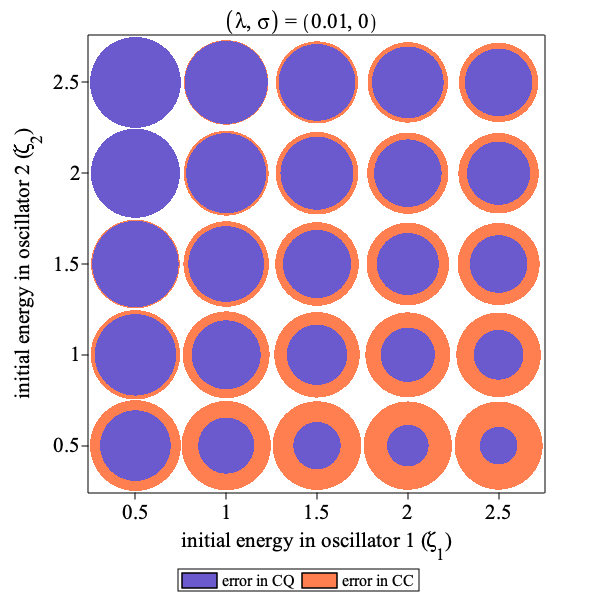} 
\caption{Errors in the CC and CQ schemes over the time interval $\tau \in [0,6\pi]$ as a function of parameters. All results assume $\nu = 2$ and use $n_\text{max} = 35$.}\label{fig:errors}
\end{figure*}

We have calculated the errors in the CC and CQ schemes over a rectangular region of the $(\lambda,\sigma,\zeta_{1},\zeta_{2})$ parameter space for the $\nu = 2$ coupling. Some of the results are shown in figure \ref{fig:errors}. From this figure and results not shown, we make the following qualitative observations:
\begin{itemize}
    \item At sufficiently small coupling, the errors in both the CC and CQ scheme are proportional to $\lambda$.
	\item For most parameters, the CQ scheme results in smaller errors than the CC scheme. Exceptions occur when the frequency asymmetry $\sigma$ approaches 1 or when the initial energy in oscillator 2 is large compared to the initial energy in oscillator 1.
	\item The error in the CQ scheme tends to increase with increasing $\sigma$, while the error in the CC scheme is fairly insensitive to the frequency asymmetry.
	\item The accuracy of the CQ scheme improves when the initial energy of oscillator 2 is less that the initial energy in oscillator 1.
	\item The accuracy of the CC scheme improves when the initial energies in each oscillator become large.
\end{itemize}

\subsection{Parametric resonance}

\begin{figure*}
\begin{center}
\includegraphics[width=0.24\textwidth]{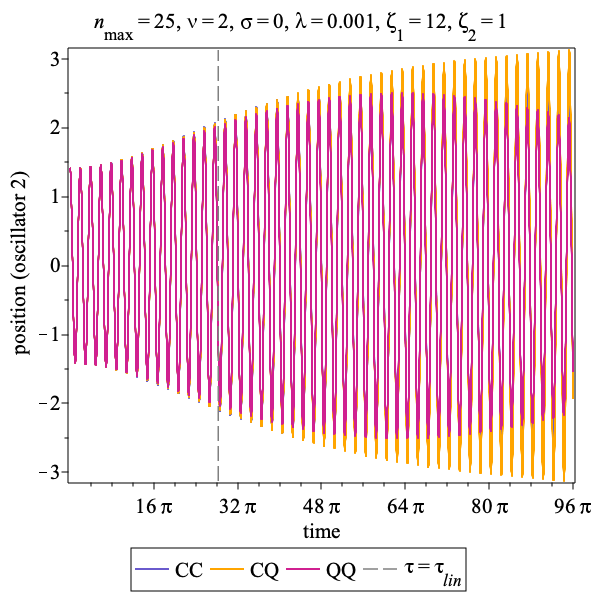} \hfill
\includegraphics[width=0.24\textwidth]{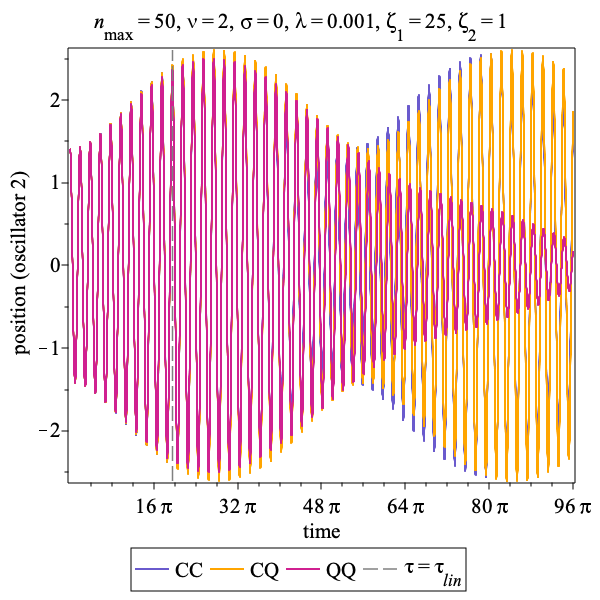} \hfill
\includegraphics[width=0.24\textwidth]{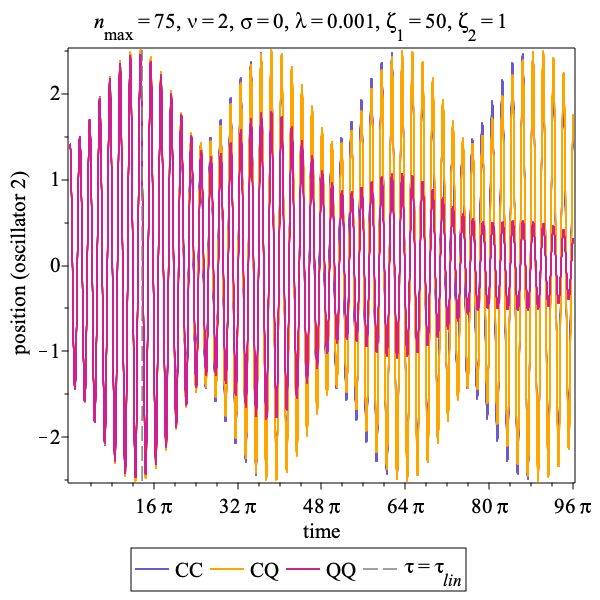} \hfill
\includegraphics[width=0.24\textwidth]{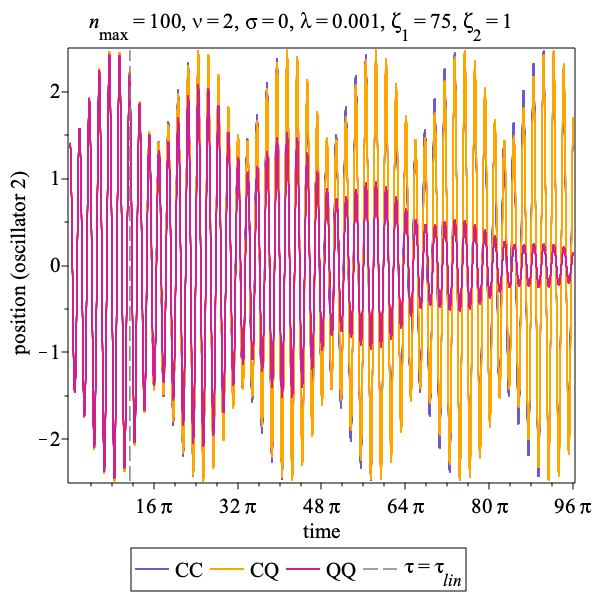} \\
\includegraphics[width=0.24\textwidth]{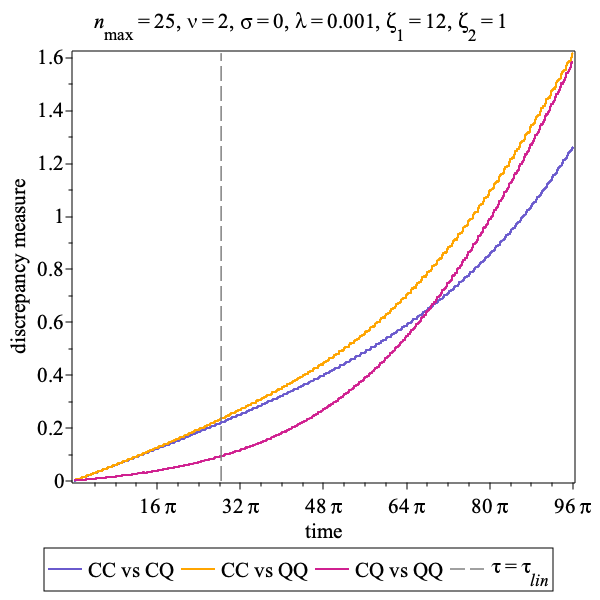} \hfill
\includegraphics[width=0.24\textwidth]{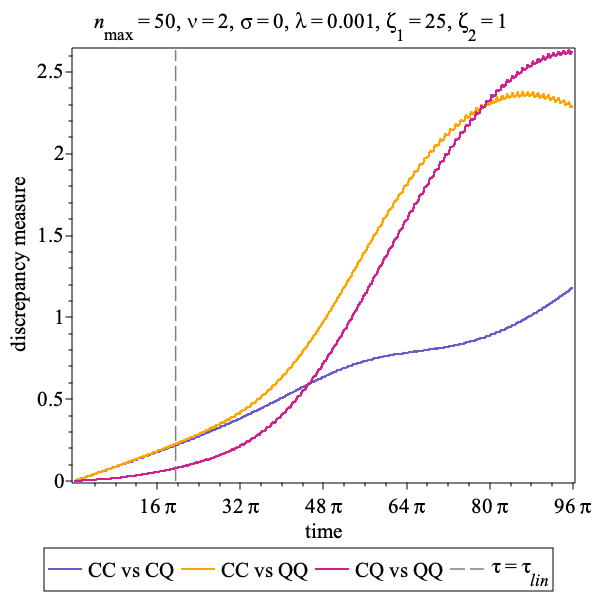} \hfill
\includegraphics[width=0.24\textwidth]{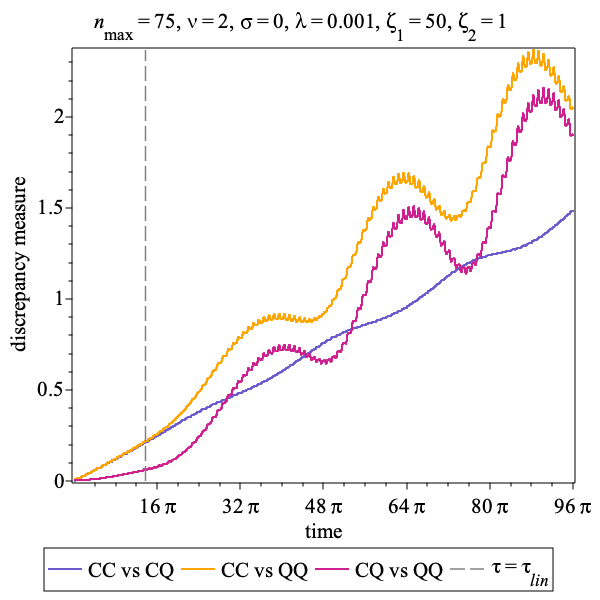} \hfill
\includegraphics[width=0.24\textwidth]{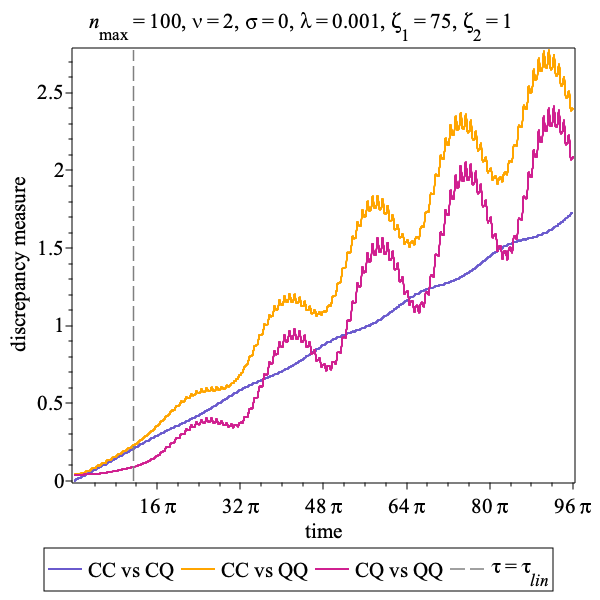} \\
\includegraphics[width=0.24\textwidth]{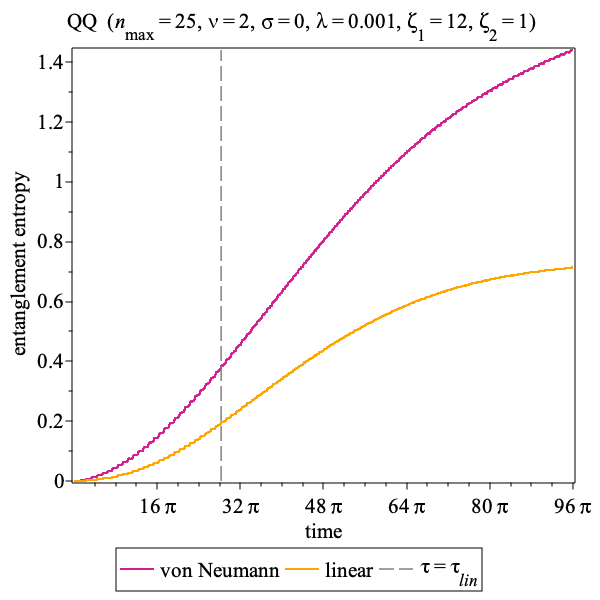} \hfill
\includegraphics[width=0.24\textwidth]{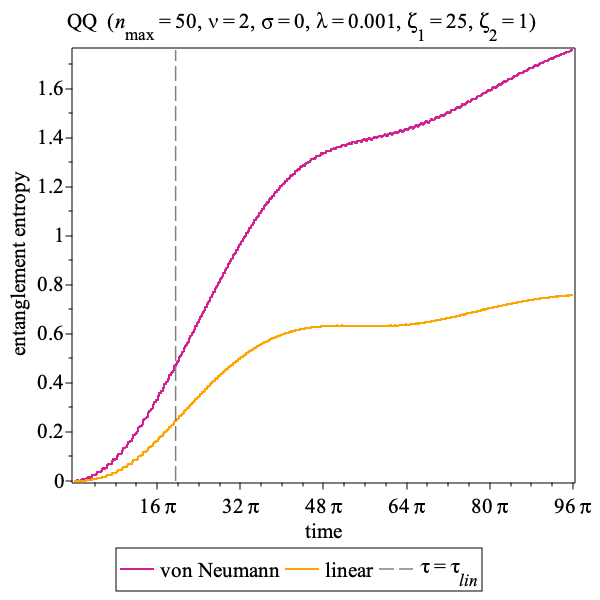} \hfill
\includegraphics[width=0.24\textwidth]{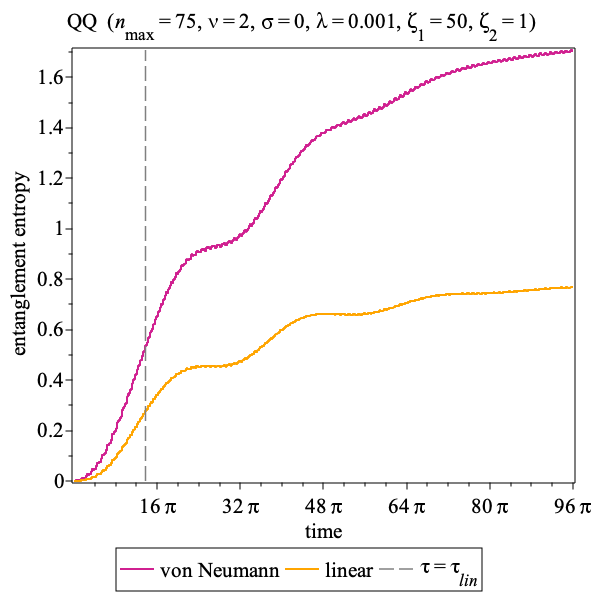} \hfill
\includegraphics[width=0.24\textwidth]{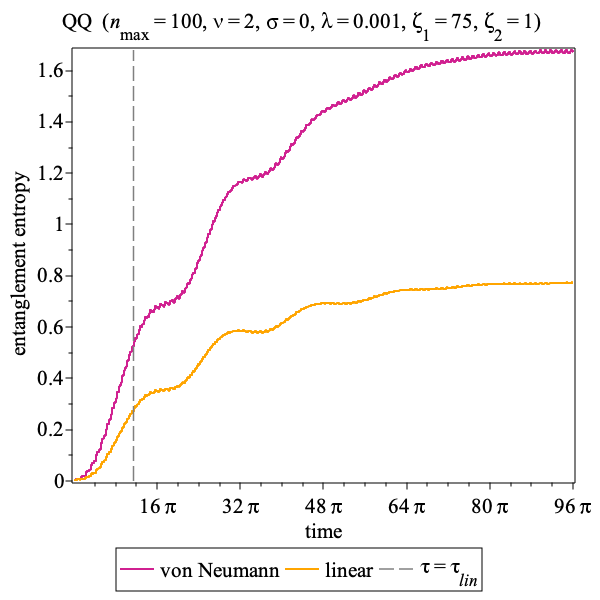}
\end{center}
\caption{Simulations illustrating parametric resonance. In each case, we have selected parameters such that the classical solution exhibits parametric resonance; i.e. the efficient periodic transfer of energy between the oscillators. The parameters for each column are the same except for the initial energy $\zeta_{1}$ of oscillator 1 and the occupation number cutoff $n_\text{max}$. We see that as $\zeta_{1}$ is increased, the QQ results match the classical prediction for a greater number of resonant cycles. This is consistent with our expectation that the ratio of the scrambling to resonant timescales satisfies $\tau_{\lin}/\tau_{\text{\tiny RES}} \sim \sqrt{\zeta_{1}}/\zeta_{2}$.}\label{fig:parametric resonance}
\end{figure*}

Parametric resonance is an important classical phenomena that this system can exhibit when $\nu = 2$. This effect allows for the efficient exchange of energy between the oscillators at small coupling. To see how this comes about, consider the classical equation of motion (\ref{eq:classical EOMs}) under the assumption that the amplitude of one of the oscillators (say, oscillator 2) is very small:
\begin{equation}
	\lambda |b| \ll 1.
\end{equation}
Under this assumption, the equation of motion for the other oscillator is easily solvable:
\begin{equation}
	i\dot{a} \approx (1+\sigma)a \quad \Rightarrow \quad a = |\alpha|e^{i(1+\sigma)t}.
\end{equation}
If we substitute this into the equation of motion for the other oscillator, we can derive a second order equation for $q_{2} = \sqrt{2} \,\text{Re} \, b$ under the assumption that
\begin{equation}
	\lambda |\alpha|^{2} \ll 1-\sigma.
\end{equation}
We find that
\begin{equation}\label{eq:Mathieu equation}
	\frac{d^{2}q_{2}}{dT^{2}} + (1+h\cos\omega T) q_{2} = 0,
\end{equation}
with
\begin{equation}
	T = (1-\sigma)\tau, \quad \omega = 2\left(\frac{1+\sigma}{1-\sigma}\right), \quad h = \frac{2|\alpha|^{2}\lambda}{1-\sigma}.
\end{equation}
Equation (\ref{eq:Mathieu equation}) is the Mathieu equation, and the properties of its solutions are well-known. In particular, when $h \ll 1$, we have exponentially growing solutions whenever $\omega \sim 2/k$ with $k = 1,2,3\ldots$. The $k=1$ case is the strongest resonance with the fastest rate of exponential growth. In that case, exponentially growing solutions exist when
\begin{equation}
	|\omega^{2}-4| < 2h \quad \Rightarrow \quad 4|\sigma| < |\alpha|^{2}\lambda,
\end{equation}
assuming that $|\sigma| \ll 1$. This assumption also implies that the growing solutions for $q_{2}$ are of the form
\begin{equation}
 	q_{2} = \text{const} \times e^{|\alpha|^{2}\lambda \tau} \cos(\tau-\phi),
\end{equation}
where $\phi$ is an arbitrary phase.  We can read off the timescale for the resonant growth of $q_{2}$:
\begin{equation}
	\tau_\text{\tiny RES} = \frac{1}{|\alpha|^{2}\lambda}. 
\end{equation} 
Comparing this to our estimate for the scrambling timescale (\ref{eq:scrambling time estimate}) when $\nu = 2$ and recalling $|b|$ is supposed to be small such that $\text{max}(|\alpha|,|\beta|) = |\alpha|$, we see that
\begin{equation}
	\frac{\tau_{\lin}}{\tau_{\text{\tiny RES}}} \sim \frac{|\alpha|}{|\beta|^{2}} \sim \frac{\zeta_{1}^{1/2}}{\zeta_{2}}.
\end{equation} 
In order for parametric resonance to occur, we need that that the scrambling time be at least the same magnitude as the resonance time. Furthermore, we would expect ``more'' parametric resonance to occur for larger values of the ratio $\tau_{\lin}/\tau_{\text{\tiny RES}}$. To illustrate what this means, we show several long time numeric simulations for small coupling and $\sigma = 0$ in figure \ref{fig:parametric resonance}. In each simulation, the classical solution consists of a sinusoidal wave with a periodically modulated amplitude. We see that when the initial energy $\zeta_{1}$ of oscillator 1 is increased, the QQ simulation matches the CC results for a greater number of cycles of amplitude modulation.  

\subsection{Wigner quasiprobability distributions}\label{sec:Wigner}

\begin{figure}
\includegraphics[width=\columnwidth]{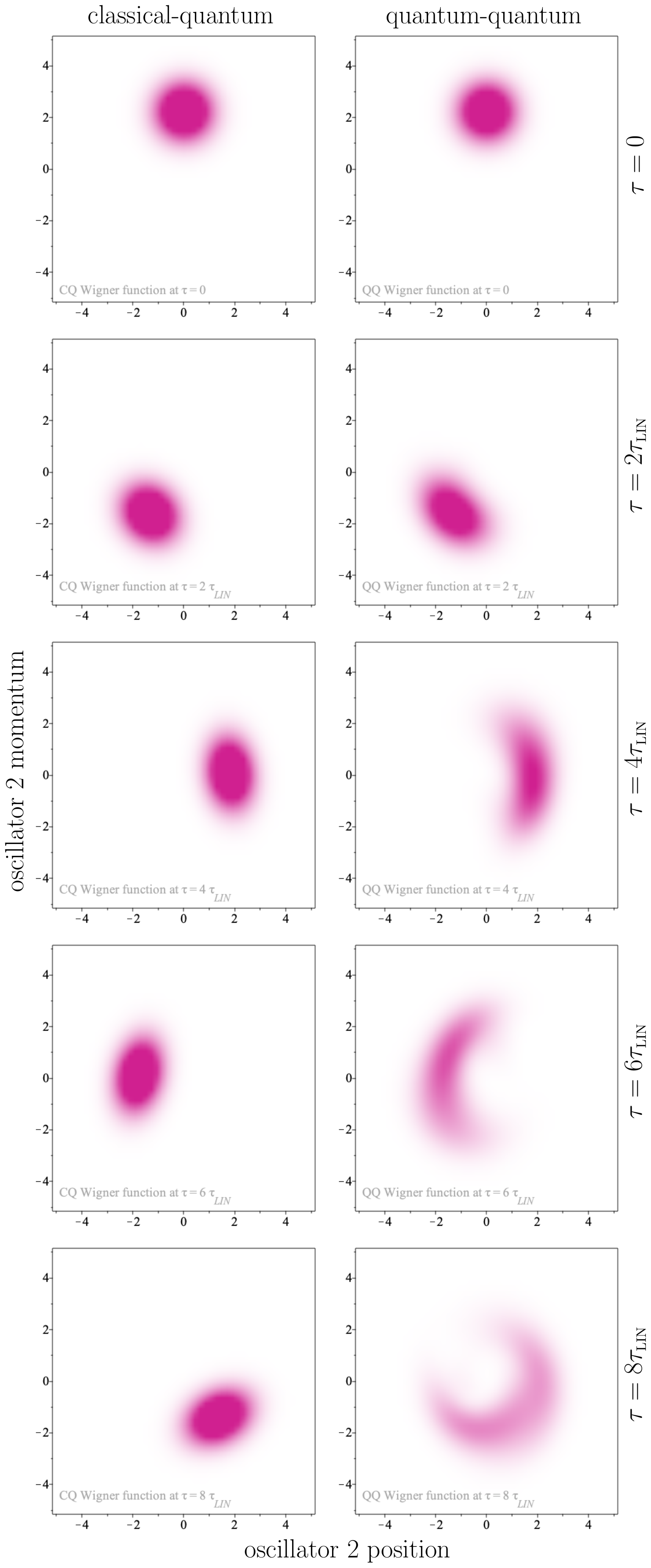}
\caption{Comparison of typical classical-quantum and quantum-quantum Wigner quasi-probability distributions for oscillator 2. Simulation parameters are $n_\text{max}=25$, $\nu=2$, $\sigma=0.2$, $\lambda = 0.01$, $\zeta_{1} = 12$, and $\zeta_{2}=2$.}\label{fig:Wigner}
\end{figure}
The Wigner quasiprobability distribution is a useful quantity for visualizing the quantum uncertainty in the phase space trajectory of systems like a simple harmonic oscillator. In this section, we consider the behaviour of Wigner distribution of oscillator 2 in both the classical-quantum and quantum-quantum calculations.

For a single harmonic oscillator in one-dimension (or any particle moving in a one-dimensional potential), the Wigner distribution is defined as
\begin{equation}
	W(q,p;\tau) = \frac{1}{\pi} \int_{-\infty}^{\infty} dy \langle q - y | \hat{\rho}(\tau) | q+y \rangle e^{-2ipy},
\end{equation}
where $\hat{\rho}(\tau)$ is the density matrix operator and $|q\pm y\rangle$ are position eigenstates.  Making use of the completeness of the energy eigenfunction basis,
\begin{equation}
	1 = \sum_{n} |n\rangle \langle n|,
\end{equation}
we get
\begin{equation}\label{eq:Wigner formula}
	W(q,p;\tau) = \text{Tr}\left[ \rho(\tau) \mathcal{W}(q,p) \right].
\end{equation}
where
\begin{gather}\nonumber
	\mathcal{W}_{nn'}(q,p) = \frac{1}{\pi} \int_{-\infty}^{\infty} dy \, \psi_{n'}(q-y) \psi_{n}^{*}(q+y) e^{-2ipy}, \\
	[\rho(\tau)]_{{n'n}} = \langle n' | \hat{\rho}(\tau) | n \rangle, \quad
	\psi_{n}(q) = \langle q | n \rangle.
\end{gather}
The functions $\psi_{n}(q)$ are merely the ordinary energy eigenfunctions of the simple harmonic oscillator:
\begin{equation}
	\psi_{n}(q) = \frac{1}{\pi^{1/4}\sqrt{2^{n}n!}} e^{-q^{2}/2} H_{n}(q),
\end{equation}
where $H_{n}(q)$ are the Hermite polynomials. Fortunately, it is straightforward to analytically calculate the entries of the $\mathcal{W}$ matrix using symbolic algebra. Once we have $\mathcal{W}$, the Wigner distribution for the CQ and QQ schemes is readily obtained from (\ref{eq:Wigner formula}) after substituting in the correct density matrix (as shown in Table \ref{tab:observables}). As elsewhere in this paper, we need to work in a finite mode truncation to get numerical values for the Wigner function. Due to limitations in available computational resources, the results in this subsection are obtained using $n,n' \lesssim 15$. 

We show the time evolution of the oscillator 2 CQ and QQ Wigner distributions for a typical simulation in figure \ref{fig:Wigner}.  We see that the phase space profile of oscillator 2 is almost constant over the time interval $[0,8\tau_{\lin}]$.   That is, the initial compact distribution associated with the initial coherent state is mostly preserved under time evolution with minor deformation.  The situation is much different for the QQ scheme: one can see appreciable blurring of the initial profile at $\tau = 2\tau_{\lin}$, and by $\tau=8\tau_{\lin}$ there has been a significant degree of decoherence.

It is interesting to observe that for many of the simulations we have presented up to this point, the CC and CQ results tend to agree with one another much more than they agree with the QQ calculation for times greater than $\tau_{\lin}$.  This is very striking in figure \ref{fig:parametric resonance}, for example.  A plausible reason for this is suggested by the Wigner distributions in figure \ref{fig:Wigner}: since the phase space profile of the initial coherent state is preserved for longer in the CQ calculation and coherent states are known to closely mimic classical dynamics, it makes intuitive sense that the CQ and CC match over longer timescales.

In addition to \ref{fig:Wigner}, we have prepared four video clips showing the evolution of the Wigner distributions, entanglement entropy, and density matrices for both oscillators for several different choices of parameters; these are viewable online \cite{youtube}.  We have also superimposed the classical phase space position of each oscillator on the Wigner distributions (as caclulated from the CC formalism) for comparison. For these simulations, we have assumed that subsystem 2 has very low initial energy and that each oscillator has the same frequency.  

\begin{description}

\item{\href{https://youtube.com/watch?v=J5U1gyCBAVI}{Video 1}:}  This video shows the moderate coupling case ($\lambda = 0.01$) over the time interval $[0,6\,\tau_\lin]$. One can see that the phase space profiles of each oscillator start to become deformed just after one scrambling time.  Interestingly, we can see that the reduced density matrix of oscillator 1 starts to become more diagonally dominant at late times, as might be expected in a system that is undergoing decoherence due to interactions with an environment.  However, the reduced density matrix of subsystem 2 does not appear to be approaching a diagonal matrix at late times.  We also remark that the late time Wigner profiles in the QQ calculation exhibit a much richer structure than the CQ case.

\item{\href{https://youtube.com/watch?v=g2xWyTnznZE}{Video 2}:}  This video has the same parameters as video 1, but it is over a shorter time interval $[0,2\,\tau_\lin]$. In this video, we see that the CQ and QQ Wigner distributions of oscillator 2 show good qualitative agreement for early times $\tau \lesssim \tau_\lin$, but start to deviate from each other after.  Throughout these simulations, the evolution of each Wigner distribution tracks the classical orbits reasonably well.

\item{\href{https://youtube.com/watch?v=lM8VA8udQhA}{Video 3}:}  This video assumes large coupling ($\lambda =1$) and a relatively long timescale $[0,18\,\tau_\lin]$.  In this scenario, the Wigner functions are significantly deformed before either oscillator undergoes a single complete period.  Also, there is little qualitative agreement between the CQ and QQ results for oscillator 2 for all $\tau > 0$.  The CQ-CC discrepancy is visibly larger than the QQ-CC discrepancy for $\tau \gtrsim \tau_\lin$.

\item{\href{https://youtube.com/watch?v=tPkiVaadQgg}{Video 4}:}  This video assumes small coupling ($\lambda =0.001$) and evolution up to the scrambling time.  One can see that in this case, each oscillator undergoes many coherent oscillations before the Wigner functions begin to lose their initial shape.  For all times, the centers of the Wigner distributions coincide with the classical orbits even as they spread out in phase space.

\end{description}

\subsection{Long time evolution of entanglement entropy}

\begin{figure*}
\foreach \i in {1,...,9}{
  \includegraphics[width=0.32\textwidth]{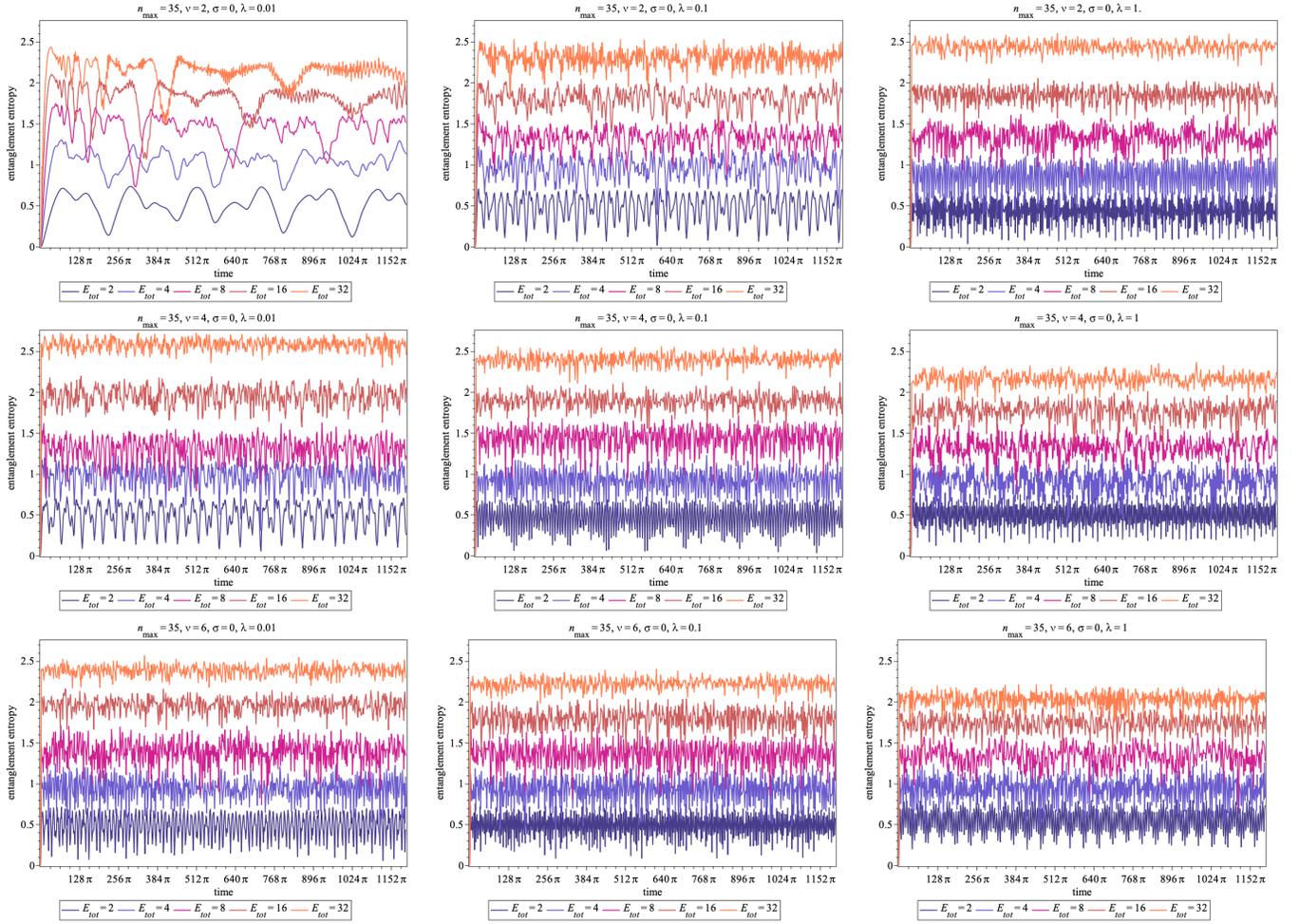} \hfill
}
\caption{Entanglement entropy for long time QQ simulation of the system with coherent state initial data for a variety of couplings.  In each case, we assume identical oscillators ($\sigma =0$) and equal initial energies for each oscillator ($\zeta_{1}=\zeta_{2}$) such the total energy of system is $E_\text{tot} = 2\zeta_{1}+1 = 2\zeta_{2}+1$.  The top row has coupling power $\nu = 2$ (quadratic coupling), the middle row has $\nu = 4$ (quartic coupling), while the bottom row has $\nu=6$.}\label{fig:longterm entanglement simulations}
\end{figure*}

\begin{figure}
\includegraphics[width=\columnwidth]{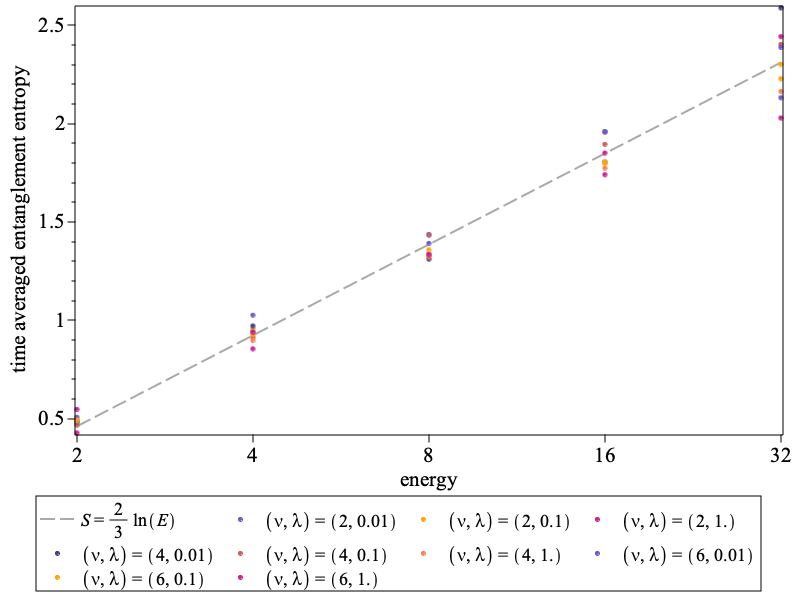}
\caption{Time averaged entanglement entropy versus energy of the simulations shown in figure \ref{fig:longterm entanglement simulations}.}\label{fig:longterm entanglement summary}
\end{figure}

In figure \ref{fig:longterm entanglement simulations}, we show the von Neumann entropy as a function of time from a number of quantum-quantum simulations with different choices of coupling power $\nu$.  These plots show the system's behaviour over timescales much longer than considered in previous sections.  After an initial transient period, we see that the von Neumann entropy for all simulations seems to oscillate erratically about an average value.  In figure \ref{fig:longterm entanglement summary}, we plot the longterm average value of the von Neumann entropy versus the total energy of the oscillators $E_\text{tot}$.  We observe a rough scaling relationship:
\begin{equation}
	S_{\VN} \sim \frac{2}{3} \ln \left( \frac{E_\text{tot}}{\bar\Omega} \right).
\end{equation}
This does seem to hold for several different choices of $\nu$.  However, at higher energies it appears to become less and less accurate.  It would be very interesting to probe this relation for different examples of bipartite system as well for significantly higher energy.  We defer such an investigation to future work.

\section{Conclusions and Discussion}\label{sec:conclusions}

For a broad class of bipartite systems, we presented derivations of the product state and classical-quantum approximations along with their ranges of validity.  We did this by providing a weak coupling analysis of the dynamics of the fully quantum system starting from an initial product state.  This demonstrated that the product state approximation remains valid for times less than the ``scrambling time," which we defined and calculated explicitly.  We also showed that the discrepancy between observables in the CQ and QQ cases was directly related to the degree of classicality exhibited by subsystem 1.

We demonstrated this general framework by numerically studying the system of two coupled oscillators interacting via monomial potentials. This example included an application to energy transfer between the two oscillators exhibiting parametric resonance at the fully quantum level, a result in agreement with classical parametric resonance with a suitable choice of initial data and parameters.  In the CQ case, this calculation bears some resemblance to the phenomenon of particle creation in dynamical spacetimes where subsystem 1 plays the role of a classical geometry driving creation of quanta in subsystem 2.

Overall, this work provides a deeper understanding of the emergence of a hybrid intermediate regime where a quantum and classical system interact with full nonperturbative ``backreaction," together with its range of validity, namely weak coupling and sufficiently short evolution time before increasing entanglement in the QQ system leads to deviation from the CQ dynamics.

Our approach to the CQ dynamics relies crucially on the choice of product state initial data consistent with semiclassical dynamics of subsystem 1 as well as weak coupling between the two systems. In contrast, the Born-Oppenheimer approximation assumes a large ``mass" parameter in the Hamiltonian that separates the bipartite system into ``heavy" and ``light''  subsystems, a measure that is \emph{prima facie} and thus qualitatively distinct from the hypotheses underlying the calculations presented above.  

Lastly, the CQ system as formulated here may be viewed as providing a ``continuous'' monitoring of the quantum system by a classical system \cite{Diosi:1997mt}, at least within the time frame that the first system's description by a semiclassical state holds and the mutual entanglement is negligible.  Beyond this time, as entanglement increases and the state of the first system  deviates from semiclassicality, the notion of classical monitoring of a quantum system ceases to be true.  Because the degree to which the classical system's trajectory is influenced by the quantum system's behaviour and the scrambling time are positively and negatively correlated with the interaction strength, respectively, it is not clear to what extent the classical subsystem is ``measuring'' the quantum subsystem.  That is, we expect that by the time the classical dynamics have been appreciably altered by the quantum dynamics, the entanglement in the system would have grown to the stage where the CQ approximation is no longer valid.

Our analysis is potentially useful for ``semiclassical approximations" postulated for gravity-matter systems where gravity is classical and matter is quantum. Such CQ models have been studied in the context of homogeneous isotropic cosmology with a scalar field \cite{Husain:2018fzg,Husain:2021rnf}. However, without a quantum theory of gravity, there is no full QQ system beyond simple symmetry reduced models, such as homogeneous cosmology. Nevertheless, our results indicate that there are regimes and states where the CQ system provides a good approximation to QQ dynamics. Thus, application to gravity-matter systems might provide a way to access at least a small sector of the dynamics of quantum gravity beyond the simplest models. 

To derive a CQ approximation from quantum gravity, a possible starting point is the Wheeler-deWitt equation (WDE). In its full generality, a derivation of a CQ approximation from the WDE would be technically difficult but may be more accessible in homogeneous cosmological settings where the quantum constraint corresponding to spatial diffeomorphisms is absent and the Hamiltonian constraint is simpler in algebraic form. The main difference from the class of Hamiltonians considered in this paper is the nature of the coupling between matter and gravity---there is matter momentum-metric coupling in addition to matter configuration-metric coupling; a related issue is whether it is even possible to have solution of the Hamiltonian constraint that is a product state. One approach to address the latter problem is to fix a time gauge (c.f.\ \citet{Husain:2011tk}) and work with the corresponding physical Hamiltonian.  This has been done in the fully quantum setting for homogeneous cosmology coupled to a dust and scalar field in the dust time gauge \cite{Husain:2019nym}; in this setting it was demonstrated that a variety of initial states, product and entangled, lead to entropy saturation at late times. This is a setting where a derivation of the CQ approximation from the QQ system appears to be possible under the conditions we have discussed.

Our work does not address the question of ``emergence" of classicality from quantum theory. That is, we do not expect nearly classical dynamics to be achieved for arbitrary choices of initial data.  The notion of a quantum system naturally behaving ``classically'' at late times possibly requires decoherence, whereby one subsystem may be viewed as an ``environment" provided it satisfies the condition that it is approximately nondynamical while the other subsystem's density matrix becomes diagonal dynamically \cite{Zurek:2003zz}. An environment by its nature should not be a system with one degree of freedom. Thus, the type of bipartite system we discuss is not large enough to address emergence. Whether a QFT coupled to a small quantum system with a few degrees of freedom can induce decoherence of the latter through the type of approximation we deploy is a potential topic for further study. 

Another area of further study is the application of the CQ approximation as described here to the gravity-scalar field system in spherical symmetry. This model is well studied classically \cite{Choptuik:1992jv} but remains to be fully studied quantum mechanically \cite{Husain:2009vx}. In the QC setting with classical gravity and quantum scalar field, this model could yield interesting insights into semiclassical black hole evolution. 

Finally, our derivation of the classical-quantum approximation from the full quantum theory, together with its regime of validity, provides in principle the possibility of an experimental probe and comparison with postulated stochastic models of classical-quantum interaction (see, e.g., \cite{Layton:2022sku} for a recent study). Our work may also be compared to the effective approach where the equations of motion of a truncated set of expectation values of products of phase space variables is used to approximate quantum evolution \cite{Bojowald:2005cw,Baytas:2018gbu}.

\appendix

\section{Frobenius inner product}\label{sec:Frobenius}

For two $n \times m$ complex matrices $A$ and $B$, we define the Frobenius inner product as
\begin{equation}
	\frobinner{A}{B} = \Tr(AB^{\dagger}).
\end{equation}
The Frobenius norm of an $n \times m$ matrix is
\begin{equation}
	\frob{A} = \sqrt{\frobinner{A}{A}} = \sqrt{\Tr(AA^{\dagger})}.
\end{equation}
Useful properties include
\begin{subequations}
\begin{align}
	0 & \le \frob{A}, \\ 
	\frob{A} & = \frob{A^{\dagger}}, \\ 
	\label{eq:frob inequality} \frob{AC} & \le \frob{A}\frob{C}, \\
	|\frobinner{A}{B}| & \le  \frob{A}\frob{B}, \\
	\frob{UMU^{\dagger}} & = \frob{M},\\
	\frob{A+B} & \le \frob{A}+\frob{B},\label{eq:unitary frob}
\end{align}
\end{subequations}
where $C$ is an $m \times k$ complex matrix, $M$ is an $n \times n$ complex matrix, and $U$ is an $n \times n$ unitary matrix.

\section{Classical-quantum approximation for one-dimensional potential problems}\label{sec:CQ in 1D potential}

\subsection{Arbitrary potentials}

In this appendix, we demonstrate explicitly how the classical-quantum approximation can emerge from the product state approximation in the scenario when the Hamiltonian of subsystem 1 takes the form of a particle moving in a one-dimensional potential.  In particular, we assume that
\begin{equation}\label{eq:H1 and V1 choices}
    \hat{H}_1 = \frac{\hat{p}_1^2}{2m} + \mathcal{U}(\hat{q}_1), \quad \hat{V}_1 = \mathcal{V}_1(\hat{q}_1),
\end{equation}
where, as usual, $[\hat{q}_1,\hat{p}_1] = i$ and the main $\mathcal{U}$ and interaction $\mathcal{V}_1$ potential functions are real. In the product state approximation described in \S\ref{sec:product state approximation}, the quantum state $|\varphi_1\rangle$ of subsystem 1 satisfies a Schr\"odinger equation, $i \,\di_t |\varphi_1\rangle = \hat{H}_1^\eff |\varphi_1 \rangle$, which can be expressed as wave equation for the wavefunction $\varphi_1(t,q_1)$:
\begin{equation}\label{eq:wave equation}
    i \frac{\di\varphi_1}{\di t} = -\frac{1}{2m}\frac{\di^2\varphi_1}{\di q_1^2} + [\mathcal{U}(q_1)+ \lambda \mathcal{V}_1(q_1) \langle \hat{V}_2 \rangle ] \varphi_1. 
\end{equation}
In this expression, $\langle \hat{V}_2 \rangle = \langle \hat{V}_2(t) \rangle$ is the expectation value of $\hat{V}_2$ as obtained from the solution of effective Schr\"odinger equation for subsystem 2; i.e., $i \,\di_t |\varphi_2\rangle = \hat{H}_2^\eff |\varphi_2 \rangle$.  In this appendix, we will make no special assumptions about the form of $\hat{H}_2$; that is, we do not assume that subsystem 2 is also a particle moving in a one-dimensional potential.

We now assume that $\varphi$ is a sharply peaked function centered about $q_1 = Q_1(t)$.  More specifically, we adopt the same \emph{ansatz} as in the seminal work of \citet{doi:10.1063/1.430620} on time-dependent semiclassical dynamics:
\begin{equation}\label{eq:varphi1 ansatz}
    \varphi_1 = \left( \frac{2}{\pi} \right)^{1/4} e^{i\gamma(t)}e^{iP_1(t)[q_1-Q_1(t)]} e^{i\Sigma(t)[q_1-Q_1(t)]^2}.
\end{equation}
Here, $Q_1(t)$ and $P_1(t)$ are real while $\gamma(t)$ and $\Sigma(t)$ are complex.  In order to satisfy the sharply peaked requirement, we assume that $\text{Im}\,\Sigma$ is a large quantity (in a sense we will define more precisely below).  In order to ensure that $\langle \varphi_1 | \varphi_1 \rangle = \int_{-\infty}^{\infty} dq_1 \varphi_1^* \varphi_1 = 1$, we impose the condition
\begin{equation}
    0 < \text{Im}\,\Sigma = \exp(-4 \, \text{Im} \, \gamma).
\end{equation}
For this wavefunction, the expectation values of $q_1$ and $p_1$ are given exactly as
\begin{equation}\label{eq:q and p exp value}
    \langle q_1 \rangle = Q_1(t), \quad \langle p_1 \rangle = P_1(t).
\end{equation}
Furthermore, the expectation value of any smooth function $f$ of $q_1$ can be expressed in terms of the derivatives of $f$ evaluated at $Q_1(t)$:
\begin{multline}\label{eq:f(q) exp value}
    \langle f(q_1) \rangle = f(Q_1)\left\{ 1 + \frac{f^{[2]}(Q_1)}{8 f(Q_1)\, \Im\Sigma} + \right. \\ \left. \frac{ f^{[4]}(Q_1)}{128 f(Q_1) (\Im\Sigma)^2}  + \mathcal{O} \left[ \frac{f^{[6]}(Q_1)}{f(Q_1)(\Im\Sigma)^3} \right] \right\},
\end{multline}
where $f^{[n]}(Q_1(t))$ is the $n^\text{th}$ derivative of $f$ evaluated at $Q_1(t)$.

We can apply the Ehrenfest theorem to subsystem 1, something which yields equations of motion for the expectation values of $q_1$ and $p_1$:
\begin{subequations}
\begin{align}
    \frac{d\langle q_1 \rangle}{dt} & = \frac{\langle p_1 \rangle}{m}, \\ \frac{d\langle p_1 \rangle}{dt} & = -\left\langle \frac{d\mathcal{U}}{dq_1} \right\rangle - \lambda \langle \hat{V}_2 \rangle \left\langle \frac{d\mathcal{V}_1}{dq_1} \right\rangle.
\end{align}
\end{subequations}
Making use of equations (\ref{eq:q and p exp value}) and (\ref{eq:f(q) exp value}), we could rewrite these as
\begin{subequations}\label{eq:Ehrenfest EOMs}
\begin{align}
    \frac{dQ_1}{dt}  = & \frac{P_1}{m},\label{eq:Q dot eq} \\ \nonumber \frac{dP_1}{dt}  = & -\mathcal{U}^{[1]}(Q_1)\left\{ 1 + \mathcal{O}\left[\frac{\mathcal{U}^{[3]}(Q_1)}{\mathcal{U}^{[1]}(Q_1) \,\Im\Sigma}\right] \right\} \\ &  - \lambda  \mathcal{V}^{[1]}_1(Q_1) \langle \hat{V}_2 \rangle \left\{ 1 + \mathcal{O}\left[\frac{\mathcal{V}^{[3]}(Q_1)}{\mathcal{V}^{[1]}(Q_1) \,\Im\Sigma}\right] \right\}.\label{eq:P1 Ehrenfest}
\end{align}
\end{subequations}
These equations are exact in the sense that they hold if $\varphi_1$ is an exact solution of \ref{eq:wave equation}.  We now demand that the width of the state $\propto (\text{Im}\,\Sigma)^{-1/2}$ is small compared to the third and higher derivatives of the potential and interaction functions
\begin{equation}\label{eq:Ehrenfest SC conditions 1}
    \left| \frac{\mathcal{U}^{[2n-1]}(Q_1)}{\mathcal{U}^{[1]}(Q_1)}\right| \ll (\text{Im}\,\Sigma)^{n}, \quad \left| \frac{\mathcal{V}_1^{[2n-1]}(Q_1)}{\mathcal{V}_1^{[1]}(Q_1)}\right| \ll (\text{Im}\,\Sigma)^{n}
\end{equation}
for $n = 1,2,3\ldots$. Under these circumstances, we can clearly see the central values of our wave packet will evolve according to a dynamical system generated by the effective classical Hamiltonian
\begin{align}\label{eq:CQ classical effective Hamiltonian 1D potential}
\mathcal{H}_{\CQ}^{\eff} & = \frac{P_1^2}{2m} + \mathcal{U}(Q_1) + \lambda \mathcal{V}_1(Q_1) \langle\hat{V}_{2} \rangle.
\end{align}
This is exactly the subsystem 1 classical-quantum Hamiltonian (\ref{eq:CQ classical effective Hamiltonian}) introduced in \S\ref{sec:classical-quantum} with our assumed form of $\hat{H}_1$ and $\hat{V}_1$.

Now, while we can always select initial data for $\varphi_1$ such that $\text{Im}\,\Sigma$ is large for some period, we generally expect the wavepacket to spread such that the error terms in (\ref{eq:P1 Ehrenfest}) become nonnegligible at late times.  In order to determine the rate of spreading, we need to impose the requirement that (\ref{eq:varphi1 ansatz}) is an approximate solution of the wave equation when the width of the wave packet is much smaller than the scale of variation of $\mathcal{U}(q_1)$ and $\mathcal{V}(q_1)$.  More specifically, we assume that it is legitimate to describe $U$ and $V_1$ with a quadratic approximation in some neighbourhood of size $\sim (\text{Im}\,\Sigma)^{-1/2}$ about the peak value of $\varphi_1$.  We begin by substituting the wavefunction ansatz (\ref{eq:varphi1 ansatz}) directly into the wave equation (\ref{eq:wave equation}) as well as series expansions of $\mathcal{U}$ and $\mathcal{V}_1$ about $q_1=Q_1$; i.e.,
\begin{multline}
    \mathcal{U}(q_1) = \mathcal{U}(Q_1) + \mathcal{U}^{[1]}(Q_1)(q_1-Q_1) + \\ \tfrac{1}{2!} \mathcal{U}^{[2]}(Q_1)(q_1-Q_1)^2+ \cdots
\end{multline}
with a similar expression for $\mathcal{V}_1$.  Then, equating coefficients of $(q_1-Q_1)^0$, $(q_1-Q_1)^1$, and $(q_1-Q_1)^2$ on the left- and right-hand sides of the resulting expression, we obtain \emph{approximate} ordinary differential equations for $\gamma$, $\Sigma$, and $P_1$.  Along with equation (\ref{eq:Q dot eq}), this gives a complete dynamical system for $\{Q_1,P_1,\gamma,\Sigma\}$:
\begin{subequations}\label{eq:SC dynamical system}
\begin{align}
    \frac{dQ_1}{dt} & = \frac{P_1}{m}, \\
    \frac{dP_1}{dt} & = -\mathcal{U}^{[1]}(Q_1) - \lambda \langle V_2 \rangle \mathcal{V}_1^{[1]}(Q_1),\\
    \frac{d\gamma}{dt} & = \frac{P_1^2}{2m} + \frac{i\Sigma}{m} - \mathcal{U}(Q_1) - \lambda \mathcal{V}_1(Q_1) \langle V_2 \rangle , \label{eq:wavepacket phase evolution} \\
    \frac{d\Sigma}{dt} & = - \frac{2\Sigma^2}{m} - \frac{\mathcal{U}^{[2]} (Q_1) + \lambda \mathcal{V}_1^{[2]}(Q_1) \langle V_2 \rangle}{2}.\label{eq:wavepacket width evolution}
\end{align}
\end{subequations}
If $\{Q_1,P_1,\gamma,\Sigma\}$ satisfy these equations and the wavepacket width is small, then we expect that (\ref{eq:varphi1 ansatz}) is a good approximate solution to (\ref{eq:wave equation}).  To confirm this, we can calculate the norm of the difference between the left- and right-hand sides of (\ref{eq:wave equation}) when (\ref{eq:SC dynamical system}) is enforced:
\begin{multline}\label{eq:SC solution error}
    \int\limits_{-\infty}^\infty \left| \left\{ i \di_t +\frac{\di_{q_1}^2}{2m} - \mathcal{U}(q_1) - \lambda \mathcal{V}_1(q_1) \langle V_2 \rangle \right\} \varphi_1 \right|^2 dq_1 = \\ \frac{5}{768} \frac{[\mathcal{U}^{[3]}(Q_1) + \lambda \langle V_2 \rangle \mathcal{V}_1^{[3]}(Q_1)]^2}{(\text{Im}\,\Sigma)^3} + \cdots.
\end{multline}
Here, the ``$\cdots$'' on the right-hand side indicates terms involving higher derivatives of $U$ and $V_1$ as well as higher inverse powers of $\text{Im}\,\Sigma$.  Hence, if $\Im \, \Sigma$ is sufficiently large, $\varphi_1$ will be a good approximate solution to the wave equation (\ref{eq:wave equation}). In this case, the evolution of the width of the wavepacket will be given by (\ref{eq:wavepacket width evolution}).  For particular choices of $\mathcal{U}$ and $\mathcal{V}$, one would need to solve the dynamical system \ref{eq:SC dynamical system}) for $\Sigma(t)$ to determine if and when the conditions (\ref{eq:Ehrenfest SC conditions 1}) break down.  If the conditions do indeed fail over a characteristic timescale when the initial wavepacket width is small, that timescale is known as the ``Ehrenfest time.''

We now turn our attention to subsystem 2.  In the product state approximation, its effective Hamiltonian is
\begin{equation}\label{eq:subsystem 2 H eff 1D potential}
    \hat{H}^{\eff}_{2}  = \hat{H}_{2} + \lambda \langle V_{1}(q_1) \rangle \hat{V}_{2}.
\end{equation}
Making use of (\ref{eq:f(q) exp value}), we can rewrite this as
\begin{equation}
    \hat{H}^{\eff}_{2}  = \hat{H}_{2} + \frac{\lambda \mathcal{V}_1(Q_1) \hat{V}_{2}}{1-\varepsilon},
\end{equation}
where, as in section \S\ref{sec:classical-quantum}, $\varepsilon$ is given by
\begin{align}
    \nonumber \varepsilon & = \frac{\langle V_1(q_1) \rangle - \mathcal{V}_1(Q_1)}{\langle V_1 (q_1) \rangle } \\ & =  \frac{\mathcal{V}_1^{[2]}(Q_1)}{8\mathcal{V}_1(Q_1) \,\text{Im}\,\Sigma} + \mathcal{O} \left[ \frac{\mathcal{V}_1^{[4]}(Q_1)}{\mathcal{V}_1(Q_1) \,(\text{Im}\,\Sigma)^2} \right].
\end{align}
Hence, if
\begin{equation}\label{eq:Ehrenfest SC conditions 2}
    \left| \frac{\mathcal{V}_1^{[2k]}(Q_1)}{\mathcal{V}_1(Q_1)}\right| \ll (\text{Im}\,\Sigma)^{k}, \quad k = 1,2,3\ldots,
\end{equation}
then the $\varepsilon$ term in (\ref{eq:subsystem 2 H eff 1D potential}) can be neglected and subsystem 2 will be governed by the effective Hamiltonian
\begin{align}
\hat{H}_{\CQ}^{\eff} = \hat{H}_{2} + \lambda \mathcal{V}_1(Q_1) \hat{V}_{2}.
\end{align}
We have hence recovered the subsystem 2 classical-quantum Hamiltonian operator (\ref{eq:CQ classical effective Hamiltonian 2}) of \S\ref{sec:classical-quantum} specialized to our choice of $\hat{V}_1$.

To summarize, for the particular choices (\ref{eq:H1 
and V1 choices}) the sharply-peaked wave-function \emph{ansatz} (\ref{eq:varphi1 ansatz}) leads us to the conclusion that subsystems 1 and 2 are well described by the coupled classical and quantum Hamiltonians
\begin{subequations}
\begin{align}
    \mathcal{H}_{\CQ}^{\eff} & = P_1^2/(2m) + \mathcal{U}(Q_1) + \lambda \mathcal{V}_1(Q_1) \langle\hat{V}_{2} \rangle, \\
    \hat{H}_{\CQ}^{\eff} & = \hat{H}_{2} + \lambda \mathcal{V}_1(Q_1) \hat{V}_{2}
\end{align}
\end{subequations}
provided that the conditions (\ref{eq:Ehrenfest SC conditions 1}) and (\ref{eq:Ehrenfest SC conditions 2}) hold.  Furthermore, the evolution of the width of subsystem 1's wavefunction will be governed by (\ref{eq:SC dynamical system}) as long as the width $(\Im\,\Sigma)^{-1/2}$ is sufficiently small.

\subsection{Quadratic potentials}

In this subsection, we specialize to the case when both the main $\mathcal{U}$ and interaction $\mathcal{V_1}$ potentials are quadratic monomial functions:
\begin{equation}
    \mathcal{U}(q_1) = \tfrac{1}{2} \Omega_1 q_1^2, \quad \mathcal{V}_1(q_1) = \sqrt{\bar{\Omega}} q_1^2, \quad m = \frac{1}{\Omega_1}.    
\end{equation}
This is similar to the $\nu=2$ coupled oscillator case considered in \S\ref{sec:nonlinearly coupled oscillators} except that we do not necessarily assume that subsystem 2 is also an oscillator.

This case is interesting because several of the exact formulae from the previous subsection simplify considerably.\footnote{We should stress that in this appendix, the phrase ``exact'' is meant to be applied to for formulae that hold in the product state approximation without any additional assumptions.}  For example, the Ehrenfest equations of motion (\ref{eq:Ehrenfest EOMs}) become exactly
\begin{subequations}\label{eq:Ehrenfest EOMs quadratic}
\begin{align}
    \di_t Q_1 & =  \Omega_1 P_1, \\ \di_t P_1 & =  -\left(\Omega_1 + \lambda  \sqrt{\bar{\Omega}} \langle \hat{V}_2 \rangle \right) Q_1.
\end{align}
\end{subequations}
That is, the exact expectation values of $\hat{q}_1$ and $\hat{p}_1$ evolve according to the CQ Hamiltonian
\begin{align}
\mathcal{H}_{\CQ}^{\eff} & = \tfrac{1}{2}\Omega_1(Q_1^2 + P_1^2) + \lambda \sqrt{\bar{\Omega}} Q_1^2 \langle\hat{V}_{2} \rangle.
\end{align}
Also, the right-hand side of (\ref{eq:SC solution error}) is identically equal to zero, implying that (\ref{eq:varphi1 ansatz}) is an exact solution to the wave equation provided that (\ref{eq:Ehrenfest EOMs}) are satisfied in addition to
\begin{align} \nonumber
    \di_t\gamma & = \tfrac{1}{2} \Omega_1 P_1^2 + i\Omega_1\Sigma - \tfrac{1}{2} \left(\Omega_1 + \lambda \sqrt{\bar{\Omega}} \langle V_2 \rangle\right )Q_1^2 , \\
    \di_t \Sigma & = - 2\Omega_1\Sigma^2 - \tfrac{1}{2} \left( \Omega_1 + \lambda \sqrt{\bar\Omega} \langle V_2 \rangle \right).\label{eq:SC dynamical system quadratic}
\end{align}
Another exact relation is
\begin{equation}
    \langle V_1(q_1) \rangle = \frac{\sqrt{\bar{\Omega}}} {2} \left( Q_1^2 + \frac{1}{4\Im\,\Sigma} \right),
\end{equation}
which gives that the effective Hamiltonian of subsystem 2 is
\begin{equation}
    \hat{H}^{\eff}_{2}  = \hat{H}_{2} + \frac{\lambda\sqrt{\bar{\Omega}}} {2} \left( Q_1^2 + \frac{1}{4\Im\,\Sigma} \right) \hat{V}_{2}.
\end{equation}
In this scenario, the classical-quantum approximation will be realized if one assumes $Q_1^2 \gg |4\,\text{Im}\,\Sigma|^{-1}$ so that
\begin{equation}
    \hat{H}^{\eff}_{2} \approx \hat{H}_{2} + \frac{1} {2} \lambda\sqrt{\bar{\Omega}} Q_1^2 \hat{V}_{2},
\end{equation}
which then implies that equations (\ref{eq:SC dynamical system quadratic}) need not be explicitly solved for $\Sigma$ and $\gamma$ in order to solve for $Q_1$, $P_1$, and $|\varphi_2\rangle$.

Of course, to actually determine if $Q_1^2 \gg |4\,\text{Im}\,\Sigma|^{-1}$ is a reasonable assumption, one does need to solve (\ref{eq:SC dynamical system quadratic}) for $\Sigma$.  When the interaction is small, $\lambda \ll 1$, we can easily obtain
\begin{subequations}\label{eq:Q1 and Sigma sols}
\begin{align}
    Q_1(t) & = \sqrt{2}|\alpha|\cos(\Omega_1 t + \phi_1) + \mathcal{O}(\lambda), \\
    \Sigma(t) & = \frac{2\Sigma_0-\tan\Omega_1 t }{2(1+2\Sigma_0\tan\Omega_1 t)} + \mathcal{O}(\lambda),
\end{align}
\end{subequations}
where $\Sigma_0 = \Sigma(0)$ with $\text{Im}\,\Sigma_0 >0$ and we have parametrized the solution for $Q_1(t)$ in a manner consistent with \S\ref{sec:quantifying CC and CQ errors} and \S\ref{sec:coherent}.  Notice that if $\Sigma_0 = i/2$, we obtain that $\Sigma(t) = i/2 + \mathcal{O}(\lambda)$ for all time; i.e., the wavepacket has a nearly constant width.  Other than the $\mathcal{O}(\lambda)$ correction, these are the well-known coherent state solutions of the simple harmonic oscillator.  When $\Sigma_0 \ne i/2$, we recover the oscillator squeezed state solutions subject to $\mathcal{O}(\lambda)$ corrections.

From (\ref{eq:Q1 and Sigma sols}), we see that the condition $Q_1^2 \gg |4\,\text{Im}\,\Sigma|^{-1}$ can be satisfied for almost all $t$ if $|\alpha|^2 \gg |\Sigma_0|^{-1}$.  This is essentially the same as demanding that subsystem 1 be initially prepared in a high energy state.  However, it should be noted that because there are times for which $Q_1(t)=0$, it is impossible to satisfy $Q_1^2 \gg |4\,\text{Im}\,\Sigma|^{-1}$ for all $t$.  However, the time intervals over which the condition fails to hold will be very short if $|\alpha|^2 \gg |\Sigma_0|^{-1}$.  Outside of these intervals, the effective Hamiltonian of subsystem 2 will reduce down to the classical-quantum form
\begin{equation}
    \hat{H}^{\eff}_{2}  \approx \hat{H}^\eff_\CQ = \hat{H}_{2} + \tfrac{1}{2} \lambda\sqrt{\bar{\Omega}} Q_1^2\hat{V}_{2}.
\end{equation}

\section{Born-Oppenheimer approximation for one-dimensional potential problems}\label{sec:Born-Oppenheimer}

In this appendix, we consider the Born-Oppenheimer approximation to our bipartite system when subsystems 1 and 2 each represent particles moving in one-dimensional potentials. Our treatment is somewhat similar to a calculation presented in \citet{singh1989notes}, but there are some key differences.  While \cite{singh1989notes} analyzes both time-independent (via the WKB approximation) and time-dependent Schr\"odinger equations (via sharply peaked states), we will only work on the latter.  There are also some technical assumptions in \cite{singh1989notes} that we prefer to avoid; these are discussed in further detail below.

For this appendix, we assume that
\begin{align}\nonumber
    \hat{H}_1 & = {\hat{p}_1^2}/{2m_1} + m_1 \mathcal{U}_1(\hat{q}_1), &  \hat{V}_1 & = \mathcal{V}_1(\hat{q}_1), \\
    \hat{H}_2 & = {\hat{p}_2^2}/{2m_2} + m_2 \mathcal{U}_2(\hat{q}_2), &  \hat{V}_2 & = \mathcal{V}_2(\hat{q}_2), \label{eq:BO choices}
\end{align}
where $[\hat{q}_i,\hat{p}_j]=i\,\delta_{ij}$ and the $\mathcal{U}_{1,2}$ and $\mathcal{V}_{1,2}$ potential functions are real. Notice that the $\mathcal{U}_{1,2}$ potentials each have a prefactor of the particle masses $m_{1,2}$ for later convenience.  As elsewhere in this paper, subsystem 1 is expected to behave classically, which, in the context of the Born-Oppenheimer approximation, means that the mass of particle 1 should be large.  That is, we will work in the $m_1 \rightarrow \infty$ limit.

The complete time dependent Schr\"odinger equation for the wavefunction $\psi = \psi(t,q_1,q_2)$ is
\begin{multline}\label{eq:BO Schrodinger}
    i \frac{\di\psi}{\di t} = -\frac{1}{2m_1}\frac{\di^2\psi}{\di q_1^2} + m_1 \mathcal{U}_1(q_1) \psi -\frac{1}{2m_2}\frac{\di^2\psi}{\di q_2^2} \\ + m_2 \mathcal{U}_2(q_2) \psi + \lambda \mathcal{V}_1(q_1) \mathcal{V}_2(q_2). 
\end{multline}
We will make an explicit product state \emph{ansatz} for $\psi$:
\begin{equation}\label{eq:BO ansatz}
    \psi(t,q_1,q_2) = \psi_1(t,q_1)\psi_2(t,q_2).
\end{equation}
Since subsystem 1 is meant to exhibit classical properties, we assume that its wavefunction is a Gaussian sharply peaked about $q_1=Q_1(t)$:
\begin{multline}
    \psi_1 = \left( {2}/{\pi} \right)^{1/4} e^{im_1\Gamma(t)}e^{im_1 v_1(t)[q_1-Q_1(t)]} \\ \times e^{im_1^2 \varsigma(t)[q_1-Q_1(t)]^2}.
\end{multline}
This is the same \emph{ansatz} (\ref{eq:varphi1 ansatz}) we made use of in the previous appendix to derive the classical-quantum approximation for a similar potential problem with a few changes of notation:
\begin{equation}
    \gamma = m_1\Gamma, \quad P_1 = m_1 v_1, \quad \Sigma = m_1^2 \varsigma.
\end{equation}
The $m_1$ scalings appearing on the right-hand sides of these definitions are required to make the $m \to \infty$ Born-Oppenheimer limit work out correctly.

The normalization of $\psi$ requires that
\begin{equation}
    1 = \int_{-\infty}^\infty |\psi_1|^2 dq_1 = \int_{-\infty}^\infty |\psi_2|^2 dq_2,
\end{equation}
which yields that
\begin{gather}\nonumber
    0 < m_1^2 \, \text{Im}\,\varsigma = \exp(-4 m_1 \, \text{Im} \, \Gamma) \\
    \langle q_1 \rangle = Q_1(t), \quad \langle p_1 \rangle = m_1 v_1(t).
\end{gather}
Using these and our assumed Hamiltonian, we can use the Ehrenfest theorem to easily show that
\begin{equation}
    v_1 = \frac{dQ_1}{dt}.
\end{equation}

As in the previous appendix, we assume that $\Im\,\Sigma$ is large such that the width of the wavefunction in the $q_1$ direction is much smaller than the characteristic variation scale of the potentials $\mathcal{U}_1$ and $\mathcal{V}_1$.  We therefore seek a solution of (\ref{eq:BO Schrodinger}) valid in some neighbourhood of $q_1 = Q_1(t)$.  More specifically, we substitute (\ref{eq:BO ansatz}) into (\ref{eq:BO Schrodinger}) and expand the result in powers of $\delta q_1 = q_1-Q_1(t)$.  We obtain
\begin{equation}\label{eq:BO expansion of Schrodinger}
    0 = c_0 + c_1 \delta q_1 + c_2\delta q_1^2 + \mathcal{O}(\delta q_1^3).
\end{equation}
Here,
\begin{align} \nonumber
     c_0 = & \,\, m_1 \left[ \frac{d\Gamma}{dt} - i\varsigma - \frac{v_1^2}{2} + \mathcal{U}_1(Q_1) \right] -i \frac{1}{\psi_2} \frac{\di\psi_2}{\di t} \\ \nonumber & \,\, - \frac{1}{2m_2}  \frac{1}{\psi_2} \frac{\di^2\psi_2}{dq_2^2} + m_2 \mathcal{U}_2(q_2) + \lambda \mathcal{V}_1(Q_1) \mathcal{V}_2(q_2), \\ \nonumber c_1 = & \,\, m_1 \frac{dv_1}{dt} + \frac{\di}{\di q_1} \left[ m_1 \mathcal{U}_1(q_1) + \lambda \mathcal{V}_1(q_1) \mathcal{V}_2(q_2) \right]_{q_1=Q_1},  
     \\ \nonumber c_2 = & \,\, m_1^2 \frac{d\varsigma}{dt} + 2m_1^3 \varsigma^2 \\ & \,\, + \frac{\di^2}{\di q_1^2} \left[ m_1 \mathcal{U}_1(q_1) + \lambda \mathcal{V}_1(q_1) \mathcal{V}_2(q_2) \right]_{q_1=Q_1}.
\end{align}

As in the previous appendix, we demand that each of the coefficents of $\delta q_1^0$,  $\delta q_1^1$, and $\delta q_1^2$ vanish independently in (\ref{eq:BO expansion of Schrodinger}).\footnote{At this stage, our approximation deviates from that of \citet{singh1989notes}.  In that paper, the authors assume that $\delta q_1 = m_1^{-1/2}y$ and then demand that the coefficient of each power of $m_1$ vanish.  This results in approximate equations of motion involving an arbitrary function $y = y(t,q_1)$; i.e., an underdetermined system of equations.}  Furthermore, let us make the approximation that in the $m_1 \rightarrow \infty$ limit
\begin{equation}\label{eq:BO supplementary approximation}
    m_1 \mathcal{U}_1(q_1) + \lambda \mathcal{V}_1(q_1) \mathcal{V}_2(q_2) \approx m_1 \mathcal{U}_1(q_1).
\end{equation}
Finally, we demand that the coefficients of $m_1^0$ and $m_1^1$ in the above expression for $c_0$ individually vanish.  This gives us a complete set of equations for all of the unknown functions:
\begin{subequations}
    \begin{align}
        \label{eq:BO EOM 1} \frac{dQ_1}{dt} & =  v_1, \\
        \label{eq:BO EOM 2} \frac{dv_1}{dt} & = -\mathcal{U}_1^{[1]}(Q_1), \\
        \label{eq:BO EOM 3} \frac{d\Gamma}{dt} & =  i\varsigma + \frac{v_1^2}{2} - \mathcal{U}_1(Q_1), \\  \label{eq:BO EOM 4} \frac{d\varsigma}{dt} &  = - 2m_1 \varsigma^2 - \frac{1}{m_1} \mathcal{U}^{[2]}_1(Q_1), \\ \label{eq:BO EOM 5} \nonumber
         i  \frac{\di\psi_2}{\di t} & = - \frac{1}{2m_2} \frac{\di^2\psi_2}{dq_2^2} + \\ & \quad\quad  [m_2 \mathcal{U}_2(q_2) + \lambda \mathcal{V}_1(Q_1) \mathcal{V}_2(q_2)]\psi_2.
    \end{align}
\end{subequations}
We see that (\ref{eq:BO EOM 1}) and (\ref{eq:BO EOM 2}) are the classical equations of motion for subsystem 1 neglecting subsystem 2 (or, in other words, in the $\lambda \to 0$ limit).  Apart from a change in notation, equations (\ref{eq:BO EOM 3}) and (\ref{eq:BO EOM 4}) are the same as evolution equations for the phase (\ref{eq:wavepacket phase evolution}) and width (\ref{eq:wavepacket width evolution}) of a Gaussian wavepacket in the $\lambda \to 0$ limit of the classical-quantum approximation.  Equation (\ref{eq:BO EOM 5}) is exactly the time-dependent Schr\"odinger equation for subsystem 2 in the classical-quantum approximation for the system studied in this appendix.  Therefore, we have essentially recovered the classical-background approximation (\ref{eq:CB EOM}) for the system described by the choices (\ref{eq:BO choices}).

Before moving on, we make note of something interesting.  Suppose that we do not make the approximation (\ref{eq:BO supplementary approximation}).  Rather, let us replace the equations $c_1 = 0$ and $c_2 = 0$ with their expectation values in the state (\ref{eq:BO ansatz}); i.e.,
\begin{equation}
    \langle \psi |c_1 | \psi \rangle = \langle \psi |c_2 | \psi \rangle = 0,
\end{equation}
or, more explicitly,
\begin{equation}
    \int_{-\infty}^\infty dq_2 \, \psi_2^* c_1 \psi_2 = \int_{-\infty}^\infty dq_2 \, \psi_2^* c_2 \psi_2 = 0.
\end{equation}
We then find that equations (\ref{eq:BO EOM 2}) and (\ref{eq:BO EOM 4}) get modified to
\begin{subequations}
    \begin{align}
        \frac{dv_1}{dt} & = - \frac{
        m_1 \mathcal{U}_1^{[1]}(Q_1) + \lambda \mathcal{V}_1^{[1]}(Q_1) \langle V_2 \rangle}{m_1}, \\
       \frac{d\varsigma}{dt} &  = - 2m_1 \varsigma^2 - \frac{m_1 \mathcal{U}^{[2]}_1(Q_1) + \lambda \mathcal{V}_1^{[2]}(Q_1) \langle V_2 \rangle}{m_1^2}.
    \end{align}
\end{subequations}
These are the analogues of the classical-quantum equations of motion in the previous appendix.  Perhaps this is not overly surprising as the strategy of replacing equations like $c_1=c_2=0$ with their expectation values is not dissimilar to the leap made from the classical Einstein equations to the semiclassical system (\ref{eq:semiclassical system 2}).  However, this discussion does provide an intriguing counterpoint to our derivation of the classical-quantum equations of motion using density matrices in \S\ref{sec:classical-quantum}.

\section{Properties of coherent states}\label{sec:coherent}

In this appendix, we review some basic properties of coherent states $|\alpha\rangle$ of the simple harmonic oscillator.  The Hamiltonian of the system is, as usual,
\begin{equation}
	\hat{H} = \Omega (\hat{a}^{\dagger} \hat{a} + \tfrac{1}{2}), \quad [\hat{a},\hat{a}^{\dagger}] = 1.
\end{equation}
In the Heisenberg representation, coherent states are eigenstates of the $\hat{a}$ operator with eigenvalue $\alpha$,
\begin{equation}
	\hat{a} |\alpha \rangle = \alpha |\alpha \rangle,
\end{equation}
where $\alpha$ is a complex constant.  Energy eigenstates of the Hamiltonian are, as usual,
\begin{align}
	\hat{H}|n\rangle = \Omega(n+\tfrac{1}{2}) |n\rangle, \quad n = 0,1,2\ldots.
\end{align}
The expansion of an arbitrary coherent state in the energy eigenbasis reads
\begin{equation}
	|\alpha \rangle = \sum_{n} \langle n |\alpha\rangle |n\rangle = \sum_{n} \frac{\alpha^{n} e^{-|\alpha|^{2}/2} }{\sqrt{n!}} |n\rangle.
\end{equation}
The position and momentum operators are
\begin{equation}
	\hat{q} = \frac{\hat{a}+\hat{a}^{\dagger}}{\sqrt{2}}, \quad \hat{p} = \frac{\hat{a}-\hat{a}^{\dagger}}{\sqrt{2}i}.
\end{equation}
We have the expectation values
\begin{subequations}
\begin{align}
	\langle \hat{q} \rangle & = +\sqrt{2} |\alpha| \cos(\Omega t - \phi), & \langle \hat{q}^{2} \rangle & = \langle \hat{q} \rangle^{2} + \tfrac{1}{2},  \\
	\langle \hat{p} \rangle & = -\sqrt{2} |\alpha| \sin(\Omega t - \phi), & \langle \hat{p}^{2} \rangle & = \langle \hat{p} \rangle^{2} + \tfrac{1}{2}, 
\end{align}
\end{subequations}
where $\phi = \text{arg}(\alpha)$.  Expectation values of higher powers of $\hat{q}$ can be obtained by making use of the identity
\begin{equation}
	[\hat{A}^{n},\hat{B}] = n\hat{A}^{n-1}[\hat{A},\hat{B}], 
\end{equation}
which holds if $[\hat{A},[\hat{A},\hat{B}]]=[\hat{B},[\hat{A},\hat{B}]]=0$.  We can use this to derive the following recursion relation for the epxection values of $\hat{q}^n$ for coherent states:
\begin{equation}
	\langle \hat{q}^{n} \rangle = \langle \hat{q} \rangle \langle \hat{q}^{n-1} \rangle + \tfrac{1}{2}(n-1) \langle \hat{q}^{n-2} \rangle.
\end{equation}
This yields
\begin{subequations}
\begin{align}
\langle \hat{q}^{3} \rangle & = \langle \hat{q} \rangle^{3} + \tfrac{3}{2}\langle \hat{q} \rangle, \\
\langle \hat{q}^{4} \rangle & = \langle \hat{q} \rangle^{4} + 3\langle \hat{q} \rangle^{2} + \tfrac{3}{4}, \\
\langle \hat{q}^{5} \rangle & = \langle \hat{q} \rangle^{5} + 5\langle \hat{q} \rangle^{3} + \tfrac{15}{4}\langle \hat{q} \rangle,
\end{align}
\end{subequations}
and so on.  A very similar derivation results in a recursion relation for the the central moments of $\hat{q}$ for coherent states:
\begin{equation}
	\langle (\hat{q} - \langle \hat{q} \rangle )^{n} \rangle = \tfrac{1}{2} (n-1) \langle (\hat{q} - \langle \hat{q} \rangle )^{n-2} \rangle.
\end{equation}
Making note of $\langle (\hat{q} - \langle \hat{q} \rangle )^{0} \rangle = 1$ and $\langle (\hat{q} - \langle \hat{q} \rangle )^{1} \rangle = 0$, this yields
\begin{equation}
	\langle (\hat{q} - \langle \hat{q} \rangle )^{n} \rangle = \begin{cases} 0, & n\text{ odd}, \\ 2^{-n/2}(n-1)!!, & n\text{ even}. \end{cases}
\end{equation}

\bibliography{Osc-Osc}

\begin{thebibliography}{39}
\expandafter\ifx\csname natexlab\endcsname\relax\def\natexlab#1{#1}\fi
\expandafter\ifx\csname bibnamefont\endcsname\relax
  \def\bibnamefont#1{#1}\fi
\expandafter\ifx\csname bibfnamefont\endcsname\relax
  \def\bibfnamefont#1{#1}\fi
\expandafter\ifx\csname citenamefont\endcsname\relax
  \def\citenamefont#1{#1}\fi
\expandafter\ifx\csname url\endcsname\relax
  \def\url#1{\texttt{#1}}\fi
\expandafter\ifx\csname urlprefix\endcsname\relax\def\urlprefix{URL }\fi
\providecommand{\bibinfo}[2]{#2}
\providecommand{\eprint}[2][]{\url{#2}}

\bibitem[{\citenamefont{Carlip et~al.}(2015)\citenamefont{Carlip, Chiou, Ni,
  and Woodard}}]{Carlip:2015asa}
\bibinfo{author}{\bibfnamefont{S.}~\bibnamefont{Carlip}},
  \bibinfo{author}{\bibfnamefont{D.-W.} \bibnamefont{Chiou}},
  \bibinfo{author}{\bibfnamefont{W.-T.} \bibnamefont{Ni}}, \bibnamefont{and}
  \bibinfo{author}{\bibfnamefont{R.}~\bibnamefont{Woodard}},
  \bibinfo{journal}{Int. J. Mod. Phys. D} \textbf{\bibinfo{volume}{24}},
  \bibinfo{pages}{1530028} (\bibinfo{year}{2015}), \eprint{1507.08194}.

\bibitem[{\citenamefont{Loll}(2020)}]{Loll:2019rdj}
\bibinfo{author}{\bibfnamefont{R.}~\bibnamefont{Loll}},
  \bibinfo{journal}{Class. Quant. Grav.} \textbf{\bibinfo{volume}{37}},
  \bibinfo{pages}{013002} (\bibinfo{year}{2020}), \eprint{1905.08669}.

\bibitem[{\citenamefont{Aastrup and Grimstrup}(2012)}]{Aastrup:2012jj}
\bibinfo{author}{\bibfnamefont{J.}~\bibnamefont{Aastrup}} \bibnamefont{and}
  \bibinfo{author}{\bibfnamefont{J.~M.} \bibnamefont{Grimstrup}},
  \bibinfo{journal}{SIGMA} \textbf{\bibinfo{volume}{8}}, \bibinfo{pages}{018}
  (\bibinfo{year}{2012}), \eprint{1203.6164}.

\bibitem[{\citenamefont{Bodendorfer}(2016)}]{Bodendorfer:2016uat}
\bibinfo{author}{\bibfnamefont{N.}~\bibnamefont{Bodendorfer}}
  (\bibinfo{year}{2016}), \eprint{1607.05129}.

\bibitem[{\citenamefont{Parker}(1969)}]{Parker:1969aa}
\bibinfo{author}{\bibfnamefont{L.}~\bibnamefont{Parker}},
  \bibinfo{journal}{Phys. Rev.} \textbf{\bibinfo{volume}{183}},
  \bibinfo{pages}{1057} (\bibinfo{year}{1969}),
  \urlprefix\url{https://link.aps.org/doi/10.1103/PhysRev.183.1057}.

\bibitem[{\citenamefont{DeWitt}(1975)}]{DEWITT1975295}
\bibinfo{author}{\bibfnamefont{B.~S.} \bibnamefont{DeWitt}},
  \bibinfo{journal}{Physics Reports} \textbf{\bibinfo{volume}{19}},
  \bibinfo{pages}{295} (\bibinfo{year}{1975}), ISSN \bibinfo{issn}{0370-1573},
  \urlprefix\url{https://www.sciencedirect.com/science/article/pii/0370157375900514}.

\bibitem[{\citenamefont{Birrell and Davies}(1982)}]{birrell_davies_1982}
\bibinfo{author}{\bibfnamefont{N.~D.} \bibnamefont{Birrell}} \bibnamefont{and}
  \bibinfo{author}{\bibfnamefont{P.~C.~W.} \bibnamefont{Davies}},
  \emph{\bibinfo{title}{Quantum Fields in Curved Space}}, Cambridge Monographs
  on Mathematical Physics (\bibinfo{publisher}{Cambridge University Press},
  \bibinfo{year}{1982}).

\bibitem[{\citenamefont{Fulling}(1989)}]{fulling_1989}
\bibinfo{author}{\bibfnamefont{S.~A.} \bibnamefont{Fulling}},
  \emph{\bibinfo{title}{Aspects of Quantum Field Theory in Curved Spacetime}},
  London Mathematical Society Student Texts (\bibinfo{publisher}{Cambridge
  University Press}, \bibinfo{year}{1989}).

\bibitem[{\citenamefont{Ford}(1997)}]{Ford:1997hb}
\bibinfo{author}{\bibfnamefont{L.~H.} \bibnamefont{Ford}}, in
  \emph{\bibinfo{booktitle}{{9th Jorge Andre Swieca Summer School: Particles
  and Fields}}} (\bibinfo{year}{1997}), pp. \bibinfo{pages}{345--388},
  \eprint{gr-qc/9707062}.

\bibitem[{\citenamefont{Brout et~al.}(1995)\citenamefont{Brout, Massar,
  Parentani, Popescu, and Spindel}}]{Brout:1995aa}
\bibinfo{author}{\bibfnamefont{R.}~\bibnamefont{Brout}},
  \bibinfo{author}{\bibfnamefont{S.}~\bibnamefont{Massar}},
  \bibinfo{author}{\bibfnamefont{R.}~\bibnamefont{Parentani}},
  \bibinfo{author}{\bibfnamefont{S.}~\bibnamefont{Popescu}}, \bibnamefont{and}
  \bibinfo{author}{\bibfnamefont{P.}~\bibnamefont{Spindel}},
  \bibinfo{journal}{Phys. Rev. D} pp. \bibinfo{pages}{1119--1133}
  (\bibinfo{year}{1995}).

\bibitem[{\citenamefont{Hawking}(1975)}]{Hawking:1975vcx}
\bibinfo{author}{\bibfnamefont{S.~W.} \bibnamefont{Hawking}},
  \bibinfo{journal}{Commun. Math. Phys.} \textbf{\bibinfo{volume}{43}},
  \bibinfo{pages}{199} (\bibinfo{year}{1975}), \bibinfo{note}{[Erratum:
  Commun.Math.Phys. 46, 206 (1976)]}.

\bibitem[{\citenamefont{Baumann}(2011)}]{Baumann:2009ds}
\bibinfo{author}{\bibfnamefont{D.}~\bibnamefont{Baumann}}, in
  \emph{\bibinfo{booktitle}{{Theoretical Advanced Study Institute in Elementary
  Particle Physics}: {Physics of the Large and the Small}}}
  (\bibinfo{year}{2011}), pp. \bibinfo{pages}{523--686}, \eprint{0907.5424}.

\bibitem[{\citenamefont{Abele et~al.}(2010)\citenamefont{Abele, Jenke, Leeb,
  and Schmiedmayer}}]{Abele:2009dw}
\bibinfo{author}{\bibfnamefont{H.}~\bibnamefont{Abele}},
  \bibinfo{author}{\bibfnamefont{T.}~\bibnamefont{Jenke}},
  \bibinfo{author}{\bibfnamefont{H.}~\bibnamefont{Leeb}}, \bibnamefont{and}
  \bibinfo{author}{\bibfnamefont{J.}~\bibnamefont{Schmiedmayer}},
  \bibinfo{journal}{Phys. Rev. D} \textbf{\bibinfo{volume}{81}},
  \bibinfo{pages}{065019} (\bibinfo{year}{2010}), \eprint{0907.5447}.

\bibitem[{\citenamefont{Born and
  Oppenheimer}(1927)}]{https://doi.org/10.1002/andp.19273892002}
\bibinfo{author}{\bibfnamefont{M.}~\bibnamefont{Born}} \bibnamefont{and}
  \bibinfo{author}{\bibfnamefont{R.}~\bibnamefont{Oppenheimer}},
  \bibinfo{journal}{Annalen der Physik} \textbf{\bibinfo{volume}{389}},
  \bibinfo{pages}{457} (\bibinfo{year}{1927}),
  \urlprefix\url{https://doi.org/10.1002/andp.19273892002}.

\bibitem[{\citenamefont{Lapchinski and
  Rubakov}(1979)}]{lapchinski1979canonical}
\bibinfo{author}{\bibfnamefont{V.}~\bibnamefont{Lapchinski}} \bibnamefont{and}
  \bibinfo{author}{\bibfnamefont{V.}~\bibnamefont{Rubakov}},
  \bibinfo{journal}{Acta Phys. Pol., B;(Poland)} \textbf{\bibinfo{volume}{10}}
  (\bibinfo{year}{1979}).

\bibitem[{\citenamefont{Kiefer and Singh}(1991)}]{PhysRevD.44.1067}
\bibinfo{author}{\bibfnamefont{C.}~\bibnamefont{Kiefer}} \bibnamefont{and}
  \bibinfo{author}{\bibfnamefont{T.~P.} \bibnamefont{Singh}},
  \bibinfo{journal}{Phys. Rev. D} \textbf{\bibinfo{volume}{44}},
  \bibinfo{pages}{1067} (\bibinfo{year}{1991}),
  \urlprefix\url{https://link.aps.org/doi/10.1103/PhysRevD.44.1067}.

\bibitem[{\citenamefont{Padmanabhan}(1989)}]{TPadmanabhan_1989}
\bibinfo{author}{\bibfnamefont{T.}~\bibnamefont{Padmanabhan}},
  \bibinfo{journal}{Classical and Quantum Gravity}
  \textbf{\bibinfo{volume}{6}}, \bibinfo{pages}{533} (\bibinfo{year}{1989}),
  \urlprefix\url{https://dx.doi.org/10.1088/0264-9381/6/4/012}.

\bibitem[{\citenamefont{Singh and Padmanabhan}(1989)}]{singh1989notes}
\bibinfo{author}{\bibfnamefont{T.}~\bibnamefont{Singh}} \bibnamefont{and}
  \bibinfo{author}{\bibfnamefont{T.}~\bibnamefont{Padmanabhan}},
  \bibinfo{journal}{Annals of Physics} \textbf{\bibinfo{volume}{196}},
  \bibinfo{pages}{296} (\bibinfo{year}{1989}).

\bibitem[{\citenamefont{Lichnerowicz and
  Tonnelat}(1962)}]{Lichnerowicz:1962gnh}
\bibinfo{editor}{\bibfnamefont{M.~A.} \bibnamefont{Lichnerowicz}}
  \bibnamefont{and} \bibinfo{editor}{\bibfnamefont{M.~A.}
  \bibnamefont{Tonnelat}}, eds., \emph{\bibinfo{title}{{Proceedings, Les
  th\'eories relativistes de la gravitation : actes du colloque international
  (Relativistic Theories of Gravitation)}: {Royaumont, France, June 21-27,
  1959}}} (\bibinfo{publisher}{CNRS}, \bibinfo{address}{Paris},
  \bibinfo{year}{1962}).

\bibitem[{\citenamefont{Rosenfeld}(1963)}]{Rosenfeld}
\bibinfo{author}{\bibfnamefont{L.}~\bibnamefont{Rosenfeld}},
  \bibinfo{journal}{Nucl Phys} \textbf{\bibinfo{volume}{40}},
  \bibinfo{pages}{353} (\bibinfo{year}{1963}).

\bibitem[{\citenamefont{Page and Geilker}(1981)}]{Page:1981aj}
\bibinfo{author}{\bibfnamefont{D.~N.} \bibnamefont{Page}} \bibnamefont{and}
  \bibinfo{author}{\bibfnamefont{C.~D.} \bibnamefont{Geilker}},
  \bibinfo{journal}{Phys. Rev. Lett.} \textbf{\bibinfo{volume}{47}},
  \bibinfo{pages}{979} (\bibinfo{year}{1981}).

\bibitem[{\citenamefont{Kiefer}(1992)}]{PhysRevD.45.2044}
\bibinfo{author}{\bibfnamefont{C.}~\bibnamefont{Kiefer}},
  \bibinfo{journal}{Phys. Rev. D} \textbf{\bibinfo{volume}{45}},
  \bibinfo{pages}{2044} (\bibinfo{year}{1992}),
  \urlprefix\url{https://link.aps.org/doi/10.1103/PhysRevD.45.2044}.

\bibitem[{\citenamefont{Husain et~al.}(2022)\citenamefont{Husain, Javed, and
  Singh}}]{Husain:2022kaz}
\bibinfo{author}{\bibfnamefont{V.}~\bibnamefont{Husain}},
  \bibinfo{author}{\bibfnamefont{I.}~\bibnamefont{Javed}}, \bibnamefont{and}
  \bibinfo{author}{\bibfnamefont{S.}~\bibnamefont{Singh}},
  \bibinfo{journal}{Phys. Rev. Lett.} \textbf{\bibinfo{volume}{129}},
  \bibinfo{pages}{111302} (\bibinfo{year}{2022}), \eprint{2205.06388}.

\bibitem[{\citenamefont{Kofman et~al.}(1997)\citenamefont{Kofman, Linde, and
  Starobinsky}}]{Kofman:1997yn}
\bibinfo{author}{\bibfnamefont{L.}~\bibnamefont{Kofman}},
  \bibinfo{author}{\bibfnamefont{A.~D.} \bibnamefont{Linde}}, \bibnamefont{and}
  \bibinfo{author}{\bibfnamefont{A.~A.} \bibnamefont{Starobinsky}},
  \bibinfo{journal}{Phys. Rev. D} \textbf{\bibinfo{volume}{56}},
  \bibinfo{pages}{3258} (\bibinfo{year}{1997}), \eprint{hep-ph/9704452}.

\bibitem[{\citenamefont{Rathorea et~al.}(2022)\citenamefont{Rathorea, Dhayalb,
  and Venkataratnamc}}]{rathorea2022validity}
\bibinfo{author}{\bibfnamefont{M.}~\bibnamefont{Rathorea}},
  \bibinfo{author}{\bibfnamefont{R.}~\bibnamefont{Dhayalb}}, \bibnamefont{and}
  \bibinfo{author}{\bibfnamefont{K.}~\bibnamefont{Venkataratnamc}},
  \bibinfo{journal}{Eur. Phys. J. C} \textbf{\bibinfo{volume}{82}},
  \bibinfo{pages}{333} (\bibinfo{year}{2022}).

\bibitem[{\citenamefont{Manzano}(2020)}]{Manzano_2020}
\bibinfo{author}{\bibfnamefont{D.}~\bibnamefont{Manzano}},
  \bibinfo{journal}{{AIP} Advances} \textbf{\bibinfo{volume}{10}},
  \bibinfo{pages}{025106} (\bibinfo{year}{2020}),
  \urlprefix\url{https://doi.org/10.1063%2F1.5115323}.

\bibitem[{\citenamefont{Seahra}(2023)}]{youtube}
\bibinfo{author}{\bibfnamefont{S.}~\bibnamefont{Seahra}},
  \emph{\bibinfo{title}{Classical-quantum approximation for bipartite systems}}
  (\bibinfo{year}{2023}),
  \urlprefix\url{https://youtube.com/playlist?list=PLDRdFkFA2uqfp-CGfbhrfZDmfoC-Ng1xR}.

\bibitem[{\citenamefont{Diosi and Halliwell}(1998)}]{Diosi:1997mt}
\bibinfo{author}{\bibfnamefont{L.}~\bibnamefont{Diosi}} \bibnamefont{and}
  \bibinfo{author}{\bibfnamefont{J.~J.} \bibnamefont{Halliwell}},
  \bibinfo{journal}{Phys. Rev. Lett.} \textbf{\bibinfo{volume}{81}},
  \bibinfo{pages}{2846} (\bibinfo{year}{1998}), \eprint{quant-ph/9705008}.

\bibitem[{\citenamefont{Husain and Singh}(2019)}]{Husain:2018fzg}
\bibinfo{author}{\bibfnamefont{V.}~\bibnamefont{Husain}} \bibnamefont{and}
  \bibinfo{author}{\bibfnamefont{S.}~\bibnamefont{Singh}},
  \bibinfo{journal}{Phys. Rev. D} \textbf{\bibinfo{volume}{99}},
  \bibinfo{pages}{086018} (\bibinfo{year}{2019}), \eprint{1811.03673}.

\bibitem[{\citenamefont{Husain and Singh}(2021)}]{Husain:2021rnf}
\bibinfo{author}{\bibfnamefont{V.}~\bibnamefont{Husain}} \bibnamefont{and}
  \bibinfo{author}{\bibfnamefont{S.}~\bibnamefont{Singh}},
  \bibinfo{journal}{Phys. Rev. D} \textbf{\bibinfo{volume}{104}},
  \bibinfo{pages}{124048} (\bibinfo{year}{2021}), \eprint{2109.12752}.

\bibitem[{\citenamefont{Husain and Pawlowski}(2012)}]{Husain:2011tk}
\bibinfo{author}{\bibfnamefont{V.}~\bibnamefont{Husain}} \bibnamefont{and}
  \bibinfo{author}{\bibfnamefont{T.}~\bibnamefont{Pawlowski}},
  \bibinfo{journal}{Phys. Rev. Lett.} \textbf{\bibinfo{volume}{108}},
  \bibinfo{pages}{141301} (\bibinfo{year}{2012}), \eprint{1108.1145}.

\bibitem[{\citenamefont{Husain and Singh}(2020)}]{Husain:2019nym}
\bibinfo{author}{\bibfnamefont{V.}~\bibnamefont{Husain}} \bibnamefont{and}
  \bibinfo{author}{\bibfnamefont{S.}~\bibnamefont{Singh}},
  \bibinfo{journal}{Class. Quant. Grav.} \textbf{\bibinfo{volume}{37}},
  \bibinfo{pages}{15LT01} (\bibinfo{year}{2020}), \eprint{1907.03776}.

\bibitem[{\citenamefont{Zurek}(2003)}]{Zurek:2003zz}
\bibinfo{author}{\bibfnamefont{W.~H.} \bibnamefont{Zurek}},
  \bibinfo{journal}{Rev. Mod. Phys.} \textbf{\bibinfo{volume}{75}},
  \bibinfo{pages}{715} (\bibinfo{year}{2003}), \eprint{quant-ph/0105127}.

\bibitem[{\citenamefont{Choptuik}(1993)}]{Choptuik:1992jv}
\bibinfo{author}{\bibfnamefont{M.~W.} \bibnamefont{Choptuik}},
  \bibinfo{journal}{Phys. Rev. Lett.} \textbf{\bibinfo{volume}{70}},
  \bibinfo{pages}{9} (\bibinfo{year}{1993}).

\bibitem[{\citenamefont{Husain and Terno}(2010)}]{Husain:2009vx}
\bibinfo{author}{\bibfnamefont{V.}~\bibnamefont{Husain}} \bibnamefont{and}
  \bibinfo{author}{\bibfnamefont{D.~R.} \bibnamefont{Terno}},
  \bibinfo{journal}{Phys. Rev. D} \textbf{\bibinfo{volume}{81}},
  \bibinfo{pages}{044039} (\bibinfo{year}{2010}), \eprint{0903.1471}.

\bibitem[{\citenamefont{Layton et~al.}(2022)\citenamefont{Layton, Oppenheim,
  and Weller-Davies}}]{Layton:2022sku}
\bibinfo{author}{\bibfnamefont{I.}~\bibnamefont{Layton}},
  \bibinfo{author}{\bibfnamefont{J.}~\bibnamefont{Oppenheim}},
  \bibnamefont{and}
  \bibinfo{author}{\bibfnamefont{Z.}~\bibnamefont{Weller-Davies}}
  (\bibinfo{year}{2022}), \eprint{2208.11722}.

\bibitem[{\citenamefont{Bojowald and Skirzewski}(2006)}]{Bojowald:2005cw}
\bibinfo{author}{\bibfnamefont{M.}~\bibnamefont{Bojowald}} \bibnamefont{and}
  \bibinfo{author}{\bibfnamefont{A.}~\bibnamefont{Skirzewski}},
  \bibinfo{journal}{Rev. Math. Phys.} \textbf{\bibinfo{volume}{18}},
  \bibinfo{pages}{713} (\bibinfo{year}{2006}), \eprint{math-ph/0511043}.

\bibitem[{\citenamefont{Bayta\c{s} et~al.}(2019)\citenamefont{Bayta\c{s},
  Bojowald, and Crowe}}]{Baytas:2018gbu}
\bibinfo{author}{\bibfnamefont{B.}~\bibnamefont{Bayta\c{s}}},
  \bibinfo{author}{\bibfnamefont{M.}~\bibnamefont{Bojowald}}, \bibnamefont{and}
  \bibinfo{author}{\bibfnamefont{S.}~\bibnamefont{Crowe}},
  \bibinfo{journal}{Phys. Rev. A} \textbf{\bibinfo{volume}{99}},
  \bibinfo{pages}{042114} (\bibinfo{year}{2019}), \eprint{1811.00505}.

\bibitem[{\citenamefont{Heller}(1975)}]{doi:10.1063/1.430620}
\bibinfo{author}{\bibfnamefont{E.~J.} \bibnamefont{Heller}},
  \bibinfo{journal}{The Journal of Chemical Physics}
  \textbf{\bibinfo{volume}{62}}, \bibinfo{pages}{1544} (\bibinfo{year}{1975}),
  \eprint{https://doi.org/10.1063/1.430620},
  \urlprefix\url{https://doi.org/10.1063/1.430620}.

\end{thebibliography}

\end{document}